\documentclass[conference, compsoc]{IEEEtran}

\IEEEoverridecommandlockouts

\usepackage{hyperref}

\usepackage{tikz}
\usepackage{tikz-network}
\usepackage{algorithmic}
\usepackage[linesnumbered]{algorithm2e}
\usepackage{subcaption}
\usetikzlibrary{arrows,chains,positioning,scopes,quotes,bending,calc,intersections}
\usepackage{breqn}
\usepackage{url}
\usepackage{balance}

\tikzset{
	block/.style={draw,minimum width=1em,minimum height=1em,align=center,fill=blue!30},
	arrow/.style={->},
	line/.style={-}
}

\newcommand{\todobox}[3]{%
	\colorbox{#1}{\textcolor{white}{\sffamily\bfseries\scriptsize #2}}%
	~\textcolor{red}{#3} %
	\textcolor{#1}{$\triangleleft$}%
}
\newcommand{\new}[1]{\textcolor{black}{#1}}
\newcommand{\todo}[1]{\todobox{red}{TODO}{#1}}

\newcommand{\ct}[1]{\todobox{magenta}{ct}{#1}}

\newcommand{\nn}{\textbf{N}} %neighbourgs list
\newcommand{\mup}[2]{\Theta_{#1}^{#2}}

\newcommand{\isom}[2]{\mathring{\Theta}_{#1}^{#2}}

\newcommand{\parabf}[1]{\vspace{1mm}\noindent\textbf{#1}}
\newcommand{\parait}[1]{\vspace{1mm}\noindent\textit{#1}}

\newcommand{\oot}{\frac{1}{2}}
\newcommand{\loss}{\textbf{L}}
\newcommand{\ie}{{i.e.,~}}
\newcommand{\eg}{{e.g.,~}}
\newcommand{\attck}{\mathcal{A}}
\newcommand{\victm}{v}
\newcommand\myeq{\mkern1.5mu{=}\mkern1.5mu}
\newcommand{\TT}[1]{``\textit{#1}''}
\newcommand{\etal}{{et~al.}}

\newcommand{\jaggi}{Koloskova2020Decentralized}

\newcommand{\dap}{\mathcal{A}^{\text{DL}}}
\newcommand{\daa}{\bar{\mathcal{A}}^{\text{DL}}}

\newcommand{\fup}{\mathcal{A}^{\text{FL}}_{\text{user}}}
\newcommand{\fsp}{\mathcal{A}^{\text{FL}}_{\text{server}}}
\newcommand{\fua}{\bar{\mathcal{A}}^{\text{FL}}_{\text{user}}}
\newcommand{\fsa}{\bar{\mathcal{A}}^{\text{FL}}_{\text{server}}}

\newcommand{\advf}[2]{\mathit{p_{leak}}(#1,#2) }

\begin{document}
\title{On the (In)security of\\ Peer-to-Peer Decentralized  Machine Learning} 

\author{\IEEEauthorblockN{Dario Pasquini}
	\IEEEauthorblockA{SPRING Lab; EPFL, Switzerland \\
		dario.pasquini@epfl.ch}
	\and
	\IEEEauthorblockN{Mathilde Raynal}
	\IEEEauthorblockA{SPRING Lab; EPFL, Switzerland \\
		mathilde.raynal@epfl.ch}
	\and
	\IEEEauthorblockN{Carmela Troncoso}
	\IEEEauthorblockA{SPRING Lab; EPFL, Switzerland \\
	carmela.troncoso@epfl.ch}
}

\maketitle
\footnotetext[1]{This paper appears in the proceedings of the 44nd IEEE Symposium on Security and Privacy S\&P 2023.}%

\begin{abstract}
In this work, we carry out the first, in-depth, privacy analysis of Decentralized Learning---a collaborative machine learning framework aimed at addressing the main limitations of federated learning. We introduce a suite of novel attacks for both passive and active decentralized adversaries. We demonstrate that, contrary to what is claimed by decentralized learning proposers, decentralized learning does not offer any security advantage over federated learning. Rather, it increases the attack surface enabling \textit{any} user in the system to perform privacy attacks such as gradient inversion, and even gain full control over honest users' local model. We also show that, given the state of the art in protections, privacy-preserving configurations of decentralized learning require fully connected networks, losing any practical advantage over the federated setup and therefore completely defeating the objective of the decentralized approach.
\end{abstract}


\section{Introduction}
Collaborative machine learning is gaining traction as a way to train machine learning models while respecting the privacy of users' local training dataset~\cite{fedavg}.
There are two main approaches to collaborative machine learning: \emph{federated learning}~\cite{fedavg} and \emph{decentralized learning}~\cite{can_dec}.

In \emph{federated learning}, the iterative learning process is orchestrated by a central parameter server. This server intermediates communication in-between users and maintains the global state of the system.
Such central component can become a communication bottleneck as the number of users grows, and, due to its full control on the learning process, can perform a number of security and privacy attacks on users~\cite{papernotclone, fowl2022robbing, evading_sec, wen2022fishing, invg2, invg1}.
\par

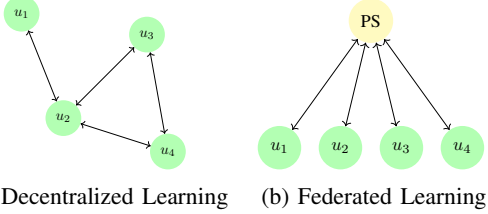
\begin{figure}[t]
	\centering
	\begin{subfigure}{.2\textwidth}
			\centering
		\resizebox{0.7\textwidth}{!}{%
			\begin{tikzpicture}
		
			\tikzstyle{user} = [fill=green!30, circle,  text width=5mm, align=center]
			\node[user] (1) {$u_1$}; 
			\node[user, yshift=-2.5cm, xshift=1cm] (2) {$u_2$}; 
			\node[user, yshift=-3.3cm, xshift=3.5cm] (3) {$u_4$}; 
			\node[user, xshift=3cm, yshift=-.5cm] (4) {$u_3$}; 
			\draw[<->] (1) to (2);
			\draw[<->] (2) to (3);
			\draw[<->] (2) to (4);
			\draw[<->] (4) to (3);
		\end{tikzpicture}
		}
	\caption{Decentralized Learning}
	\label{fig:dl}
	\end{subfigure}~\begin{subfigure}{.2\textwidth}
			\centering
		\resizebox{0.87\textwidth}{!}{%
		\begin{tikzpicture}
			\tikzstyle{user} = [fill=green!30, circle,  text width=5mm, align=center]
			\node[user, fill=yellow!30] (1) {PS}; 
			
			\node[user, yshift=-2.5cm, xshift=-1.8cm] (2) {$u_1$}; 
			\node[user, yshift=-2.5cm, xshift=-0.6cm] (3) {$u_2$}; 
			\node[user, yshift=-2.5cm, xshift=+0.6cm] (4) {$u_3$}; 
			\node[user, yshift=-2.5cm, xshift=+1.8cm] (5) {$u_4$}; 
				
			\draw[<->] (2) --  (1);
			\draw[<->] (3) -- (1);
			\draw[<->] (4) -- (1);
			\draw[<->] (5) -- (1);
		\end{tikzpicture} 
	}
		\caption{Federated Learning}
		\label{fig:fl}
	\end{subfigure}
	\caption{Schematic representation of the decentralized learning and federated learning protocols. \TT{PS} stands for Parameter Server.}
	\label{fig:cml}
\end{figure}

\emph{Decentralized machine learning}, also known as fully-decentralized machine learning, peer-to-peer machine learning, or gossip learning, aims at addressing these issues by \emph{eliminating the central server}. Instead, the learning takes place via peer-to-peer communication, see Figure~\ref{fig:cml}.
Proponents of decentralized learning argue that decentralization: \emph{(a) }reduces bandwidth consumption, \emph{(b)} provides users with control on who they communicate with, and \emph{(c)} increases privacy of users in the system by eliminating the central server. 
A large body of theoretical studies, empirical evaluations, and model extensions attest to \emph{(a)} and \emph{(b)}~\cite{dp2, hegedHus2019gossip, hu2019decentralized, \jaggi, pmlr-v119-koloskova20a, pmlr-v97-koloskova19a, lalitha2019peer, lalitha2018fully, 9086196, pappas2021ipls, pei2021decentralized, braint, vogels2021relaysum, can_dec, xing2020decentralized, ying2021exponential, ying2021bluefog, 9123209, he2019central}. However, these works do not assess \emph{(c)}.
%%, i.e., the impact of decentralization on users' privacy
Either they state that decentralized learning offers a higher level of privacy compared to the centralized approach without any evidence~\cite{dp2, vogels2021relaysum, he2019central, marfoq2021federated, 9850408}, or simply do not provide any privacy argument~\cite{hu2019decentralized,  pmlr-v97-koloskova19a, lalitha2019peer, lalitha2018fully, can_dec, xing2020decentralized, ying2021exponential, 9123209, pmlr-v162-dai22b}.
\par

In this work, we thoroughly evaluate the privacy offered by decentralized learning, against both passive and active adversaries.
%We show that there is a strong relation between the underlying communication topology of the decentralized system and the level of privacy that can be achieved by any decentralized learning protocol. 
%We characterize the factors of decentralization that lead to privacy leakage.
We propose novel attacks that demonstrate that in a decentralized setting: \textbf{(1)} A passive adversarial user can successfully (i) \emph{infer membership of samples} with better accuracy than in the federated setting and (ii) \emph{perform reconstruction attacks} on the training set of arbitrary honest users. \textbf{(2)} An active adversarial user can (i) \emph{influence the update process of honest users} in arbitrary ways and (ii) \new{and perform effective privacy attacks such active gradient inversion~\cite{wen2022fishing, papernotclone}.}
\par

We show that these attacks are possible because decentralization \emph{increases} the inference power of users, as well as their influence on other users' status. 
\textbf{This leads to adversarial users in decentralized learning becoming as powerful as the parameter server in federated learning.}

We study the effectiveness of mitigation techniques against our attacks. Our findings are two-fold.
We first show that the potential protections against our attacks are in conflict: trying to eliminate one leakage factor augments another, leaving little space to eventually develop truly privacy-preserving decentralized learning. 
%% they are actually unable to prevent all of our attacks simultaneously, and that different system configurations can only trade-off protection against one attack for vulnerability against another --breaking (3).
Second, we show that, while it is possible to reduce the attack surface resulting from decentralization, e.g., by changing the underlying topology and using expensive aggregation techniques, the privacy provided by decentralized learning will always be less (or equal at best) than the one provided by the federated counterpart. This invalidates claim \emph{(c)}.
Moreover, achieving protection comparable to federated learning comes at a huge cost in efficiency that destroys any remaining advantage of decentralization, invalidating claims \emph{(a)} and \emph{(b)}.
%Equivalently, users participating in a DL protocol are always going to be more vulnerable to privacy attacks than federated users, and will pay a higher cost.
%%The introduced attacks are general and move orthogonal to the current line of research, laying the foundation for a sound and more principled privacy analysis of decentralized machine learning systems.
%%{decentralization does not offer any privacy advantage over more the federated approach}. 

In summary, in contrast to common belief, in collaborative learning decentralization does not increase privacy. Instead, it inherently boosts the capabilities of privacy attackers and thus, {decentralized learning tends to degrade users' privacy compared to the federated setting}.
This disadvantage cannot be overcome by existing mitigations without sacrificing the gains of decentralization over federated learning. 
%%While it is possible to configure the underlying topology and use expensive aggregation techniques so that the privacy is comparable, we highlight that it would destroy any advantage of using decentralized learning over the federated alternative.
%While it is possible to reach the same level of privacy as FL, it would imply using a complete topology and[...] which would defeat the benefits coming from decentralization}, decentralized learning has to be built on a fully-connected communication topology and use expensive aggregation techniques, essentially destroying any advantage of decentralization over the federated approach.

%% CT - removed, nothing new with respect to the above. Am I missing something?
%More importantly, our attacks demonstrate that, despite decentralization, every user in decentralized learning is allowed to become as powerful as an adversarial parameter server in federated learning whether the underlying communication topology allows it. However, unlike federated learning, decentralized learning enables multiple users in the system to reach the same adversarial capabilities of a central server simultaneously, multiplying adversarial power rather than distributing it. Eventually, these results bring us to fundamentally question the applicability of decentralized learning in the privacy-preserving collaborative machine learning setting.

Our work makes the following contributions: 

\vspace{1mm} \noindent $\bullet$ We provide the first in-depth privacy evaluation of decentralized learning, in which users exchange updates in a peer-to-peer fashion. We characterize the key factors that contribute to privacy leakage within the protocol. We explore how these elements interact with each other to comprehend their effects on the privacy of decentralized users.

\vspace{1mm} \noindent $\bullet$ We introduce a suite of novel privacy attacks designed specifically for the decentralized setting, covering both the passive and active security models. Our attacks demonstrate that decentralized users can reach the same adversarial capabilities of a parameter server in federated learning.

\vspace{1mm} \noindent $\bullet$ We show that the capabilities that decentralization grants to adversaries results in a degradation of users' privacy compared to federated learning. Moreover, we show that no existing protection can bridge this privacy gap without eliminating the advantages of decentralization.

\iffalse
\paragraph*{Organization}
We start in Section~\ref{sec:cl}, where we briefly survey collaborative machine learning and define the notation we use within the paper. Section~\ref{sec:setup} follows by formalizing our evaluation setup. In Section~\ref{sec:localgen_and_systemk}, we abstract the main sources of privacy leakage of decentralized learning protocols. In Section~\ref{sec:hbc}, we analyze the security of the decentralized protocol against passive (honest-but-curious) adversaries. Section~\ref{sec:active} extends the analysis to active (malicious) adversaries. Section~\ref{sec:opm} surveys  the main, privacy-related, open problems in decentralized learning.
Section~\ref{sec:conclusion} concludes the paper, with Appendices containing additional material. In the paper, background and analysis of previous works are provided, when necessary, within the respective sections.
\fi

\section{Preliminaries and  Setup}
\label{sec:setup}
%Our privacy evaluation of decentralized learning is designed as a comparative analysis with federated learning given equivalent setup. In this section, we formalize the privacy notion we evaluate in this paper. We also introduce the decentralized and federated learning protocols targeted by our analysis and describe how we parametrize them in our evaluation. We provide more details about our setup in Appendix~\ref{app:setup}.
In this section, we provide the necessary background on decentralize learning, we introduce the notation used in the paper, and we define our evaluation setup. 
We start in Section~\ref{sec:cml} by covering the Decentralized and Federated Learning protocols. In Section~\ref{sec:privacy}, we formalize the privacy attacks upon which we build our analysis and comparison.

\subsection{Collaborative Machine Learning}
\label{sec:cml}
We define Collaborative Machine Learning (CML) as the class of learning algorithms that enable a set $V$ of $n$ distributed \textit{users} to train a shared model $f$ defined by a set of parameters $\Theta$. Each client~$v \in V$ participates to the protocol with a local training dataset $X_v$. The local training sets must be kept private during the training. Thus, not only the training data cannot be shared directly, but the parties involved in the CML training must not be able to learn anything about each other's training set from their interactions, besides the information obtainable from the final model. 
Any additional information that an adversary can learn from observing or participating in the collaborative training constitutes a \textit{privacy leakage}.
  %In this paper, we limit our privacy definition to the privacy of this training set, and not other attributes that could be learned about users (e.g., location~\cite{locationprivacy} or device type).

In CML, the model $f$ is typically trained using a distributed version of Stochastic Gradient Descent (SGD), where users iteratively propagate \textit{model updates}. These updates are intermediate outputs of each user's local optimization process such as gradients or updated parameters~\cite{fedavg, can_dec}, obtained after one (or more) local SGD step computed using their training set. A CML protocol defines the way in which model updates are shared among the parties.

\begin{algorithm}[b]
	\SetAlgoLined

	\KwData{\footnotesize Initial parameters: $\Theta_v^0$, local training set: $X_v$}
	\scriptsize
	%\tcc{Select neighbors nodes}
	%$\nn(v)  = \texttt{pick\_neighbors}(v)$
	\For{$t \in [0, 1, \dots]$}{
		\tcc{Local optimization step}
		$\xi_v^{t}\sim X_v$\;
		$\mup{v}{t+\oot} \myeq \mup{v}{t}-\eta \nabla_{ \mup{v}{t}}(\xi_v^{t}, \mup{v}{t})$\;
		\tcc{Communication with neighbors}
		\For{$u \in \nn(v) / \{v\}$}{
			\texttt{send} $\mup{v}{t+\oot}$ \texttt{to} $u$\;
			\texttt{receive} $\mup{u}{t+\oot}$ \texttt{from} $u$\;
		}
		\tcc{Model updates aggregation}
		$\mup{v}{t+1} \myeq \frac{1}{|\nn(v)|}\underset{u \in \nn(v)}{\Sigma} \mup{u}{t+\oot}$\;
	}
	\caption{Training protocol for every decentralized user~$v \in V$.}
	\label{alg:dl}
\end{algorithm}
\subsubsection{Decentralized Learning}
\par

In Decentralized Learning (DL)~\cite{can_dec}, users connect to each other in a peer-to-peer fashion. During the protocol, every user $v \in V$ connects with a non-empty set of other users which we call \textit{neighbors}, $\nn(v)$, where~$v \in \nn(v)$. This set is typically small compared to the set of all users, and can either be fixed at the beginning of the execution or dynamically change across iterations. 
We model the communication links shared among users as an \textit{undirected} graph~$G \myeq \{V, \cup_{v\in V} \nn(v) \}$, where users are nodes and communication links are edges (see Figure~\ref{fig:dl}). Hereafter, we refer to this graph as \textit{communication topology} or \textit{topology} for short. 

During the DL training, users share model updates only with their set of neighbors~$\nn(v)$.
% which is typically only a small subset of the whole available population $V$. 
Through a gossip-like propagation mechanism, DL protocols ensure that users indirectly receive and benefit from the model updates produced by non-neighbor users. 

In contrast to federated learning (see below) where connections are fixed and proxied by a server, DL protocols tend to not constrain the users' choice of neighbors. This flexibility is claimed as an advantage of DL as it enables users to cluster according to arbitrary criteria, e.g., data similarity or computational capabilities~\cite{bellet2018personalized, dladaptive}. In the general case, therefore, users participating in DL protocols can arbitrarily pick their neighbor nodes and autonomously define the underlying topology~$G$. %\todo{We note that a large body of work~\cite{} aimed to study the effect of fixed topologies in DL do exist. However, these works do not discuss or propose solutions on how to enforce topological constraints at deployment time, making their application on the real-world dubious.} 
\new{It is worth mentioning that a significant body of research has been dedicated to examining the effects of fixed and pre-defined topologies in decentralized learning. However, these studies primarily focus on theoretical aspects and do not provide insights into the practical challenges of enforcing topological constraints during the deployment of the system.}
%\paragraph{DL on pre-defined topologies}
%\todo{While there is extensive research on the effects of fixed topologies in decentralized learning~\todo{\cite{}}, there has been a dearth of discourse and viable solutions regarding the implementation of such topological restrictions during deployment.\footnote{Excluding the use of a trusted orchestrator or central server, which would essentially nullify the benefits of decentralized learning.} Without a viable solution for constraining user connectivity, DL protocols that rely on fixed topologies may prove challenging to apply in practical scenarios.}
\par

%\paragraph{\textbf{The protocol}}
In this paper, we target our privacy analysis on the \texttt{D-PSGD} protocol proposed by Lian~\etal~\cite{can_dec}. This protocol provides the same core functionality and properties as the bulk of DL protocols in the literature~\cite{bellet2018personalized, el2020collaborative, hu2019decentralized, \jaggi, lalitha2019peer, lalitha2018fully, 9086196, Shokri15, xing2020decentralized, Yin18, yuan2021defed}. Thus, it is representative of the decentralized learning state-of-the-art. Its similarity to \texttt{FedAVG} (see Section~\ref{sec:fl}) permits a direct comparison between decentralized and federated learning.

In \texttt{D-PSGD}, summarized in Algorithm~\ref{alg:dl}, $n$ users start with common model parameters~$\mup{}{0}$ and iterate over the following three steps until a stop condition is met:
\begin{enumerate}
\item \textit{Local training:} Users sample a mini-batch $\xi$ from their  (private) local training set and apply gradient descent on their local view of the model parameters. This results in an intermediate model~$\mup{v}{t+\oot}$ that we refer to as \textit{model update}.

\item \textit{Communication:} Users share their model updates $\mup{v}{t+\oot}$ with their neighbors, and receive their neighbors' updates (line 4 in Algorithm~\ref{alg:dl}). 

\item \textit{Aggregation:} Users compute their new model by aggregating \textit{all} their neighbor's updates with their local one.
The aggregation is the average of the model parameters.\footnote{We assume model updates to have equal weights, though, in general, the aggregation of line $8$ can be expressed via a weighted sum.}
\end{enumerate}

In contrast to the federated setting, at each round in DL, \textbf{users' local set of parameters can be arbitrarily different}. After a suitable number of communication rounds, users find \textit{consensus} on a global state; that is, users' local parameters become equal. In this paper, we measure how close users are from reaching consensus using the \textit{consensus distance}~$C$. This distance is computed as the pairwise discrepancy among local parameters at time~$t$:%\ct{would be nice to include here a layman
\begin{equation}
		\label{eq:consesus}
		C(t) =\frac{\sum_{v\in V} \sum_{u\in V / \{v\}} \|\mup{v}{t}-\mup{u}{t}\|^2}{|V|^2-|V|}.
\end{equation}
Intuitively, large values of $C$ indicate that there is a large discrepancy among users' local parameters, whereas small values indicate that users have similar local models. We say that the system has found consensus when $C$ approaches zero.\looseness=-1
\par
The \texttt{D-PSGD} protocol does not specify how users select their neighbors nor how they agree on a training setup (including initial parameters of the local models~$\mup{}{0}$).
Following the most relevant works in decentralized learning~\cite{Blanchard17, DamaskinosMGPT18, el2020collaborative, he2022byzantine, \jaggi, vogels2021relaysum, can_dec, Yin18}, we assume those decisions happened during a honest setup and focus on the fixed communication graph setting; i.e., the graph $G$ does not change over time and users do not drop-off in-between rounds. Nevertheless, our attacks and the result of our analysis apply also to cases in which the topology changes dynamically and users initialization is arbitrary.

We consider three communication topologies:

\parait{\textbf{Torus}:} A regular topology where every user is connected with four other nodes. This topology represents the best-case scenario for decentralized learning given its good mixing properties and higher spectral gap that allow for fast convergence and efficient communication~\cite{\jaggi}. We consider torus graphs with different number of nodes and we refer to them as \textit{torus-$n$}, where $n$ is the number of nodes. %; for instance,  \textit{torus-36}.

\parait{\textbf{Random regular}:} A regular topology in which all nodes have $d$ random neighbors. Random regular topologies enable us to analyze the impact of the density of connections ($d$) on the privacy of DL. We refer to these graphs as \textit{regular-$(n, d)$}, where $n$ is the number of nodes and $d$ the density.

\parait{\textbf{Davis Southern women social network}:} An unstructured topology which represents a more realistic case, e.g., users communicating in a cross-device setting, used by Koloskova et al~\cite{\jaggi}. This social network has $32$ users with diverse degree. The average degree is $5.74$. We refer to this topology as \textit{social-32}.
\par

\new{We provide additional results for different topologies and configurations in Appendix~\ref{app:add_res}.}\\
 
\subsubsection{Federated Learning}
\label{sec:fl}
In Federated Learning (FL)~\cite{fedavg} users perform the distributed training process with the support of a central \textit{parameter server} that aggregates and synchronizes model updates among users (Figure~\ref{fig:fl}).
 %The server manages users' training process and maintains a master copy of the global model's parameters within the training process.
%In the initial setup phase, users and server choose a training task and define a machine learning model.
%This model is initialized and hosted by the parameter server.
At each iteration, users download the global model from the server and locally apply one or more local training steps. Users send their model updates back to the server. The server aggregates these updates and applies the result to the global parameters, completing a training round.\looseness=-1

%as the server has linear communication cost in the number of users.
Compared to DL, there is no direct communication among users in FL. All the communication takes place through the server. 
Following our notation, we write that every federated user $v$ has neighbors $\nn(v)\myeq \{PS\}$, where $PS$ stands for parameter server. 
As main consequence, users, by design, do not have access to each other's model updates during the training; they can only access the aggregated model update (the average of users' model updates) sent to them by the parameter server. 
Another consequence of this design is that, at each round, users always share the same set of local parameters for the model $f$. We refer to those parameters as the \emph{global parameters}.
%\label{sec:fl}

In our evaluation, we take the Federated Averaging protocol \texttt{FedAVG}~\cite{fedavg} as representative of FL algorithms.
Motivated by the small amount of users assumed in the decentralized learning literature ($n<100$)~\cite{can_dec, \jaggi, lalitha2019peer, lalitha2018fully, pappas2021ipls, braint, vogels2021relaysum}, we evaluate a cross-silo federated setting~\cite{advfl}, where all users participate in each training round.
This setting represents a lower bound for privacy compared to a cross-device federated setting in which only a subset of users participate in every round.
We also force users' local training step to be computed on a single, random batch per round to match \texttt{D-PSGD}. 
Indeed, under this configuration, \texttt{FedAVG} becomes functionally equivalent to \texttt{D-PSGD} where the topology is fully connected (\ie all users are connected to each other).
This configuration enables a fair comparison between decentralized and federated approaches. \looseness=-1

\subsubsection{Datasets and architecture}
In our experiments, we use the CIFAR-10, CIFAR-100~\cite{cifar10}, and STL10~\cite{pmlr-v15-coates11a} datasets. As in~\cite{\jaggi}, we consider the users' local training sets to be uniformly distributed among users; i.e., every user gets an uniformly sampled (without replacement) fraction~$\frac{1}{n}$ of the training set (where $n\myeq |V|$ is the number of users in the system). We use a ResNet20~\cite{resnet} architecture, with the same hyper-parameters for both the decentralized and federated settings. %For completeness, we report results obtained with a shallower architecture in Appendix~\ref{app:shallow}. 
For each comparison we also consider the same number of users, and the same local training set partition for the decentralized and federated settings. We provide more details about our setup in Appendix~\ref{app:setup}.

%\mr{Moved here from below, need to merge in text + add smthing like "By doing this, the only difference from our two representatives is whether they are centralized or not."} We use the same model, hyper-parameters, number of users, and local training set partition for both.  

\subsection{Privacy attacks} 
\label{sec:privacy}

We evaluate the privacy offered by decentralized learning using two attacks: \textit{membership inference}, in which an adversary learns whether a target sample is in the training set of a user; and \textit{gradient inversion} in which the adversary can reconstruct  samples in the training set of a user. \looseness=-1

\subsubsection{Membership Inference Attacks}
\label{sec:mia_emp}

%The most atomic privacy vulnerability of a protocol with respect to the training dataset is the one of membership inference.
In a Membership Inference Attack (MIA)~\cite{shokri2017membership}, the adversary tries to infer whether a sample is part, or not, of the training dataset. 
To make their guess, the adversary can use all information available to them: they can look at the model updates or query the trained model.
Vulnerability (or equivalently robustness) to MIA is a good privacy beacon, as membership inference connects to almost all other privacy attacks, e.g., attribute inference attacks can be reduced to MIA~\cite{yeom2018privacy, smi21}.
As a result, capturing privacy through MIA is a common choice in the CML literature~\cite{song2021systematic, yeom2018privacy, long2018understanding, shokri2017membership, imias0, imias1, gasc, melis}.

%In this paper, we quantify the privacy risk using \textit{Membership Inference Attacks} (MIAs).
\par

In our evaluation, we measure a learning protocol's vulnerability to MIA through the success of a simple metric-based attack. 
%We use as metric the \TT{label-informed} entropy introduced in \cite{song2021systematic}. 
Formally, given a set of model parameters~$\Theta$, a local training set $X$, and a test set $\mathcal{O}$ s.t. $X\cap \mathcal{O}=\emptyset$ and $|X|\myeq |\mathcal{O}|\myeq m$, we estimate membership vulnerability as the accuracy of the membership inference attack over the sets $X$ and $\mathcal{O}$: %\ct{here we call the privacy risk M, but we never use this symbol anymore. We should add it to the labels of the graphs (or think about removing the notation)}:
\begin{gather}
	M(\Theta, X, \mathcal{O})=  \frac{1}{2 \cdot m} \sum_{i=0}^{m-1}  [ MIA_\Theta(X_i) + \neg MIA_\Theta(\mathcal{O}_i) ]	\label{eq:mia} \\
	\text{with} \quad MIA_\Theta(x) =  \xi (f_\theta(x)) < \rho,
\end{gather}
where $\xi$ is the \TT{label-informed} entropy~\cite{song2021systematic} and $\rho$ is the optimal threshold. For convenience, when presenting our results we subtract the random guessing baseline $(0.5)$ from the accuracy so that the results we report are centered in $0$.
%Intuitively, $M$ measures the average margin that separates members from non-members instances. Higher values of $M$ indicate that the model $f$ behaves very differently for members and non-members, and thus the attacker can easily identify training set samples. Low values of $M$ indicate that the model $f$ behaves comparably on member and non-member inputs, hindering membership inference.
We choose to rely on this simple attack because it allows us to quantify the difference in membership vulnerability between DL and FL protocols at a low computational cost (see Section~\ref{sec:hbc}). It would be straightforward to run our evaluation with more complex and effective inference attacks including white-box attacks~\cite{gasc, imias1, imias0}. This would likely increase the vulnerability estimations in DL and FL, but we do not expect that it would significantly affect the difference in between the estimation in each setting.

\subsubsection{Gradient inversion}

Gradient inversion attacks exploit the observation that the gradient produced by one (or more) SGD steps is just a smooth function of the training data used to compute it. Thus, an attacker capable of accessing users' model updates during a CML protocol may be able to invert them and fully or partially recover the underlying users' private data~\cite{invg2, jeon2021gradient, nvidiagi, invg1}.
%\new{In the federated setting, a passive server has access to the gradient $\nabla_{\Theta}\loss(x, f_\Theta)$ and the neural network $f_\Theta$, the server can recover the (private) input $x$ used to produce the gradient.\footnote{For a supervised task, the symbol $x$ captures both input and label}
%	Typically, this is achieved by searching for a set of input instances that generates a gradient similar to the one produced by the user. Using the inherent smoothness of the neural model, this search can be solved as a second-order optimization~\cite{invg2, jeon2021gradient, nvidiagi, invg1}. 
The quality of this inversion process is heavily dependent on the configuration used to compute the gradient, the number of trainable parameters of the network and batch size being the most impactful factors.

% \ct{why is the following sentence relevant?}As a matter of fact, gradient inversion attacks are likely to fail when applied  on realistic collaborative learning deployments such as settings with large batch sizes~\cite{invg1, invg2}. 
%A more recent line of research demonstrated that the effect of gradient inversion can be greatly magnified in both reconstruction quality and applicability when the attacker has some control on the parameters used to compute the gradient. The attacker can inject the victim's network with maliciously crafted parameters that force the computed gradient to artificially memorize more information than intended about the input batches~\cite{fowl2022robbing, wen2022fishing, papernotclone}.
	
%We adapt two instances of gradient inversion (designed for FL) to the decentralized learning framework for the passive and active setting respectively. For the passive attack, we rely on the optimization-based approach proposed by Geiping~\etal~\cite{invg2}. For the active setting, we consider the attack originally proposed by Boenisch~\etal~\cite{papernotclone}; however, our technique seamlessly extends to others~\cite{wen2022fishing}.

We adapt two instances of gradient inversion (originally designed for FL) to the DL framework: the \new{passive} optimization-based approach proposed by Geiping~\etal~\cite{invg2} and the \new{active} attack proposed by Boenisch~\etal~\cite{papernotclone}.
Our setup seamlessly extends to other attacks~\cite{wen2022fishing}.

\section{The generalization and  knowledge trade-off in Decentralized Learning}
\label{sec:localgen_and_systemk}

%Decentralization seems a natural solution to alleviate the communication constraints and the trust requirements in federated learning. However, these advantages come at a cost. 
In this section, we characterize two main byproducts resulting from decentralizing FL. We refer to them as: \textit{local generalization} and adversarial \textit{system knowledge}. We demonstrate that these properties alone prevent honest users' from reaching any meaningful level of privacy in DL.

\subsection{Local generalization}
\label{sec:localg}
Generalization is pivotal to protect the privacy of the training set against attacks based on the model behavior. While well-generalized models may still leak information about the underlying training set~\cite{yeom2018privacy, long2018understanding}, it has been demonstrated that poor generalization is the root cause of the privacy risk~\cite{shokri2017membership}.
\par

Informally, good generalization in CML is achieved when the number of users participating in the learning protocol is maximized: the more users involved in training, the less information about a single individual can be inferred from model updates shared during the protocol. 

FL maximizes generalization in this respect: the central server ensures that every state of the global model is computed using \textit{all} the $n$ available model updates, and, importantly, that every model update contributes equally to this computation. %, and all users have the same model.

In the decentralized setting, this is not the case. While, as in FL, users' models are a function of the models of all other users in the system; in DL every user has a different \TT{personalized} local model to which not all users contribute equally. The contribution of user $u_i$ on user $u_j$'s local parameters depends on the distance between those users in the communication topology. %When two users are not neighbors, their updates reach each other only through other users.
The further these users are, the weaker is the influence of $u_i$'s updates on $u_j$'s model. The strength of the influence decays exponentially with the number of intermediary users, as the updates of $u_i$ get blended with those of intermediary users' models before arriving to $u_j$ due to the gossip-based propagation. %reducing its relevance in the computation of the local state of non-neighbor users.
We illustrate this phenomenon with a chain-like topology represented in Figure~\ref{fig:string_like_loss} (top). In this topology, $u_1$ (the first user in the chain) only has one direct neighbor $u_2$ and the rest of the users in the system are several hops away. Here, user $u_1$ only receives the updates produced by $u_5$ after they have been \TT{consumed}  (i.e., aggregated with other received model updates at the end of a round) and propagated by each intermediate user in the chain within the following model update. When $u_5$'s update reaches $u_1$, the strength of its signal is reduced by a factor $\frac{1}{|\nn(u_4)|}\cdot\frac{1}{|\nn(u_3)|}\cdot\frac{1}{|\nn(u_2)|} \cdot\frac{1}{|\nn(u_1)|}=\frac{1}{54}$ due to the aggregation rule (line $8$ Algorithm~\ref{alg:dl}). There is less information about $u_5$'s data compared to model updates produced by closer users (e.g. $\frac{1}{2}$ for $u_2$). In contrast, in FL every pair of users is virtually separated by a single hop: the server. This ensures that every user's contribution is weighted equally in the global model. 

\input{string_like_scheme}

%\textbf{Now, given a node $v$, this means that the local training sets of the other users in the system do not have the same relevance in defining  $v$'s local parameters during the training.}
This slow and uneven propagation of updates results in the local model of decentralized users being dominated by their own training set and the training sets of their immediate neighbors, yielding \emph{poor generalization}. We illustrate this effect in Figure~\ref{fig:string_like_loss} (bottom), where we report the average loss when applying $u_1$'s local model on the local training sets of other users. The loss increases with the distance between $u_1$ and the owner of the local training set. We call this phenomenon  {\TT{local generalization}}, in contrast to \TT{global generalization} offered by federated learning. 
%Additionally, as described in Algorithm~\ref{alg:dl}, local states are optimized via an iteration of SGD, imprinting even more information about the private set into the model without the possibility of external generalization.
%\mr{The information contained in one user’s intermediate state is mostly influenced by its own training data, and less by far away users, because of hops to others’ info to reach node, whereas an aggregate update contains information from all training data.} 

%These poorly generalized models, and their associated model updates, are available to an adversary during the protocol execution, even if after a suitable number of iterations consensus is met and the local models of all the user are the same. As we empirically demonstrate in the following sections, the ability to access those poorly generalized model updates gives a substantial advantage to adversaries compared to what can be learned by just accessing the global (aggregated) model available in FL.
After a suitable number of iterations, information will be uniformly propagated in the system and every user's local model will be the same: consensus is met and generalization is maximized. For every round before that, all intermediate user’s local models and model updates store more information about their private local training sets than the updates of the other users in the system due to the local generalization phenomenon.  As we empirically demonstrate later in the paper, the ability to access these poorly generalized model updates gives a substantial advantage to privacy adversaries compared to what can be learned by just accessing a global (aggregated) model as in FL.
\par

%This advantage can only be reduced by reducing the average distance between each pair of users in the communication graph, therefore reducing the effect of local generalization. To achieve this, it is necessary to increase the number of neighbors of each node.
This advantage can only be reduced by limiting the effect of local generalization; that is, by reducing the average distance between each pair of users in the communication graph, or, more pragmatically, increasing the number of neighbors of each node. In the extreme, when all nodes are connected to each other, local effects disappear and the protocol achieves global generalization as in FL. % and completely suppressing the local generalization phenomenon. 

%\paragraph{System knowledge}
\subsection{System knowledge}
Dense topologies, as those needed to limit the effect of local generalization, have negative implications on performance and on privacy. On the performance side, dense topologies increase the communication overhead of users, defeating one of the objectives of decentralized protocols. %: reducing the overall consumed bandwidth. 
On the privacy side, dense topologies increase adversaries' knowledge of the state of other users in the system, providing them with more information to perform privacy attacks.
\par

In FL, users can only observe the aggregated result provided by the server, i.e., the global set of parameters. In contrast, in DL, decentralized users receive the model updates of each of their neighbors (Figure~\ref{fig:dl}). Each of these updates captures a \textit{view of the system}, as it contains information coming from a different subset of nodes in the graph (the neighbors of the neighbors). We call \TT{system knowledge} the ability that a user of DL has to access multiple individual model updates per round.	
The system knowledge gained from having simultaneous access to disjoint views of the system grants an additional advantage to decentralized attackers. As we demonstrate in the next sections, they can combine the model updates they receive and use them to isolate individual contributions of other users, effectively reducing generalization and its positive effect on users' privacy. More critically, we show that an attacker with enough system knowledge can reach the same adversarial capabilities as a parameter server in FL. This effectively defeats another of the objectives of decentralizing learning.\\

\subsection{Take aways}
Local generalization and system knowledge are in direct opposition. Increasing the number of users' neighbors as a way to reduce the adversary's advantage coming from local generalization inherently increases the adversary's advantage coming from system knowledge. Conversely, topologies that limit decentralized adversary's system knowledge to prevent them from isolating individual users' contribution inherently increase local generalization, increasing the adversary's advantage.

In the following sections, we quantify the privacy loss stemming from the adversary exploiting this conflict. Our study leads to the conclusion that \textbf{the intrinsic trade-off between local generalization and system knowledge fundamentally limits the privacy achievable by users in decentralized learning}.
%In this work, we show that \textbf{this intrinsic trade-off fundamentally limits the privacy achievable by users in decentralized learning} against both passive and active decentralized adversaries. In the following sections we quantify the privacy loss  effect of those conflicting properties on decentralized users' privacy.

\section{Privacy Against Passive Adversaries}
\label{sec:hbc}

To prove the argument of Section~\ref{sec:localgen_and_systemk}, we start by comparing the privacy offered by the DL approach against FL in the semi-honest model, i.e., when there is a passive adversary in the system. 
We start by formalizing our threat model. % and then compare adversarial capabilities.

\parabf{Passive adversary threat model.}
A first type of adversaries in CML are passive adversaries. Such adversaries legitimately follow the steps of the CML protocol yet will attempt to learn all possible information from received model updates. Their goal is to infer information about the private training sets of one or more honest users in the system, which we refer to as \emph{victims} or \emph{targets} interchangeably. Passive adversaries do not forge adversarial model updates neither by changing the loss function of the model nor by tampering with their local training set~\cite{serum}. 

In this paper, we assume a \emph{weak} passive adversary who has no information about the system (\eg they know their neighbors but not the rest of the communication topology).% for DL). 
\par

\noindent\textbf{DL passive adversary.} Any user involved in a decentralized protocol can be a passive adversary, as they observe the model updates of their neighbors. 
As the communication topology is always connected in the DL setup~\cite{\jaggi, lalitha2019peer, lalitha2018fully}, {every user has at least one neighbor}. Thus, DL is required to guarantee privacy against adversarial neighbors as a cardinal property of the protocol. Note that the only scenario that allows decentralized users to rule out adversarial neighbors is when trust is introduced in system; that is, users assume that all their neighbors are fully honest.
{Current decentralized learning frameworks do not impose any limitation on user connectivity. Thus, an adversarial user can connect to a chosen victim to become their \textit{adversarial neighbor} (see Section~\ref{sec:const_top}). 
Hereafter, we denote a passive adversarial user in DL as~$\dap$.}

\noindent\textbf{FL passive adversary.}
In FL, the server has similar capabilities to an adversarial neighbor in DL, \ie it receives and sends (an aggregation of) model updates. Federated users can only observe the global model sent by the server at the end of each round. Hereafter, we denote a passive adversarial user in FL as $\fup$, and a passive adversarial parameter server as $\fsp$.

\iffalse
%However, if we consider that users (and, so, adversaries) are enabled to choose their neighbors, this option becomes unworkable as this would be equivalent at assuming that all the users in the system (\ie all the parties participating at the protocol) are honest.\footnote{The attacker can simply connect to the target.}
\fi
%
%\subsection{Decentralized  adv. user vs Federated adv. user (passive)} 

\subsection{Decentralized user vs federated user (passive)} 
\label{sec:passive_user_vs_usr}

\begin{figure*}[t!]
	\centering
	\includegraphics[trim=0mm 100mm 0mm 0mm, clip, width=1.1\columnwidth]{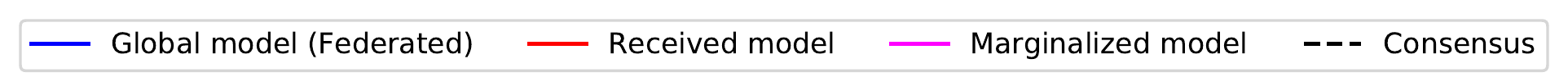}\\
	\begin{subfigure}{.5\columnwidth}
		\centering
		\includegraphics[trim = 0mm 0mm 0mm 0mm, clip, width=\columnwidth]{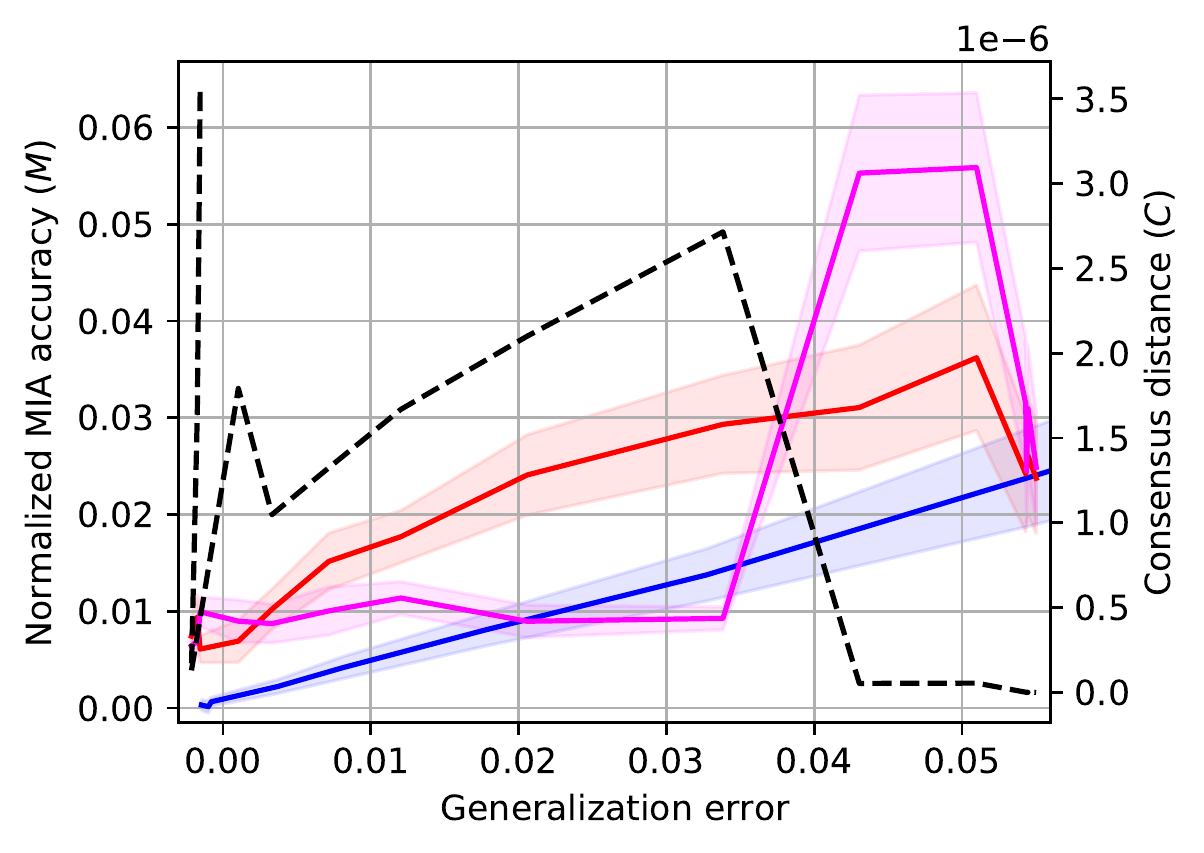}
		\caption{torus-36 \& CIFAR-10}
	\end{subfigure}\begin{subfigure}{.5\columnwidth}
		\centering
		\includegraphics[trim = 0mm 0mm 0mm 0mm, clip, width=\columnwidth]{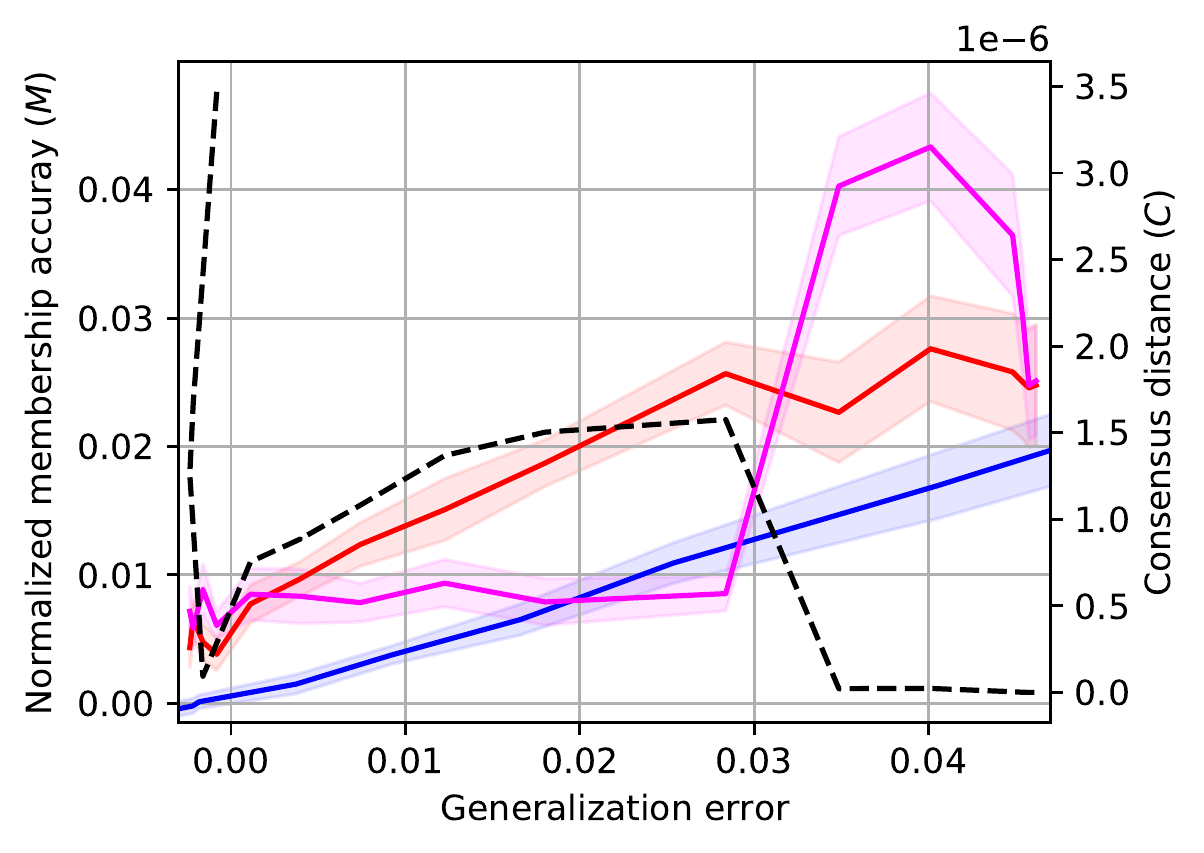}
		\caption{social-32 \& CIFAR-10}
	\end{subfigure}\begin{subfigure}{.5\columnwidth}
		\centering
		\includegraphics[trim = 0mm 0mm 0mm 0mm, clip, width=\columnwidth]{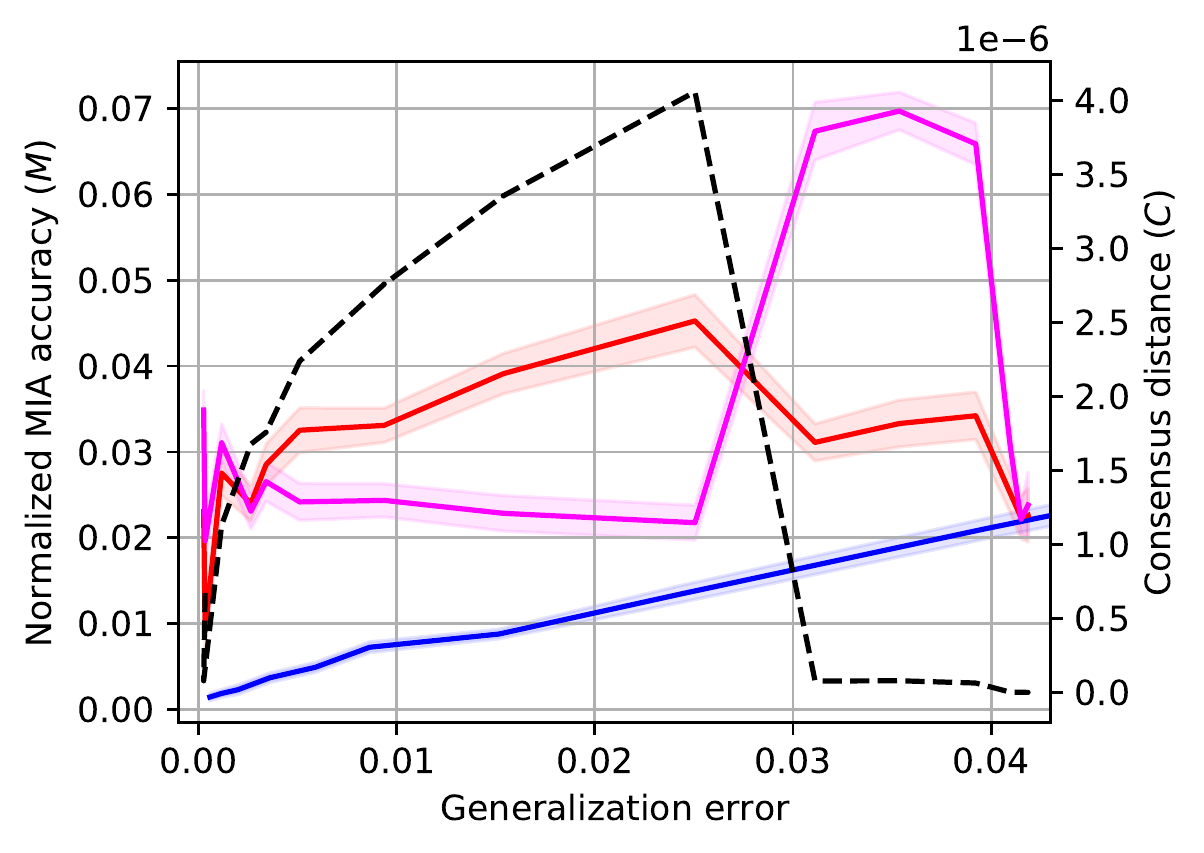}
		\caption{torus-36 \& CIFAR-100}
	\end{subfigure}\begin{subfigure}{.5\columnwidth}
		\centering
		\includegraphics[trim = 0mm 0mm 0mm 0mm, clip, width=\columnwidth]{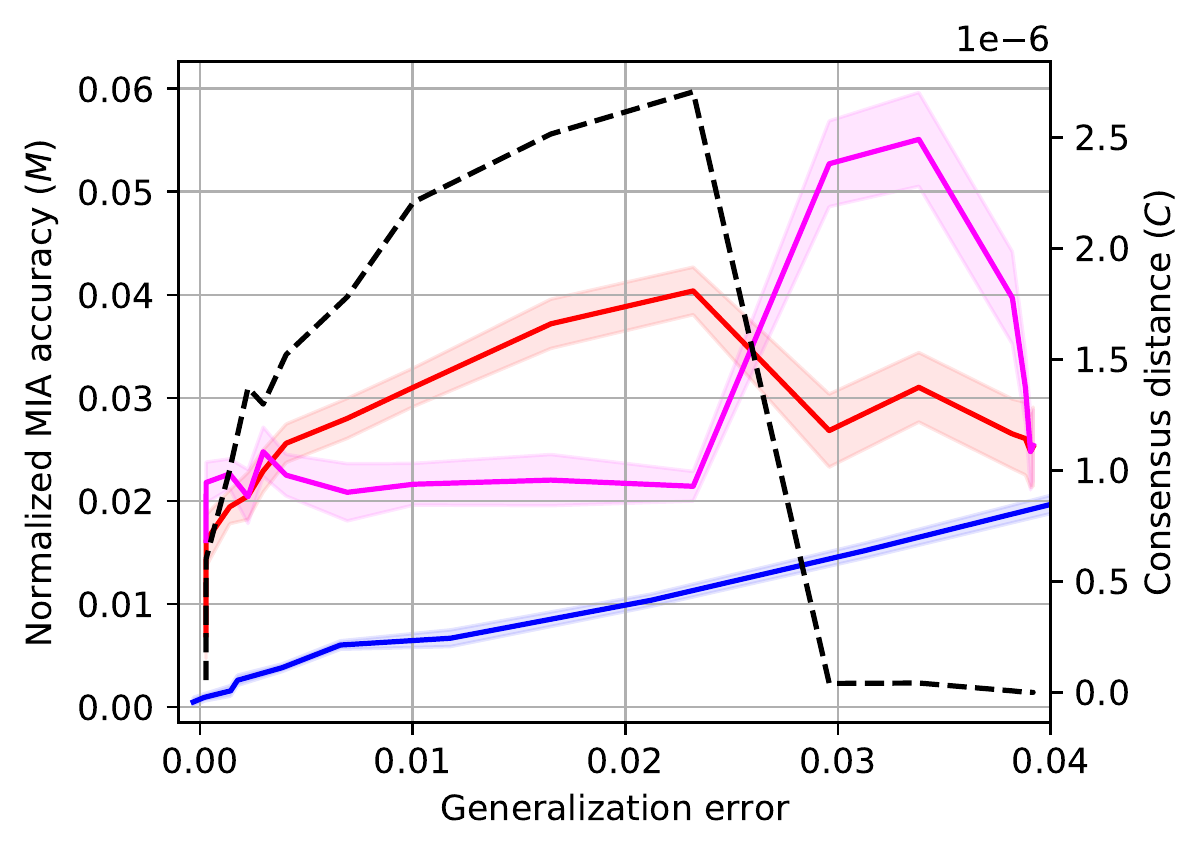}
		\caption{social-32 \& CIFAR-100}
	\end{subfigure}
	\caption{Average MIA vulnerability on four different communication topologies and datasets (DL in red and purple, and FL in blue). For each combination of topology and dataset, we report the average results over $16$ runs. In each run, we select a different adversarial user uniformly at random in the system. The halo around the curves reports the standard deviation over the various runs. The gray dotted line represents the consensus distance in DL, computed using Eq.~\ref{eq:consesus}.}
	\label{fig:mias}
\end{figure*}
\begin{figure}[t]
	\centering
	\includegraphics[trim = 0mm 3mm 0mm 0mm, clip, width=.55\linewidth]{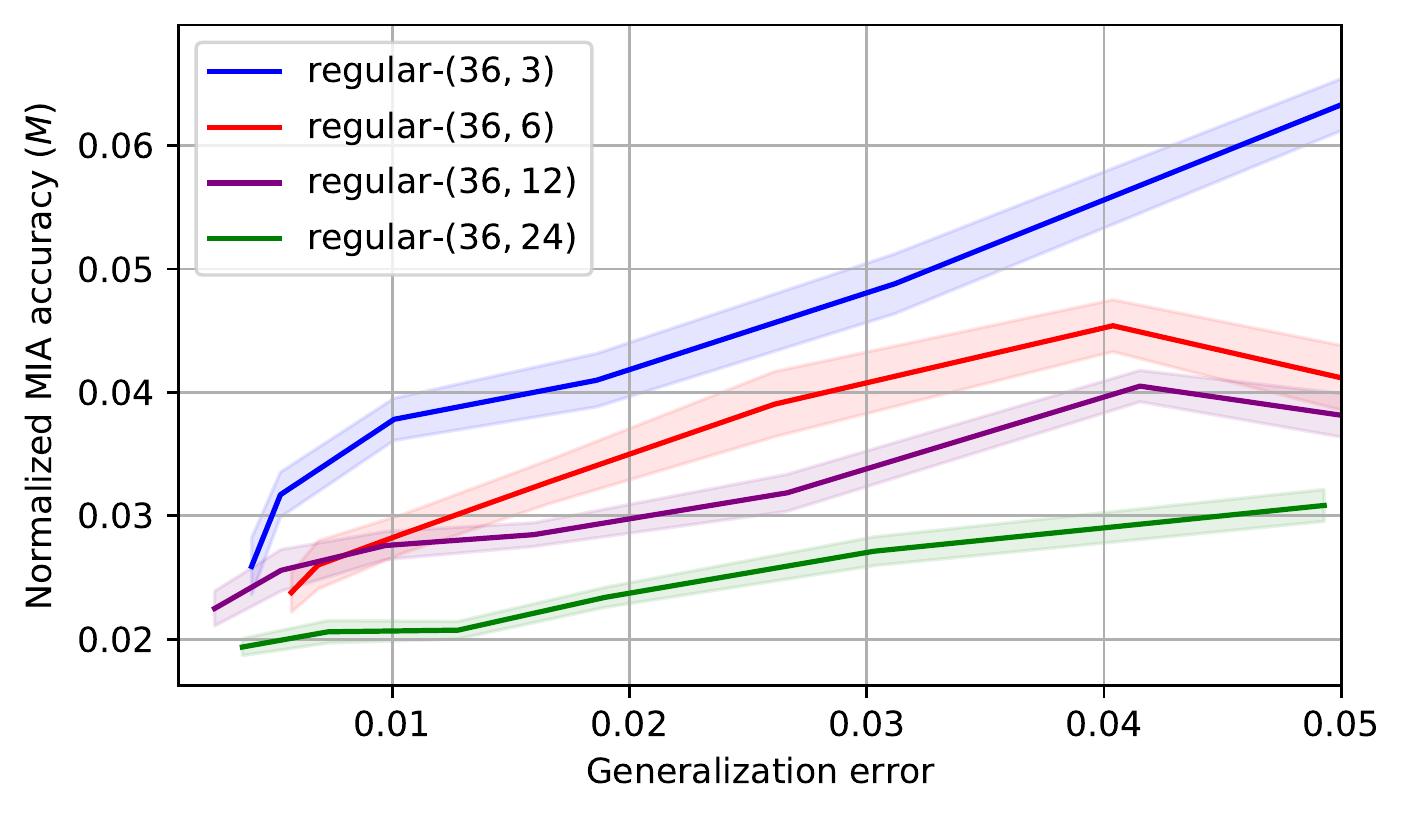}
	\caption{Average MIA vulnerability of model updates generated by decentralized users on regular graphs with increasing density (CIFAR-100).}
	\label{fig:gen_error_rec}
\end{figure}
\label{sec:passive_user_vs_user}

%\ct{I am not super thrilled by the title of this section. Let's see when we have a story}

We first compare the privacy that honest users can enjoy against an adversarial decentralized {user} ($\dap$) and an adversarial federated {user} ($\fup$). Our results demonstrate that decentralization provides an intrinsic advantage to passive adversarial users compared to the federated setting. 

% demonstrating and quantifying the inherent adversarial advantage of former over the latter.\\

%
\subsubsection{Inference on model updates} 
\label{sec:recmup}
As part of any CML protocol, users share the model updates computed on their local data. A passive adversary can use them as parameters for a model~$f_\victm$ and run arbitrary privacy attacks on the victim(s) who generated the updates.
%As discussed in Section~\ref{sec:cml}, CML users compute model updates as a function of the their local private data and share them with the other parties running the protocol. An adversary, upon receiving model updates produced by honest users, can exploit them to recover or infer information about users' private training sets. 

In the DL setting, at every round $t$ of Algorithm~\ref{alg:dl}, the attacker $\dap$ receives the model updates~$\mup{\victm}{t+\oot}$ from each of their neighbors $\victm \in \nn(\dap)$. 
In the FL setting, $\fup$ has only access to the global state of the model provided by the parameter server at the start of the round, which aggregates the updates of all the users in the system.
% a set of neighbors~$\nn(\dap)$. Thus, at every round $t$ of Algorithm~\ref{alg:dl},  $\dap$ receives a model update~$\mup{\victm}{t+\oot}$ (\ie a set of valid parameters for the model $f$) from each neighbor $\victm \in \nn(\dap)$.
%In FL setting, instead, $\fup$ has only access to a single (aggregated) model update---the global state of the model provided by the parameter server at the start of the round (by FL design, no other source of information is available to users). 
%Given access to those model updates, the two adversaries can use them as parameters for the model $f$ and run arbitrary privacy attacks on the users who generated the updates. 
\par

In Figure~\ref{fig:mias}, we compare the vulnerability of victims against membership inference attacks in DL and FL. We plot the evolution of the vulnerability of the adversary's neighbors to membership inference attacks as the training process progresses.
To capture vulnerability, y-axis represents the average MIA accuracy across all the attacker's victims $\victm$. From Eq~\ref{eq:mia}, this is computed as:
\begin{equation}
	\frac{1}{m}\sum_{\victm} M(\mup{v}{t+\oot}, X_v, \mathcal{O}),
\end{equation}
where $\attck$ is either $\dap$ (red line) or $\fup$ (blue line), and the set of victims is either all users in FL (and $m=n$), or  the neighbors of $\attck$ in DL (and $m=|\nn(\attck)/\attck|$). 
The figure also reports the progression of the consensus distance (Eq.~\ref{eq:consesus}) in DL (gray dotted line).

%For $\fup$, recall that $\nn(\fup) \myeq V $. 
%\ct{should we have a forward reference about the purple line?}

The x-axis aims at capturing the training progression. A natural choice for this axis would be to use the number of protocol iterations $t$. However, DL and FL do not converge at the same speed. Therefore, DL and FL models at the same round $t$ may be arbitrarily different. To address this limitation, we choose to compare models when they have the same generalization error $g_{err}(t)$  (\ie same level of \textit{overfitting}). 
%\mr{As consequence, comparing privacy according to the current round is not informative. Instead, we compare models based on their current generalization state. We generated the Figure 4\ref{} like that. It should be read like this: for a given generalization state, what privacy offer a moded. @Dario Food for thought, what if we were to put privacy as x-axis?} 
In DL, we compute the average generalization error of users' local parameters at iteration $t$ as:
\begin{equation}
	g_{err}(t) = \texttt{acc}(X,  \mup{}{t}) -  \texttt{acc}(\mathcal{O}, \mup{}{t}),
\end{equation}
where, $\Theta^t \myeq \frac{1}{|V|}\sum_{v\in V} \mup{v}{t}$ is the average state of the system, $X \myeq \bigcup_{v\in V} X_v$ is the union of all the local training sets, $\mathcal{O}$ is a test set completely disjointed from $X$, and \texttt{acc} is the accuracy function. 
In FL, the average state of the system is simply the global model as $\forall_{v\in V}\mup{}{t} \myeq \mup{v}{t}$.

In DL, the vulnerability is a function of both the generalization error and the consensus distance. Larger generalization error denotes overfitting, which is known to result in information leakage about the training set~\cite{shokri2017membership}. Larger consensus distance indicates that information is still not uniformly propagated in the system, thus updates carry significantly more information about the local training set than sets from other users. This is what we referred to as \emph{local generalization} in Section~\ref{sec:localgen_and_systemk}. We observe that privacy leakage due to local generalization may happen even when the generalization error is close to 0 (leftmost parts of the plots). 
%\ct{This sentence says a bit the same as the previous one. It's ok to hammer the point, I am just marking it as redundant}
That is, \textbf{even when a decentralized system has perfect generalization, decentralized model updates still contain individualized information that can be used to infer about training data.} In FL, as expected, when assessing the privacy risk at a generalization error level that is near to zero, the privacy risk also approaches zero.

When the DL system reaches consensus (rightmost part of the plots), the vulnerability in DL and FL approaches the same value. This is because when consensus is reached local generalization disappears: decentralized users share the same global model (as in FL). Thus, the victim's model updates do not carry information particular to their own training sets.  %This indicates that, regardless of their direct or indirect connectivity, the attacker is eventually able to learn the same amount of information about all the users' private training sets in the system. %We conclude that {decentralized learning does not offer better privacy than federated learning} even if neighbors are fully trusted by the user. Indeed, as soon as the network reaches consensus and the adversary has access to the final global model, non-neighbor adversaries can obtain the same amount of information on users as in federated learning. %Before reaching consensus, due to local generalization, decentralized users are always at a disadvantage with respect to adversaries. \ct{I am not sure this is vvisible in th figure on the right privacy risk is less for decentralized learning}
%Moreover, once consensus is reached, global generalization error (\ie overfitting) reaches its highest level and, therefore, the highest level of privacy risk for the global model.
\par

The harmful effect of local generalization on privacy can be reduced by increasing the density of the communication topology (see Section~\ref{sec:localgen_and_systemk}). We confirm this in Figure~\ref{fig:gen_error_rec}, where we show that increasing the density in random regular topologies (\textit{regular}-$(36, d)$ with $d \in \{3, 6, 12, 24\}$) reduces the privacy harm. 

\subsubsection{Inference on functionally marginalized model updates}
\label{sec:fun_iso}

In contrast to $\fup$, the decentralized adversary~$\dap$ has access to multiple model updates produced by different users (one per each of their neighbors). This \textit{system knowledge} can be used to further increase membership vulnerability by combining model updates and using them to isolate the local contribution to the update from honest neighbors. To demonstrate this capability, we introduce a novel attack that we call \TT{functional marginalization}.
\par

Functional marginalization exploits the fact that the local model update $\mup{v}{t}$ of user $v$ can be divided into two core components:
\begin{equation}
	\label{eq:factors}
	\mup{v}{t} \approx \isom{v}{t} + \mup{V/{v}}{t},
\end{equation}
where $\isom{v}{t}$ represents the contribution to the update computed with the local training set of the node $v$, and $\mup{V/{v}}{t}$ captures the contributions of all other nodes in the system.%This factorization holds in any collaborative machine learning setting.

With enough information about $\mup{V/{v}}{t}$, and because they know $\mup{v}{t}$, the adversary can extract the marginalized contribution $\isom{v}{t}$ from Eq~\ref{eq:factors}.
Exactly recovering the term $\mup{V/{v}}{t}$ is unfeasible. However, the adversary can compute a rough approximation of this value for a victim~$\victm$ from the model updates the adversary receives from other neighbors. %Given a neighbor victim~$\victm$, an attacker $\dap$ can approximate the global functionality of the system $\mup{V/{\victm}}{t}$ from the model updates received from other neighbors.
The adversary estimates $\mup{V/{\victm}}{t}$ as the average of all parameters they receive, excluding the victim's:
\begin{equation}
	\mup{V/{\victm}}{t} = \frac{\sum_{u\in \nn(\dap)/\victm} \mup{u}{t+\oot}}{| \nn(\dap)|}.
\end{equation}
%This aggregation represents the global functionality \textit{without} the contribution from~$v$. %(to note that we divide models parameters by $|\nn(\attck)|$, where the aggregation is composed by $| \nn(\attck)|-1$ models \ct{this note needs more justification why do we do this and why is it good?}).
%\ct{this is a bit repetitive, maybe we can merge with the above?}
Then, by removing this approximation from the victim's model update, $\dap$ isolates the victim's contribution:
\begin{equation}
	\label{eq:marg}
	\isom{\victm}{t} = | \nn(\dap)| \cdot (\mup{\victm}{t+\oot} -  \mup{V/{\victm}}{t} ).
\end{equation}
This process can also be seen as reversing the aggregation operation in line $8$ in Algorithm~\ref{alg:dl} by pulling out the term $\mup{\victm}{t+\oot}$ from the averaged model $\mup{\victm}{t+1}$, using somewhat incomplete information.
\par

The recovered \TT{functionally marginalized model} $\isom{\victm}{t}$ is a function of the local training set of $v$ only. Thus, the adversary can use it to obtain better results than when attacking directly~$\mup{v}{t+\oot}$, which contains contributions from other users. The difference between the red and purple lines in Figure~\ref{fig:mias} captures this improvement. As can be seen in the figure, the improvement is not consistent. This is because, as showed in the previous section, membership vulnerability is a function of the global generalization error and the consensus distance. 
When the consensus distance is high (leftmost part of the plots), $\dap$ cannot compute an accurate representation of the global functionality~$\mup{V/{v}}{t}$. Thus, the marginalized model~$\isom{\victm}{t}$ may not be a good representation of the victim's local training set and the attack performs worse than when performed directly on the received model. 
When the consensus distance $C$ decreases (rightmost part of the plots), $\mup{V/{\victm}}{t}$ becomes a good representation of the global state, the marginalization (Eq.~\ref{eq:marg}) becomes accurate, and membership vulnerability abruptly increases. 
Finally, when the consensus distance $C$ approaches zero and all users have the same view, and marginalization has no effect as there is no victim's contribution to be isolated.
% \TEMP{then $C\myeq0$ in Eq.~\ref{eq:marg}, $\isom{\victm}{t} \myeq \mup{\victm}{t+\oot}$, and the privacy risk is the same when the received model is attacked directly.}
At that point, Eq.~\ref{eq:marg} results in $\isom{\victm}{t} \myeq \mup{\victm}{t+\oot}$ and membership vulnerability is the same as when the received model is attacked directly.

\begin{figure}
	\centering
	\includegraphics[trim = 0mm 3mm 0mm 0mm, clip, width=1\linewidth]{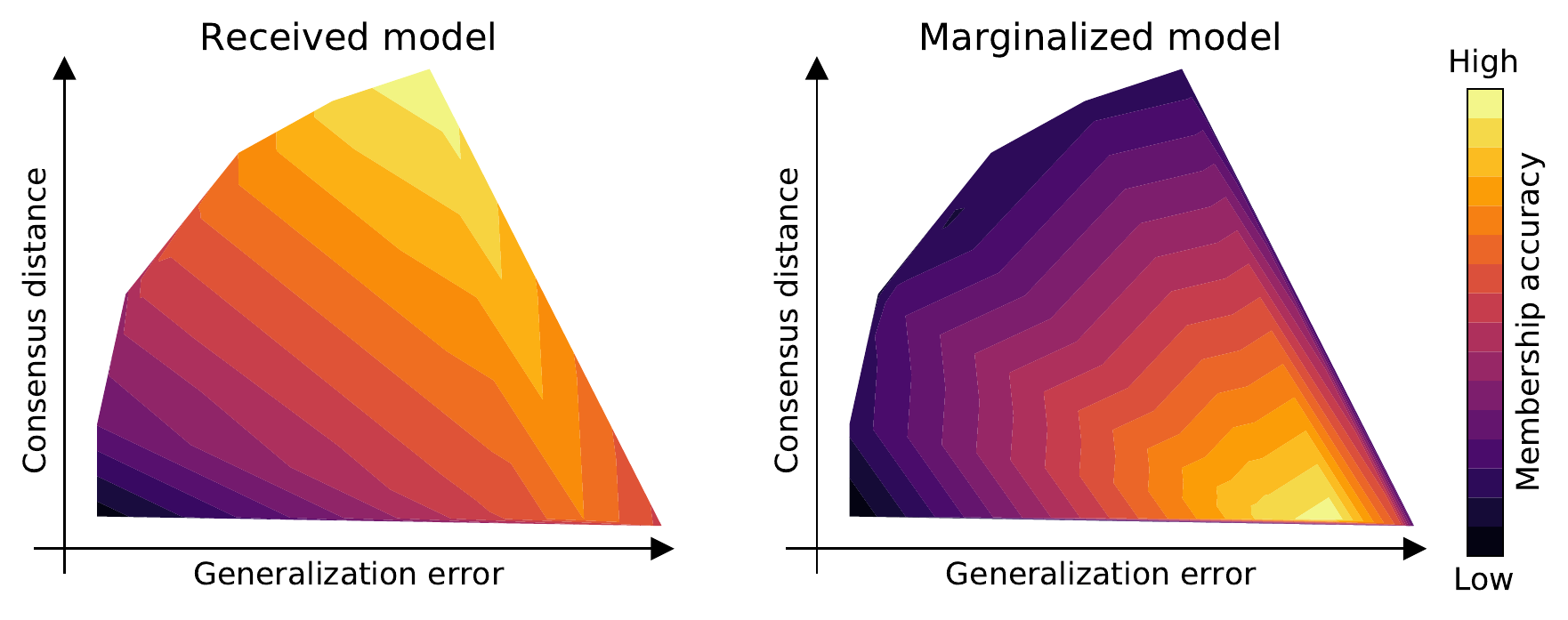}
	\caption{MIA vulnerability as a function of generalization error and consensus distance for the received and marginalized model. Setup: torus-36 on CIFAR-100.}
	\label{fig:cons_vs_generr_vs_pr_torus_cifar10}
\end{figure}

%% previous
%\ct{Do we really need this figure here? seems not much new information. If we need space maybe it can go to the appendix?}
The results in Figure~\ref{fig:mias} hint that attacks on the received model update or its functionally marginalized version are complementary: the former succeeds when the consensus distance is high, the latter succeeds when consensus distance is low. In order to understand the conditions under which attack is more powerful, we compare the two attacks in Figure~\ref{fig:cons_vs_generr_vs_pr_torus_cifar10}. This figure represents the level of vulnerability (lighter colors represent more vulnerable cases) depending on the generalization error (x-axis) and the consensus distance (y-axis). In both attacks, vulnerability is proportional to the generalization error, but they behave differently depending on the consensus distance. Attacking the received model results in high vulnerability when the distance is large (top center), while when attacking the marginalized model vulnerability is maximized when the consensus distance is low (bottom \new{right}). This means that the adversary can maximize their effectiveness by choosing the best attack after evaluating the consensus distance on the received model updates. As before, when the network reaches consensus (\ie $C(t)\myeq 0$ in the rightmost edge of the plots), the vulnerability is minimized as local datasets have the least influence on the updates. \new{In Appendix~\ref{app:torus64}, we present results obtained on a larger number of users ($64$ and $128$), along with additional topologies. Those show that increasing the number of users magnifies the local generalization phenomenon, which in turn amplifies the vulnerability of model updates. In Appendix~\ref{app:nlp}, we present an extension of our evaluation to a NLP task, yielding results that are consistent with those in Figure~\ref{fig:mias}. }
\par

\new{It should be noted that the inference attacks discussed in this section are \textit{stateless} and can be executed in a single DL round. Therefore, they can be directly applied to dynamic topologies, where the attacker's neighbors may vary during the training process.} Additionally, we highlight that the introduced functional marginalization technique is general and can be potentially extended to other CML frameworks. For instance, it can be applied to most Personalized FL frameworks, where users or the server~\cite{pmlr-v162-dai22b, zhang2021personalized} have access to multiple (personalized) versions of users' models.

\vspace{\baselineskip}
%In conclusion, {a passive adversary that only exploits information available from an honest execution of the decentralized protocol, can increase the privacy risk of users' models compared to what is achievable in federated learning for the same setup} \new{(up to three times for the reported experiments)}.
{The results in this section provide empirical support to our claims in Section~\ref{sec:localg}; local generalization in DL is an unavoidable source of leakage that does not exist in the FL setup. Ultimately, this means that, for every non-complete topology $G$ (\ie every topology that induces local generalization), a passive decentralized adversary $\dap$ would always be able to infer more information about honest users than an equivalent passive federated adversarial user $\fup$. In the next section, we show that this claim holds also for any complete topology.}
\par

%%%%%%%%%%%%%%%%

%
\subsection{Decentralized user vs federated server (pass.)}
\label{sec:passive_user_vs_server}
We now compare the adversarial capabilities of an adversarial passive decentralized user $\dap$, against an adversarial passive federated server $\fsp$. Our results demonstrate that an adversarial user in DL can have the same adversarial capabilities as a parameter server in FL.\\

In FL, the adversary $\fsp$ is in a privileged position to run privacy attacks. Unlike adversarial federated users $\fup$, who only receive aggregate model updates, the parameter server has access to user's model updates and the intermediate states of their local optimization processes -- pseudo-gradients for \texttt{FedAVG}. This position enables $\fsp$ to perform powerful privacy attacks such as accurate inference attacks on gradients~\cite{nasr2018comprehensive} or gradient inversion~\cite{invg2,  jeon2021gradient, pan2020theoryoriented, nvidiagi, rgap, invg1}.
\par
%of the form:
%\[\underset{\widehat{x}}{\arg\min} [d(\nabla^\Theta_{\widehat{x}}, \nabla^\Theta_{x}) \cdot \alpha r(\widehat{x})],\]
%where $\widehat{x}$ is the candidate solution of the attacker (\eg a set of images), $d$ is a distance function that measures the discrepancy between the two gradient signals $\nabla^\Theta_{\widehat{x}}$ and $\nabla^\Theta_{x}$, and $r$ is a regularization function defined on the input domain. The attack outputs a suitable reconstruction of the original, private input $x$. Gradient inversion can also be performed using other techniques~\cite{rgap, pan2020theoryoriented}.

%In a decentralized setting, every user can be seen as a local parameter server, and thus every user can perform gradient inversion attacks on their neighbors.
To carry out a gradient inversion attack, an attacker $\attck$ needs two pieces of information: (1)~the gradient $\nabla_{\Theta^{t}_\victm}\loss(\xi_{\victm}^{t})$, and (2)~the parameters of the network $\Theta^{t}_\victm$ used to compute such a gradient. These two components are, by design, available to an adversarial parameter server in FL. However, they are not directly accessible to an adversarial user in DL~$\dap$.

An attacker $\dap$ in DL receives the following model update $\Theta^{t+\frac{1}{2}}_\victm$ from their neighbor $\victm$: 
\begin{equation}
	\Theta^{t+\oot}_\victm=\Theta^{t}_\victm - \eta \nabla_{\Theta^{t}_\victm}\loss(\xi_v^t).
	\label{eq:model_up_gi}
\end{equation}
To extract the gradient $\nabla_{\Theta^{t}_\victm}\loss(\xi_{\victm}^{t})$ from this model update, $\dap$ needs to know $\Theta^{t}_\victm$. In principle, the exact value $\Theta^{t}_\victm$ is not available to the attacker as it is a function of the model updates from $\victm$'s neighbors. In principle, this could render gradient inversion attacks in DL unfeasible.
%As result, gradient inversion attacks are likely to fail in DL~\cite{invg1, invg2}. 
%
\subsubsection{Gradient inversion attack in decentralized learning}
\label{sec:gi}

We now show how the adversary $\dap$ can estimate the gradient  $\nabla_{\Theta^{t}_\victm}\loss(\xi_{\victm}^{t})$ in order to perform the gradient inversion attack.

There are three ways in which $\dap$ can perfectly recover the individual gradient of their neighbors.
%In two trivial cases $\Theta^{t}_\victm \myeq \Theta^{t}_\attck$: when $t\myeq0$ in the first training iteration, and when users $\attck$ and $\victm$ achieve consensus (\ie $C(t)=0$).
The first two are trivial cases in which $\Theta^{t}_\victm \myeq \Theta^{t}_\dap$: the first training iteration $t\myeq0$, and when users $\dap$ and $\victm$ achieve consensus (\ie $C(t)=0$).
In both cases, the attacker can recover the victim's gradient by computing $ \frac{1}{\eta}(\Theta^{t+\frac{1}{2}}_\victm - \Theta^{t}_\dap)$.

Gradient recovery at $t \myeq 0$ could be prevented by having users choose different initial parameters $\Theta^0$, with the caveat that this modification may impact the learning process. The second case, however, cannot be avoided. Reaching consensus is the goal of decentralized learning. Thus, eventually, the attacker will have the opportunity to recover the gradient and perform the inversion attack. 

\iffalse
We note that even when consensus is not achieved, as long as $C(t)$ is close to $0$, the adversary can approximate the model of the victim~as:
\begin{equation}
	\nabla_{\Theta^{t}_\victm}\loss(\xi_{\victm}^{t}) = \lim_{C(t) \to 0} \frac{1}{\eta} (\mup{\victm}{t+\oot}- \Theta^{t}_\attck),
\end{equation}
and use it to compute an approximation of the gradiant sufficiently good to run the attack. \ct{There is a note about a potential appendix. Do we have it? If not, honestly we can live without this approximation as it is not central to the paper and does not reduce the threat -- In fact, we could say that the second trivial case is $C(t)$ close to $0$ instead o $0$ and the story would hold.}
\fi
%\todo{In Appendix~\ref{}}, we measure the performance of noisy gradient inversion as the system approaches the consensus state.
\par
%
%Here, the quality of the recovered gradient would be inversely proportional to the consensus distance $C(t)$.  {However, even when $C(t)$ is arbitrary large, having {system knowledge} (see Section~\ref{sec:localgen_and_systemk}) enables the adversary to unconditionally recover neighbors'  gradients.}

%When $C$ is too large for this attack to be effective, the adversary can use \textit{system knowledge} (see Section~\ref{sec:localgen_and_systemk}) to perfectly recover the gradient. %\ct{I do not like this footnote, if we want to claim this lets put it above. Honestly, I would remove it. It seems to not add much}\footnote{\new{Note that also the previous configurations are consequences of  system knowledge. Those were two special cases where every user is in the same state, and, therefore, the attacker has perfect knowledge of the system.}} %\ct{it is not clear how Eq 10 leads to this. There is no gradient in this one...}
%\new{However, there is another configuration that enables a passive adversary to perfectly recover neighbors' gradients for any given time step $t$.}

%However, there is another configuration that enables a passive adversary to perfectly recover neighbors' gradients for any given time step $t$ which relies on the system knowledge of the adversary.

\parabf{Gradient recovery from system knowledge.}
%
% The third and most interesting case: CT -- I removed the qualifier, here and in other places. "more intersting" is subjective, an opinion. It is not good practice to have those in scientific writing. That is what you say in the presentation :) 
The third situation in which the adversary $\dap$ can recover neighbors' gradients is when the set of attacker's neighbors $\nn(\dap)$ is a super-set of the victim's neighbors set $\nn(\victm)$. This situation, provides $\dap$ with enough system knowledge to recover the gradient regardless of the time step $t$. 

In order to recover the gradient produced by $\victm$, an attacker $\dap$ needs to subtract the unknown set of parameters~$\Theta_\victm^{t}$ from the received model update $\Theta_\victm^{t+\oot}$. Recall that the victim's set of parameters $\Theta_\victm^{t}$ is defined as:
\begin{equation}
\Theta_\victm^{t}=\frac{1}{|\nn(\victm)|}\sum_{u \in \nn(\victm)} \Theta_u^{t-\frac{1}{2}},
\end{equation}
where, $\Theta_u^{t-\frac{1}{2}}$ is the model broadcasted by the node $u$ to all its neighbors during the previous training iteration ($t-1$). 
If the adversary has access to the updates of the neighbors of the victim, they can use these neighbours updates at time $t-1$ to reconstruct $\sum_{u \in \nn(\victm)} \Theta_u^{t-\frac{1}{2}}$, and use this to compute~$\Theta_\victm^{t}$.

%It follows that, if the set of attacker's neighbors $\nn(\dap)$ is a super-set of the victim's neighbors set $\nn(\victm)$, then the attacker can perfectly recover $\Theta_v^{t}$ using the neighbors' models updates received at time $t-1$. In other words, if the adversary has a sufficient number of neighbors, the latter can be used to recompute the unknown local state of the victim.

%
\begin{figure}[t]
	\centering
	\resizebox{.2\textwidth}{!}{
		\begin{tikzpicture}
			
			\iffalse
			\clip  (-4,-1.2) rectangle (4,3.5); 
			
			\tikzstyle{node} = [draw, circle, fill=green!30, text width=3mm, align=center,]
			\def \n {8}
			\def \radius {3cm}
			\def \margin {6} % margin in angles, depends on the radius

			\foreach \s in {1,...,\n}
			{
				\node[node] at ({360/\n * (\s - 1)}:\radius) {$u_\s$};
				\draw[>=latex] ({360/\n * (\s - 1)+\margin}:\radius) 
				arc ({360/\n * (\s - 1)+\margin}:{360/\n * (\s)-\margin}:\radius);
			}
			
			\node[node, fill=red!30] at ({360/\n * (1- 1)}:\radius) (1) {$\attck$};
			\node[node, fill=yellow!30] at ({360/\n * (2- 1)}:\radius) (2) {$v$};
			\node[node, fill=green!30] at ({360/\n * (3- 1)}:\radius) (3) {$u_3$};
			
			\draw[densely dotted, draw=red] (1) -- (3);
		\fi

			\def \angle {7}
		
		\tikzstyle{user} = [fill=green!30, circle,  text width=6mm, align=center]
		\node[user] (1) {$u_1$};

		\node[user, yshift=-2.5cm, xshift=1cm,  fill=red!30] (2) {$\dap$};

		\node[user, yshift=-3.3cm, xshift=3.5cm,  fill=yellow!30] (3) {$\victm$};

		\node[user, xshift=3cm, yshift=-.5cm] (4) {$u_3$};

		\draw[<->] (1) to node[above, xshift=-0.3cm] {} (2);
		
		\draw[<-] (2) to node[above] {$\mup{v}{t-\oot}$} (3);
	
		\draw[<-, draw=red!50] (2) to node[above,  xshift=-0.3cm] {$\mup{u_3}{t-\oot}$} (4);
		
		\draw[->] (4) to node[above, xshift=0.4cm] {$\mup{u_3}{t-\oot}$} (3);
		
		\end{tikzpicture}

	}

	\caption{$\dap$ is able to access the gradient of the victim node~$v$ because of its connection with the honest node $u_3$.}
	\label{fig:gi_top}
\end{figure}
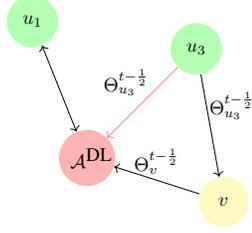
\begin{figure}[t]

	\fboxsep=0.01mm%padding thickness
	\fboxrule=0mm%border thickness
	\centering
%	\begin{subfigure}{.8\linewidth}
%		\fcolorbox{black!50}{black!50}{\includegraphics[trim = 90mm 5mm 70mm 5mm, clip, width=1\linewidth]{./images/gt_gi_mnist_32}}\\
%		\fcolorbox{red!50}{red!50}{\includegraphics[trim = 90mm 5mm 70mm 5mm, clip, width=1\linewidth]{./images/r_gi_mnist_32}}
%		\caption{MNIST \mr{maybe to rm}}
%	\end{subfigure}	\\

	\begin{subfigure}{1\linewidth}
		\fcolorbox{black!50}{black!50}{\includegraphics[trim = 30mm 1mm 23mm 1mm, clip, width=1\linewidth]{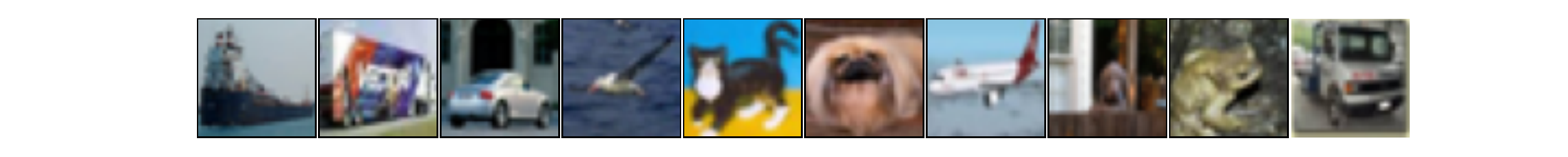}}\\
		\fcolorbox{red!50}{red!50}{\includegraphics[trim = 30mm 1mm 23mm 1mm, clip, width=1\linewidth]{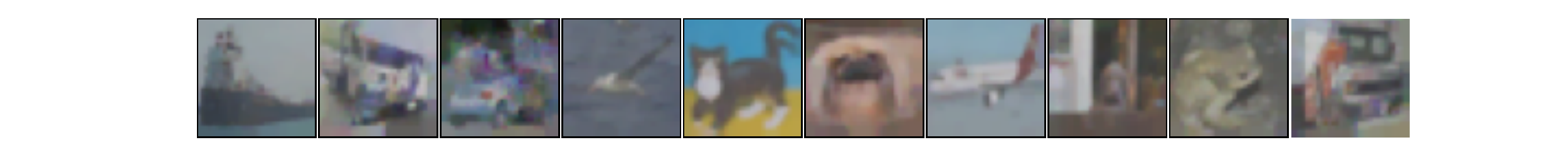}}
	\end{subfigure}	\\

	\caption{Examples of reconstruction (red panels) obtained via gradient inversion on the node $v$ (see Figure~\ref{fig:gi_top}) using~\cite{invg2} for a batch size of size $16$ on the CIFAR-10 dataset.}
	\label{fig:gi_example}
\end{figure}

Figure~\ref{fig:gi_top} illustrates this issue. The local model of the victim $\victm$ (in yellow) is:
\begin{equation}
\mup{\victm}{t} \myeq \frac{1}{3} (\Theta_{\victm}^{t-\frac{1}{2}}+\Theta_\dap^{t-\frac{1}{2}}+\Theta_{u_3}^{t-\frac{1}{2}}),
\label{eq:grex}
\end{equation}
 where $\Theta_{u_3}^{t-\frac{1}{2}}$ is a model update produced by another user who is not under the control of the attacker (\ie $u_3$). If the attacker $\dap$ (in red) also has access to $u_3$'s updates (\ie $\nn(\victm) \subset \nn(\dap)$), then they can recompute the local state $\mup{v}{t}$ as in Eq.~\ref{eq:grex} and recover the gradient signal from $v$'s model updates as $\nabla_{ \mup{\victm}{t}} \myeq \Theta_{\victm}^{t-\frac{1}{2}} - \mup{v}{t}$.

Once $\dap$ has $\nabla_{ \mup{\victm}{t}}$, they can run an arbitrary inversion attack. Figure~\ref{fig:gi_example} shows a sample of images reconstructed via gradient inversion for the topology in Figure~\ref{fig:gi_top} obtained using the optimization-based method proposed in~\cite{invg2}. Given the same underlying setup (e.g., same batch size), the result of the inversion attack is equivalent to the one that would be achieved by $\fsp$. Note that this is not a targeted attack and $\dap$ can perform gradient recovery on all the neighbors for which the condition is met simultaneously at every given round.

\parabf{The neighbors-discovery trick.}
\label{sec:nn_discovery}
A condition for $\dap$ to be able to perform the gradient recovery attack described above is that they must know the exact set of neighbors of their target $\victm$: $\nn(\victm)$. We now show how, even if the attacker does not know the global communication topology, they can learn $\nn(\victm)$ from the model updates whenever $\nn(\victm)\subseteq \nn(\dap)$.% (\ie the attacker is densely connected).  We refer to this approach as the \TT{\textbf{neighbors-discovery trick}}.\\ %The quality of the reconstruction only depends on the size of the batch used to compute the gradient, the number of parameters in the network, and the possible transformations applied on the model updates \eg compression  local differential privacy~\cite{huang2021evaluating}. % The quality of the inversion  can be improved if the adversary has auxiliary data~\cite{nvidiagi, jeon2021gradient}.

In a nutshell, the {neighbors-discovery trick} finds the model updates of the previous round that explain the victim's model update received at the current time step. This can be done by finding  $Q\subseteq \nn(\dap)$ such that:
%\begin{equation} \label{eq1}
\begin{gather}
		\
		 \text{argmin}_{Q} \| \mup{\victm}{t+\oot} - (\frac{1}{|Q|}\sum_{u\in Q}\Theta_u^{t-\frac{1}{2}}) \|.
		 \end{gather} 
%\end{equation}
When $Q \myeq \nn(\victm)$, the operation $\mup{\victm}{t+\frac{1}{2}} - \sum_{u \in Q}\Theta_u^{t-\frac{1}{2}}$ yields the gradient $\nabla_{\Theta_{\victm}^t}$ which is expected to be of low magnitude. Moreover, an attacker can iteratively apply this procedure across multiple rounds to refine their estimate of~$Q$ or directly employ gradient inversion attacks to verify if the recovered gradient leads to suitable reconstructions.
 %In Appendix~\ref{sec:mapping}, we empirically demonstrate the effectiveness of this discovery trick. 
%We note that Eq.~\ref{eq:find_nn} is linear and can be solved via linear/dynamic programming. 
The chances of successfully discovering the neighbors---and consequently the ability to perform gradient inversion---is maximized when the adversary has as many neighbors as possible.
%We recall that in current decentralized learning protocols, users (including the adversary) can choose \new{arbitrarily} whom they connect to.

\new{It is worth noting that techniques like gradient recovery and the neighbor discovery trick are effective even in a dynamic communication topology setting, where users select their neighbors on a round-by-round basis. The attacker only needs to be connected to the victim for a minimum of two consecutive rounds to execute such attacks.}
\vspace{\baselineskip}
Summing up, \textbf{a passive adversarial user in decentralized learning can be as powerful as a passive server in the federated setup}.
As we show, it is the case for all victims $v$ such that $\nn(v)\subseteq \nn(\dap)$.
Since DL allows the adversary to connect to users of their choice, hence to be connected to all users, $\dap$ eventually is as powerful as $\fsp$: just like an adversarial server, $\dap$ can (1)~observe the model update of every user in the system and (2)~isolate the individual gradients of a user. 
It is also trivially true when the DL topology is fully connected to begin with.

%Trivially, this is the case when the decentralized adversary is fully-connected. Just like an adversarial server $\fsp$, the adversarial user $\dap$ can (1)~observe all the {individual} model updates produced by every user in the system and (2)~isolate the individual gradients of a user. %regardless of their local communication patterns (by exploiting Eq.~\ref{eq:find_nn}). 
%Therefore, \textbf{any adversarial capability $\fsp$ has in an FL protocol can be simulated by an fully-connected adversarial user~$\dap$\ct{this is abuse of notation, $\dap$ does not need to be fully connected} in DL}. 
%Even if the adversary is not fully-connected, $\dap$ is as powerful as $\fsp$ for all users $v$ in the system for which $\nn(v)\subseteq \nn(\dap)$.
%\mr{It is the case for all users such that... When the adv it fully connected, this apply to all users.}

\subsection{Take aways}

In Section~\ref{sec:passive_user_vs_usr}, we show that an adversarial decentralized user can exploit the local generalization of any non-complete topology to launch membership inference attacks.
To limit this leakage, the density of the communication topology must increase, up to the complete topology, where there is no local generalization phenomenon anymore.
%, d the topology is complete, there is no local learning effects and leakage disappears. 
Increasing connectivity, however, is in conflict with the conclusions of Section~\ref{sec:passive_user_vs_server} which show that increasing connectivity increases the system knowledge of the adversary.  %the adversary's capability to collect knowledge of the system.
Giving the adversary the ability to collect additional information on the system, results in even more significant leakage, and enables powerful attacks such as gradient inversion.

We conclude that there is no topology $G$ for which DL provides better, or even equal, honest user protection against a passive adversary in the network than the one federated users enjoy. 
In other words, the adversary $\dap$ is more powerful than $\fup$ \emph{regardless} of the underlying DL setup.
%Section~\ref{sec:passive_user_vs_server} also shows that, as 
In addition, $\dap$ can acquire the same adversarial capabilities as $\fsp$, as long as there are no constraints on how users connect to each other in DL (or the topology is complete). 
As consequence, while in FL there is at most one powerful adversary: the server, in DL there may be multiple powerful adversaries: any user with enough connections. 
This means that contrary to what is claimed by its proponents, DL does not reduce the capabilities of adversaries. Rather \textbf{the power of adversaries, and so the privacy vulnerability of honest users, is multiplied}.

\section{Privacy Against Active Adversaries}
\label{sec:active}

In this section, we compare the privacy offered by the DL approach against the FL alternative in the malicious model, i.e., when there is an active adversary in the system. 
As in Section~\ref{sec:hbc}, we start by formalizing our threat model. % and then compare adversarial capabilities.

\noindent\textbf{Active adversary threat model.} Active adversaries in CML (eq. malicious) behave maliciously during the protocol execution. In this paper, we instantiate such an adversary by allowing the them to send arbitrary model updates to their neighbors in addition to their passive capabilities. We refer to an active adversarial user in DL as~$\daa$, and to a malicious user and a malicious parameter server in FL as $\fua$ and $\fsa$, respectively.
% We assume a \textit{rushing adversary}, meaning that the adversary is allowed to see the honest parties’ updates before sending its own. %\new{The attacker can also send different model updates to different neighbors during the same communication round (\ie model inconsistency~\cite{evading_sec}). 
%As in Section~\ref{sec:hbc}, we evaluate users’ privacy loss against an adversarial neighbor, under the assumption that the adversary has no auxiliary information.}
%
\subsection{Decentralized user vs federated user (active)}
\label{sec:influence_factor}
%While we already demonstrated the effect of local generalization in the honest-but-curious setting, this represents even a major concern in the malicious case. Indeed,
%As demonstrated in the federated setting~\cite{}, the effectiveness of a malicious adversarial user depends their capability to their capability of influencing the local state (local model parameters) of their victim.
%As it has been shown in the federated setting, malicious users can extract information from other users' training sets. Their effectiveness is proportional to their capability of influencing the \new{local} state, i.e., model parameters, of their victim~
In CML, the effectiveness of active attacks is proportional to the capability of adversarial users of influencing the model parameters of their victim~{\cite{deep_model_under_the_gan, melis, gasc}}.
% This is because the adversary attacks the victim's updates, which are a function of the user's local model parameters and private data.
Intuitively, this is because the adversary uses the victim model updates as input for making inferences. These updates are a function of the victim's local model parameters and training set. By influencing the victim's model parameters, the adversary can modify the model updates to leak more information about the private training set~\cite{papernotclone, fowl2022robbing, deep_model_under_the_gan, gasc, evading_sec, wen2022fishing}.
\par

In both FL and DL, a user computes their local model parameters $\victm$ as the aggregation of their own model update and the model updates of other users in the system:
\begin{equation}
	\label{eq:aggif}
	\mup{\victm}{t+1} = \frac{1}{m}\mup{\victm}{t+\oot}+ \frac{1}{m}\mup{u_1}{t+\oot}+\dots+ \frac{1}{m}\mup{u_{m-1}}{t+\oot},
\end{equation}
where $m$ is the number of users participating in the aggregation (all users in FL, and the neighbors of $\victm$ in DL). 

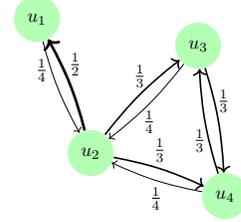
\begin{figure}[t]
	\centering
	%\begin{subfigure}{.23\textwidth}
		%	\centering
		\resizebox{.18\textwidth}{!}{%

			\begin{tikzpicture}
				\def \angle {7}
							
				\tikzstyle{user} = [fill=green!30, circle,  text width=5mm, align=center]
				\node[user] (1) {$u_1$};

				\node[user, yshift=-2.5cm, xshift=1cm] (2) {${u_2}$};

				\node[user, yshift=-3.3cm, xshift=3.5cm] (3) {${u_4}$};

				\node[user, xshift=3cm, yshift=-.5cm] (4) {$u_3$};

				\draw[->, line width=.1mm, bend right=\angle] (1) to node[above, xshift=-0.3cm] {$\frac{1}{4}$} (2);
				\draw[->, line width=.5mm, bend right=\angle] (2) to node[above, xshift=.1cm] {$\frac{1}{2}$} (1);
	
				\draw[->, line width=.3mm, bend left=\angle] (2) to node[above] {$\frac{1}{3}$} (3);
				\draw[->, line width=.1mm, bend left=\angle] (3) to node[below] {$\frac{1}{4}$} (2);
				
				\draw[->, line width=.3mm, bend left=\angle] (2) to node[above] {$\frac{1}{3}$} (4);
				\draw[->, line width=.1mm, bend left=\angle] (4) to node[below] {$\frac{1}{4}$} (2);
				
				\draw[->, line width=.3mm, bend left=\angle] (4) to node[above, xshift=0.1cm] {$\frac{1}{3}$} (3);
				\draw[->, line width=.3mm, bend left=\angle] (3) to node[below, xshift=-0.1cm] {$\frac{1}{3}$} (4);
			\end{tikzpicture} 
		}
	%\caption{Decentralized Learning}
	%\label{fig:dl}
	%\end{subfigure}

\iffalse

\begin{subfigure}{.23\textwidth}
			\centering
		\resizebox{1\textwidth}{!}{%
		\begin{tikzpicture}
			\tikzstyle{user} = [fill=green!30, circle,  text width=5mm]
			\node[user, fill=red!10] (1) {$\qquad$}; 
			\node[above of =1,  yshift=-.4cm] (1l) {P. server}; 
			
			\node[user, yshift=-2.5cm, xshift=-2cm] (2) {${\Theta^t}$};

			\node[user, yshift=-2.5cm, xshift=0cm] (3) {${\Theta^t}$};

			\node[user, yshift=-2.5cm, xshift=+2cm] (4) {${\Theta^t}$};

			\node[user, yshift=-2.5cm, xshift=+4cm] (4) {${\Theta^t}$}; 
			
			\draw[->] (2) -- node[midway, above, xshift=-.2cm] {$\Theta^{t+\oot}$} (1);
			\draw[->] (3) -- node[midway, above, xshift=-.3cm] {$\Theta^{t+\oot}$} (1);
			\draw[->] (4) -- node[midway, above, xshift=.3cm] {$\Theta^{t+\oot}$} (1);
		\end{tikzpicture} 
	}
		\caption{Federated Learning}
		\label{fig:fl}
	\end{subfigure}
\fi 
	\caption{Figure describing the \textbf{direct} influence factor (edge thickness and label) of each node on neighbors's local models for a given topology composed of four users.}
		\label{fig:if}
\end{figure}

Assuming that the model updates are bounded in norm, any user can influence at most a fraction $\frac{1}{m}$ of $v$'s model. The larger $m$ is, the smaller the influence a single adversarial user can have on $\victm$'s model. In the federated setting, by definition all users participate in the aggregation ($m=n$). Thus, the influence an adversarial user can have on their victim is the minimum possible.
In decentralized settings, the number of neighbors, and thus the level of adversarial influence, depends on the topology (see Figure~\ref{fig:if}). For the DL settings that offer a significant cost advantage with respect to FL, the topology is sparse and therefore users have a small number of neighbors ($m<<n$). In such scenario an adversarial user $\daa$ in DL always has higher \textbf{influence} over their targets' model parameters than an equivalent malicious user $\fua$ in FL. The influence is only equal to an FL adversarial user when the DL topology is fully connected, but at that point there is no advantage in decentralizing the learning process.
\par

Besides the increase of influence on each honest user, decentralization enables active adversarial users $\daa$ to send $m$ different updates to their $m$ neighbors each round. This is in contrast to the FL scenario where $\fua$ can only submit a single model update per round (to the server). This extra capability enable adversarial users in DL to carry out attacks that an adversarial FL user cannot launch.

We briefly introduce the \TT{echo attack}, as example of an active attack that can be performed by $\daa$ but not by $\fua$. (More details can be found in Appendix~\ref{sec:echo}.)
During an echo attack, the adversary `echoes' back each received model update \new{(or a variation, \eg the functionality marginalized version of the model update)} to the neighbor who sent it, in order to push them to overfit on their local training data. 
Our experimental results show a significant increase in the generalization error ($4\times$), and consequently in the privacy leakage, when performing the echo attack (see Figures~\ref{fig:gen_error_echo_iso} and~\ref{fig:mias_echo} in Appendix~\ref{sec:echo}).
The federated adversary $\fua$, which only receives the aggregated model from the server, cannot isolate users’ individual contributions and is left with one possibility: echoing the global model to the server, which will have little to no effect on the generalization error of users.  
In general, decentralized users can perform any active attack in reach for federated users (e.g.~\cite{deep_model_under_the_gan, melis, gasc}), while the converse is not true. %\ct{The paper would be stronger with some more about the echo attack in the main body to better support these claims}

%\todo{Moreover, congruently to the passive setting, decentralization also lends additional capabilities to active adversaries. For instance, while $\fua$ can only submit a single model update at each round, $\daa$ is able to provide $m$, possibly different, model updates to different users, resulting on higher degree of freedom to actively influence the system (see Section~\ref{sec:override}). Eventually, this means that the set of adversarial functionalities of $\fua$ is only a subset of $\daa$'s one. More pragmatically, this also tells us that every attack strategy in reach of federated users (\ie all currently exiting~{\cite{deep_model_under_the_gan, melis, gasc} and future active attacks) can be also performed by decentralized users, but not the opposite.}}

Adversarial influence, like local generalization, grows with sparsity of the underlying topology. Therefore, similarly to local generalization, it can be diminished by increasing the number of neighbors $m$ of the victim.
%\new{Note that the higher adversarial influence of DL attackers is nothing more than the natural byproduct of the local generalization phenomenon in the malicious setting. As such, those share the same underlying defense.} 
%In order to reduce the influence of potential malicious neighbors, an honest user~$\victm$ can only increase the number of neighbors $|\nn(\victm)| \myeq m$. 
When users are connected to all nodes in the system, $m$ is maximized and their models reach global generalization, \eg node $u_2$ in Figure~\ref{fig:if}. 
%
%
\iffalse
While increasing the number of neighbors reduces the adversary's capability to influence the victim's model, it provides the adversary with a new attack vector: the adversary can increase their influence by propagating malicious updates through those honest neighbors.
%The malicious updates broadcasted by the attacker are propagated through the other honest nodes.
For instance, in Figure~\ref{fig:if}, $u_2$ has additional influence on $u_3$ through $u_4$.
%The strength of this signal, as explained in Section~\ref{sec:localgen_and_systemk}, decays exponentially with the distance between users, and requires linear to time steps to manifest.
%
In fact, if there are no restrictions on connectivity, the attacker can create new connections to the victim's neighbors to further boost their influence on the victim's model. As in the case of passive adversaries, this results on conflicting connectivity requirements: to reduce the impact of active attacks users must be densely connected. But, to minimize the number of paths by which the adversary can influence their models, users should avoid having a large number of neighbors.
\par
\fi
\vspace{\baselineskip}

In conclusion, given that all the attacks that $\fua$ can perform can be also performed by $\daa$ and that $\daa$ has always greater influence on victims than~$\fua$, 
%an active adversarial user~$\daa$ in decentralized learning would be always capable of performing more powerful privacy attacks than an equivalent malicious user $\fua$ in FL. In other words, 
\textbf{any non-complete topology in DL offers less protection against active privacy attacks from malicious users than the equivalent federated approach}. In Section~\ref{sec:override}, we show that this result extends to the complete topology.
%Again, while point \textbf{(1)} tells us that $\fua$ can be simulated by $\daa$, the opposite does not hold, and there is a potentially large class of active privacy attacks that can be performed exclusively by $\daa$. 
%In Appendix~\ref{sec:echo}, we introduce one of them that we call \textit{echo attack}.\\

\iffalse
\subsubsection{Echo attacks}
\label{sec:echo}
\input{echo_att}
\fi
%
\subsection{Decentralized  user vs federated server (active)}
\label{sec:active_dec_vs_fed}
%When comparing the adversarial capabilities of a malicious, decentralized user against of a malicious  parameter server, things seem different.

We now compare the adversarial capabilities of active adversarial users ($\daa$) in decentralized learning against an active adversarial parameter server in FL ($\fsa$).
\par

A malicious parameter server $\fsa$ is the strongest active attacker possible: it can arbitrarily decide the local state of any user in one single iteration. This results in extremely effective privacy attacks~\cite{papernotclone,fowl2022robbing, evading_sec, wen2022fishing}.
%% CT -- why is this relevant??
%\footnote{\new{A malicious parameter server can provide arbitrary parameters to users at every iteration.}}
%
These attacks cannot be directly applied by $\daa$, as obtaining such influence on victims' models is inconceivable in the decentralized setting, regardless of the underlying topology.
% not within a limited number of rounds.\footnote{Yet, adversaries can achieve a similar\ct{similar to what? Why is this a footnote and not main text} effect using \textit{time coupled attacks}, where they influence the local state of the victim over multiple iterations~\cite{NEURIPS2019_ec1c5914}}.
Indeed, even when the attacker is the only neighbor of the victim, their influence on the victim's model is in theory at most $\frac{1}{2}$, since the victim aggregates the adversary's contribution with their own local information (see edge $(u_2, u_1)$ in  Figure~\ref{fig:if}).
%~\cite{papernotclone,fowl2022robbing, evading_sec, wen2022fishing} %\TEMP{The adversary needs multiple iterations to influence the local state of the victim~\cite{NEURIPS2019_ec1c5914}}. %However, if the adversary uses their system knowledge they can increase it completely determine the victims state using state override attacks.
%
However, we show how $\daa$ can use system knowledge to achieve full influence on a victim's state, just like a malicious federated server.
\subsubsection{State-override attack}
\label{sec:override}
In Section~\ref{sec:gi}, we show that system knowledge enables the attacker to remove the effect of generalization and isolate victims' gradients. % and the underlying consequences \eg a bounded influence factor.
This knowledge can also be used to cancel out contributions coming from honest neighbors on the victim's aggregated model.
The goal of the adversary is not to isolate information, but to increase its influence capability.
%in order to increase the influence capability of the adversary on this model. 
We now introduce the \TT{state-override attack}, in which the adversary uses this capability to \textit{override} the result of the local model aggregation computed by the victim at line $8$ of Algorithm~\ref{alg:dl}.
\par

Formally, given a target $\victm$ and an adversary $\attck$ such that $\nn(\victm)\subseteq \nn(\attck)$, the adversary can distribute the following model update to override the victim's model with parameters~$\tilde{\Theta}$ chosen by the adversary:
\begin{equation}
	\label{eq:payload}
	\mup{\attck}{t+\oot} = -(\sum_{u\in \nn(\victm)/\attck}\mup{u}{t+\oot}) + { |\nn(\victm)|} \cdot \tilde{\Theta}.
\end{equation}
This forged update contains the negated, partial aggregation in Eq.~\ref{eq:payload} of the model updates from the victim's neighbors~$\nn(\victm)/\attck$.

Upon receiving the model updates, the victim $\victm$ proceeds to aggregate the receive inputs locally:
\begin{multline}
	\label{eq:over_agg}
	\mup{\victm}{t+1} =  \frac{1}{ |\nn(\victm)|} \sum_{u\in \nn(\victm)}\mup{u}{t+\oot} = \\
	\frac{(\sum_{u\in \nn(\victm)/\attck}\mup{u}{t+\oot}) +  \mup{\attck}{t+\oot}]}{ |\nn(\victm)|}=\tilde{\Theta}.
\end{multline}
The adversary's update cancels the contribution of the neighbors, and the result of the aggregation becomes the \TT{payload} $\tilde{\Theta}$.
%That is, the local aggregation results in the adversarial parameters arbitrary chosen by the attacker, completely overriding the previous state of the node.

With this attack, the adversary can take complete control of the victim's parameters regardless the number of the victim's neighbors. As a result, the adversary can perform attacks such as~\cite{papernotclone, fowl2022robbing, evading_sec, wen2022fishing} within two iterations: one to override the model and one to extract the result. This is equivalent to a single round of federated learning.

\begin{figure}[t]
	\centering
	\includegraphics[trim = 80mm 10mm 63mm 10mm, clip, width=.95\linewidth]{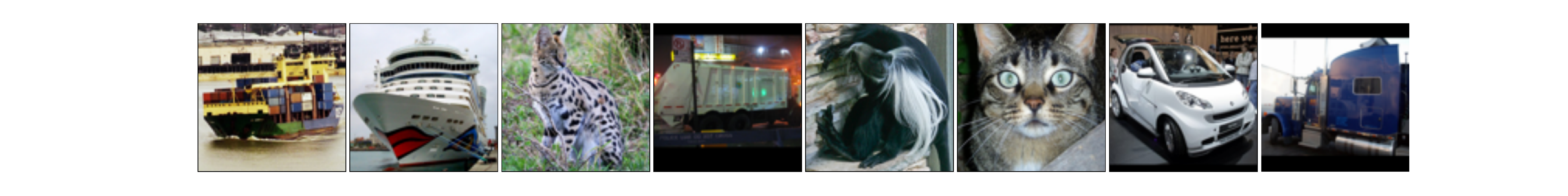}\\
	\includegraphics[trim = 80mm 10mm 63mm 10mm, clip, width=.95\linewidth]{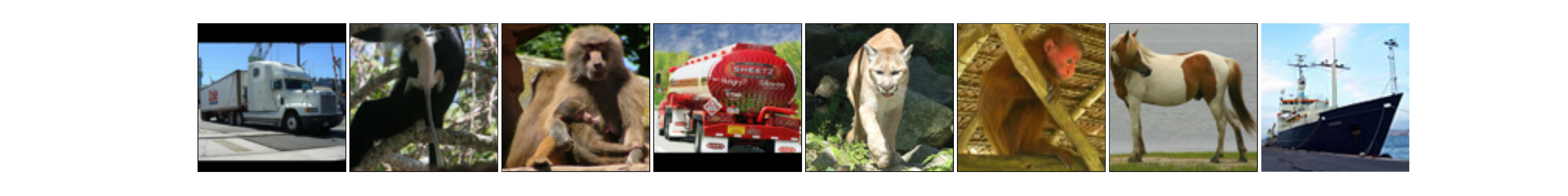}\\

	\caption{Examples of reconstruction obtained via state-override attack and gradient inversion (malicious initialization~\cite{papernotclone}) on the node $v$ (see Figure~\ref{fig:gi_top}) for a batch size of size $64$ on the STL10 dataset.}
	\label{fig:gi_example_act}

\end{figure}

To give a concrete example of the impact of the state-override attack on the privacy of users in DL, we consider active gradient inversion attacks.
Boenisch et al demonstrate in~\cite{papernotclone} that the effect of gradient inversion can be greatly magnified in both reconstruction quality and applicability when the attacker has full control on the parameters used to compute the gradient. The attacker can inject the victim's network with maliciously crafted parameters that force the computed gradient to artificially memorize more information than intended about the input batches~\cite{fowl2022robbing, wen2022fishing, papernotclone}. The state-override attack is the perfect way for the adversary to get control on the parameters used to compute the gradient.

In our setting, the attacker~$\daa$ first uses the \textit{state-override attack} to maliciously force the victim $\victm$'s local state to be the adversary-chosen parameters $\tilde{\Theta}$ created according to~\cite{papernotclone}. 
In the next round, $\daa$ first receives $\victm$'s model update. Second, recovers the gradient signal as~$\nabla_{\tilde{\Theta}}\loss(\xi_{\victm}^{t})  \myeq \mup{\victm}{t+\oot} - \tilde{\Theta}$. And third performs the inversion. We show the results of this attack in Figure~\ref{fig:gi_example_act}. %The quality of the reconstruction is undeniably good. It is actually the \textit{verbatim} copy of the original sample. %regardless the underlying data dimensionality (\eg resolution for images) and batch size which represent major limiting factors for passive gradient inversion attacks.}
It is important to note that the adversary $\attck$ can, at each round, simultaneously perform the state-override attack on all users whose neighbors are a subset of the adversary's neighbors. It suffices to send a different adversarial model update (computed according to Eq.\ref{eq:payload}) to each neighbor within the same communication round, and perform the gradient inversion steps discussed above in parallel.
% ($\{v\in \nn(\attck) | \nn(v) \subseteq \nn(\attck) \}$)

To perform the state-override attack as described above,~$\attck$ must be a \textit{rushing} adversary, \ie the last user to communicate its model update, so as to know the inputs of the victim's neighbors beforehand. If the adversary cannot do this (e.g., if the system has a broadcast schedule), $\attck$ can use model updates from the previous round and achieve comparable results, as we show in Appendix~\ref{app:override}.
%\new{\ct{I don't get the point of this sentence. Isn't this the case of all the attacks in this paper?}However, currently, this attack vector is unknown to the community and no solution has been proposed to prevent it.} 

Additionally, while in this work we focus on privacy attacks, the state-override attack can also be used as a stepping stone towards robustness attacks. Trivially, it enables the attacker to plant arbitrarily backdoor/trojan functionality in users' models~\cite{backdoor} or completely destroy the utility of their models (\eg by setting the payload to random parameters).

%Eventually, this is the prime example of how system-knowledge allows the attacker to subvert generalization, making even more evident the inherent  security trade-off discussed in Section~\ref{sec:localgen_and_systemk}.\\

\subsection{Take aways}
We show that active adversaries can gain full influence over honest users' state. This enables them to mount privacy attacks with extremely high quality of reconstruction of the users' training sets.
Like in the semi-honest setting, active adversarial users in decentralized learning can be as powerful as a malicious parameter server as long as the underlying topology allows for it. When the attacker is fully-connected, they can arbitrarily decide the parameters of \textbf{all the honest users in the system every round}, matching the functionality of the FL server.
\new{In practice, this means that, with respect to current real-world deployments of federated learning~\cite{gboard0, mcmahan2017learning, gboard1} in which the parameter server must be (semi-)honest in order to guarantee a meaningful level of privacy to users~\cite{papernotclone, fowl2022robbing,  evading_sec}, the decentralized learning paradigm increases the number of potential adversaries, and thus the number of entities that need to be (semi-)honest for users to have privacy}.

\section{Defences}
\label{sec:opm}
In this section, we discuss the suitability of different defense techniques from the CML literature and their effectiveness to prevent the attacks we introduce in this the paper.

\subsection{Secure Aggregation}
\label{sec:evade_sa}
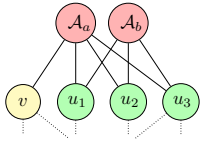
\begin{figure}[t]
	\centering

	\resizebox{.15\textwidth}{!}{%

		\begin{tikzpicture}
			\def \angle {7}
			
			\tikzstyle{user} = [draw, fill=green!30, circle,  text width=3mm, align=center]
			\node[user, fill=red!30] (aa) {$\attck_a$}; 
			
			\node[user, fill=red!30, right of=aa, ] (ab) {$\attck_b$}; 
			
			\node[user,  fill=yellow!30, yshift=-1.5cm, xshift=-1cm] (v) {$\victm$}; 
			\node[user, right of=v] (1) {$u_1$}; 
			\node[user, right of=1] (2) {$u_2$}; 
			\node[user, right of=2] (3) {$u_3$};

			\node[user, yshift= .2cm, below of=v, draw=none,  fill=green!0, text width=0mm] (iv){}; 
			\node[user, right of=iv, draw=none,  fill=green!0, text width=0mm] (i1){}; 
			\node[user, right of=i1, draw=none,  fill=green!0, text width=0mm] (i2){}; 
			\node[user, right of=i2, draw=none,  fill=green!0, text width=0mm] (i3){};

		\draw[-, densely dotted] (v) -- (iv);
		\draw[-, densely dotted] (1) -- (i1);
		\draw[-, densely dotted] (v) -- (i1);
		\draw[-, densely dotted] (2) -- (i2);
		\draw[-, densely dotted] (3) -- (i3);
		\draw[-, densely dotted] (3) -- (i2);
			
		\draw[-] (aa) -- (v);
		\draw[-] (aa) -- (1);
		\draw[-] (aa) -- (2);
		\draw[-] (aa) -- (3);
		
		\draw[-] (ab) -- (1);
		\draw[-] (ab) -- (2);
		\draw[-] (ab) -- (3);

		\end{tikzpicture} 
	}

	\caption{Minimal example of secure aggregation evasion for an aggregation threshold of $3$ users.  %The attacker controls the nodes $\attck_a$ and $\attck_b$ and recovers the exact model update produced by $\victm$.
	}
	\label{fig:esa}
\end{figure}

A way to prevent \new{system-knowledge-based} attacks is to use secure aggregation (SA) protocols~\cite{secagg}.
When using SA, users privately perform the aggregation step (line $8$ of Algorithm~\ref{alg:dl}) without revealing their model updates in \new{the} clear to each other.
Users \new{can} only access the result of the aggregation after their update has been averaged with those of other users.
Since individual model updates are not observable by the adversary, SA can eliminate attacks relying on {system knowledge} such as gradient recovery, functional marginalization, or state-override.\footnote{Although gradient recovery would unavoidably succeed when $\nn(\victm)\myeq\{\attck\}$ (always) or $\nn(\attck)\myeq\{\victm\}$ (when $C\myeq 0$).} 

\new{
Therefore, by using SA and a fully-connected topology to achieve global generalization, DL offers users the same level of privacy as federated learning against passive adversaries (see Section~\ref{sec:localgen_and_systemk}). However, this setting would impose a significant overhead: every decentralized user has the same communication complexity as a parameter server in FL, and in addition the overhead imposed by cryptographic operations needed for secure aggregation.} This overhead with respect to FL would come at no gain in privacy.

%\new{However, conversely, the security offered by secure aggregation scales with the number of participants in the protocol. With more model updates aggregated, less information can be deduced about a single user's update. Hence, sparse topologies reduce the effectiveness of SA, and decentralized secure aggregation offers less privacy to users than federated learning for non-complete communication topologies. Therefore, a sparsely connected decentralized user learns more information about their neighbors' model updates than a federated server. This suggests that communication efficiency and privacy are in tension in decentralized learning, and achieving both simultaneously may be unfeasible.}

\parabf{Evading SA in Decentralized Learning} \new{
Even if the overhead introduced by the protocol would be acceptable, securely implementing SA in decentralized learning poses a significant challenge. We demonstrate below how an attacker can always retrieve the model update of another user if they can impersonate or compromise an additional node in the system.}

Essentially, the attacker can obtain the model update of a victim~$\victm$ by calculating the difference between two aggregated values that differ only by~$\victm$'s model update. 
More formally, given $\attck_a$ and $\attck_b$ the nodes under the control of the attacker $\attck$ and a victim node $\victm$, $\attck$ can recover the victim's model update $\mup{\victm}{t}$, by choosing  $\nn(\attck_b) \myeq \nn(\attck_a) / \victm$. Once the attacker nodes received the aggregated values, these can recover $\victm$'s model update by computing: $\mup{\victm}{t}\myeq \text{SA}(\sum_{u}^{\nn(\attck_a)}) - \text{SA}(\sum_{u}^{\nn(\attck_b)})$. An example of this configuration is depicted in Figure~\ref{fig:esa}. This approach does not require any auxiliary knowledge on the victim, and $\nn(\attck_a)$ can be chosen arbitrarily by the attacker. We remark that this simple SA-evasion technique is independent from the employed aggregation protocol and they would work even under verifiable SA or SA performed via Trusted Execution Environment (TEE)~\cite{nguyen2021federated}. Assuming fault resilient SA~\cite{secagg} (which is necessary under real-world deployments), this strategy would work also in a complete topology, where $\nn(\attck_a) \myeq \nn(\attck_b)$. In this case, it is enough for $\attck_b$ to simulate the drop-off of the victim. In the general case, this technique would be remain applicable as long as the threshold for SA is greater equal to $|\nn(\attck_b)|-1$. 
\par

More research is needed to find effective and reliable topology-aware SA in decentralized learning.

\paragraph{\textbf{Differential privacy}}
\label{sec:dp}
A formal approach to achieve privacy in DL would be using Differential Privacy (DP)~\cite{DPref}, such as differential-private SGD to implement the local optimization steps. 

In decentralized learning, the lack of a trusted, centralized curator (role taken by the parameter server in federated learning) prevents the use of central-DP. Thus, DL protocols have to resort to local-DP.	
Local-DP results in a worse trade-off between privacy and utility compared to central-DP~\cite{kairouz2021distributed, naseri2020local}.	
One common way to improve this trade-off is to use distributed-DP~\cite{chen2022fundamental, kairouz2021distributed} which assumes the existence of an effective secure aggregation protocol that would only reveal the noisy sum of the local model updates to the aggregator. Distributed-DP allows to tune the local noise proportionally to the number of users $m$ participating at the aggregation~($\sim \frac{1}{m}$). {As for plain secure aggregation (see Section~\ref{sec:evade_sa})}, the success of this approach depends on the density of the topology. The lower the number of neighbors of a user, the less participants in the aggregation, and the more noise users have to add locally to achieve a desired level of privacy. Increasing the number of neighbors would solve this issue, but would also increase the communication overhead, suppressing the advantage of decentralized learning over the federated approach. Indeed, as for SA, distributed-DP matches the utility of federated learning only when the topology is complete.	
%\footnote{Indeed, as for SA, achieving the same utility/privacy trade-off  of the federated setting would require a complete topology ($m \myeq n$ for every node).}	
%, but  this also  means increasing communication overhead, defeating the underlying objective of decentralize learning.	
%Besides that, this still hinges on the assumption that SA can be soundly deployed in decentralized learning.	

In summary, decentralized learning cannot match the utility/privacy trade-off of the federated setting given existing DP techniques due to the lack of a centralized curator and the need to keep its communication overhead advantage. This gap may be reduced using differentially-private techniques tailored to decentralized learning. The community already started moving in this direction~\cite{dp2, cyffers2020privacy, xiao2019local}, so far achieving only limited results.%has still not found a suitable protocol that can solve all the problems we highlight in this paper.	

Finally, while these perturbation-based defenses may work, they can be also applied in FL. Thus, they do not result on any privacy advantage for the decentralized setting. In fact, techniques such as distributed-DP~\cite{chen2022fundamental, kairouz2021distributed} (which uses SA as a primitive) are easier, and more efficient, to apply to FL protocols compared to DL.

\subsection{Robust aggregation protocols }
Robust aggregation methods~\cite{karimireddy2021learning} aim at reducing the influence of active adversaries on the local state of users by replacing the plain average-based aggregation (line $8$ of Algorithm~\ref{alg:dl}) with more robust \new{metrics}. These techniques can neither prevent the privacy attacks we propose nor confer any advantage to decentralization.

Robust aggregation techniques trade privacy for robustness as they rely on magnifying the influence of local information (the current state of the user) over external one (model updates provided by other users). This amplifies the local generalization effect, increasing the information our attacks can exploit.
Robust aggregation can \emph{hamper} attacks such as the state-override attack, but not prevent them entirely. 
We demonstrate this is the case in Appendix~\ref{app:selfclip}, where we apply our attacks to self-centering clipping~\cite{he2022byzantine}---a robust aggregation protocol for DL. 

Furthermore, robust aggregation can also be applied to the federated setting by letting federated users maintain a consistent local state and implement any user-side robust aggregation. Therefore, they do not provide an advantage for DL with respect to the federated approach.

%In general, robust aggregation techniques can only trade privacy for robustness as they either magnify the influence of the local training, leaking information; or protect this information by letting model updates provided by neighbors influence the local training (see the effect of echo attacks on self-centering clipping~\cite{he2022byzantine} in Section~\ref{app:selfclip}).

%In general, robust aggregation techniques can only trade privacy for robustness as they  magnify the influence internal ones  (\ie the current state of the model).

%In conclusion, while the communication topology is the most relevant factor in defining the security of the decentralized learning, it is not clear how this can be chosen (and by whom) in order to guarantee some level of security to users.

\subsection{Constraining the communication topology}
\label{sec:const_top}

Decentralized learning advocates often point out that freedom to choose neighbors is a positive and unique feature of decentralization. In this paper, we thoroughly demonstrate that such freedom can be leveraged by the adversary to boost their capabilities to the point of achieving the same attack power as the parameter server in federated learning.

To address this issue, the communication topology underlying DL should be carefully designed if we wish to prevent certain attacks.
This means that systems in which users join the network without constraints are unworkable, as individual decisions are unlikely to match any pre-defined topology. In fact, it is actually hard to enforce constraints without a central orchestrator that has a global knowledge of the system as highlighted by years of research on peer-to-peer anonymous communications~\cite{herbivore:tr,MittalB09,ShiraziSABD18,SchuchardDHHK10,WangMB10}. Yet, introducing such a powerful central entity in the system would result on new security threats if this entity is malicious: Assuming a malicious central orchestrator who can arbitrary choose the communication topology is equivalent to assuming a malicious parameter server in FL. Trivially, the orchestrator can maliciously design the topology in order to grant full adversarial capability to itself (and carry the attacks in Sections~\ref{sec:passive_user_vs_server}~and~\ref{sec:active_dec_vs_fed}).

%More research is needed to create DL systems with secure joining processes that constraint users' positioning in the network to mitigate the attacks introduced in this paper.

\new{n siItuations where trust can been established among users within the system, the ability to choose neighbors can also have a positive impact. If users have the ability to differentiate between trustworthy and untrustworthy nodes, they can selectively connect with users whom they trust to be honest and reject connection requests from those who are untrusted. This approach can lead to secure configurations without the need for strict topological constraints, as it assumes that no honest user would connect to the untrusted adversarial nodes. However, this approach requires making strong assumptions about both the behavior of users and the security of the underlying implementation.}%\footnote{\new{For instance, users should also trust that no one of their chosen neighbors can be compromised or impersonated by an adversary.}}

\section{Conclusion}
\label{sec:conclusion}

\new{Fully decentralized collaborative machine learning has been proposed as a potential solution for preserving user privacy while avoiding the performance issues associated with federated learning. In this work, we introduced a series of attacks that show that existing decentralized learning protocols do not deliver on their promised privacy properties. Our results indicate that decentralized users possess adversarial capabilities that are comparable to those of a federated parameter server.}

\new{We also show that due to the increase in capabilities decentralization confers to adversaries, existing defenses cannot prevent all the attacks we propose. In order for decentralized learning to provide the same level of privacy guarantees as federated learning, it must give up on any potential performance gains. We hope that our findings can serve as benchmarks for the research community, inspiring the development of new design principles that enable truly privacy-preserving decentralized learning.}

\section*{Acknowledgements} This work was partially supported by \textit{Fondation~Botnar}. We thank Fabrice Nemo for pointing out an error in the original formalization of the neighbors discovery trick.

\bibliographystyle{IEEEtranS}
\bibliography{bib}

\appendices
\section{Echo attacks}
\label{sec:echo}
\everypar{\looseness=-1}

%In this section, we describe two new active attacks on decentralized learning rooted on the capability malicious users have to influence the local parameters of their victims, and discuss how existing attacks can be used in the decentralized scenario. We compare the capabilities of such an adversary with those of a federated parameter server in Section~\ref{sec:user_vs_server_act}.
%In the next Section, we show how an attacker node in decentralized learning can exploit its inherent influence on honest nodes to further jeopardize their privacy.
%
%

%
%
%
%
\begin{figure}[b]
	\centering
	\includegraphics[trim = 0mm 3mm 0mm 0mm, clip, width=.8\linewidth]{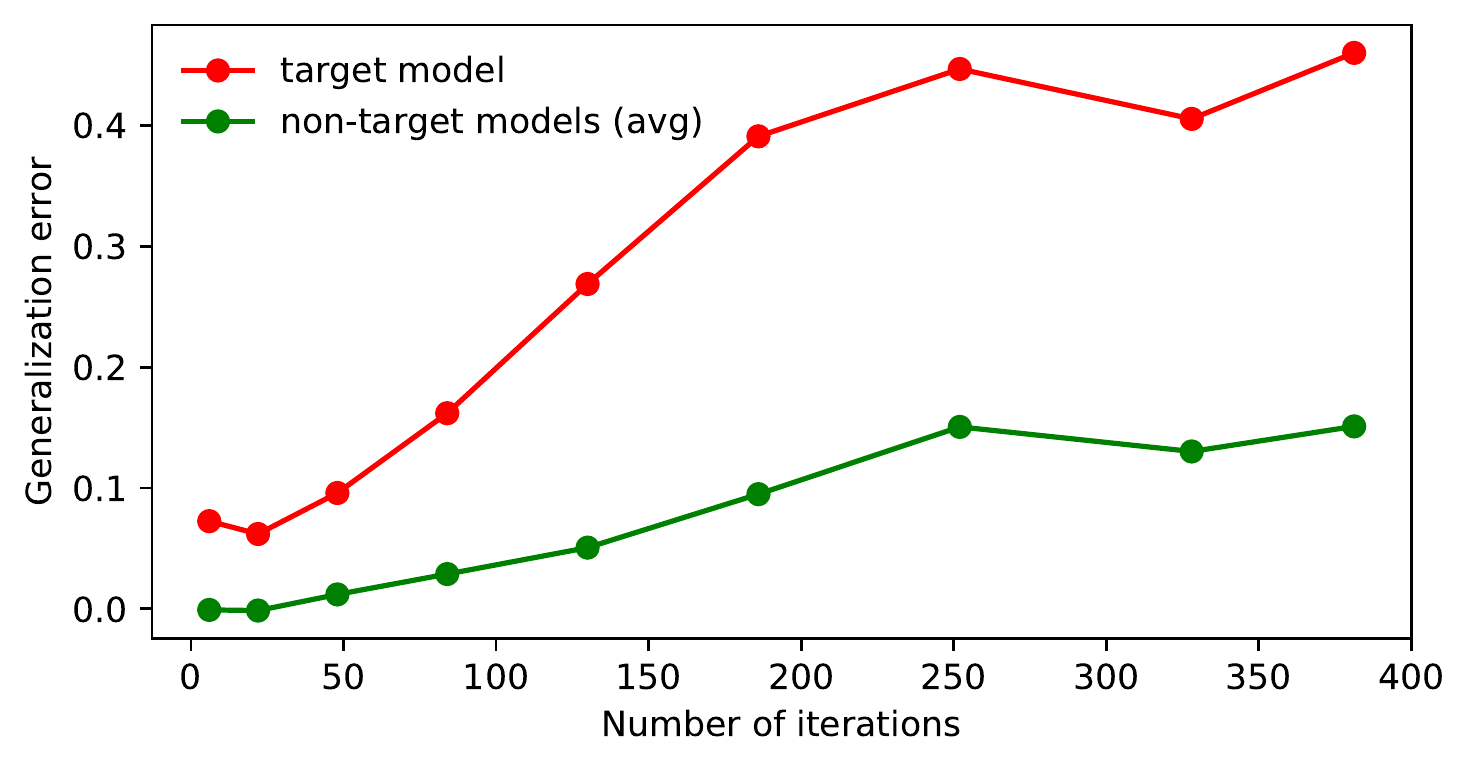}
	\caption{Generalization errors for the target user of the echo attack and the non-target ones during the protocol for the setup: torus-36, ResNet20 and CIFAR-100.}
	\label{fig:gen_error_echo_iso}
\end{figure}
%

%We now introduce the \TT{echo attack}, a new, extremely simple, active attack that improves the adversary's inference capabilities. 
\iffalse
This attack is tailored to the capabilities of an adversarial user in decentralized learning, and it adds to other active attacks on federated learning that remain fully applicable in the decentralized setting\ct{refs, which attacks?}.
We also show that current defensive techniques against active attackers in decentralized learning, fail to stop the echo attack. % highlighting their  fallacy in the privacy context.\\
\fi
 %\new{\TT{Echo attacks}, are new, extremely simple, active attacks that cannot be replicated in the federated setup. With this attack, we challenge the assumptions behind current defensive techniques aimed at hampering active attackers in decentralized learning, highlighting their limited effect in the privacy context.}

%We now introduce \TT{echo attacks}, active attacks that exploit all the adversarial capabilities enabled by decentralized learning protocols and that cannot be replicated by a malicious federated user.
Echo attacks exploit the high influence factor of a decentralized attacker and system knowledge to force the local state of a chosen neighbor victim to leak information about its private training set.
The aim of the attacker is driving the victim's local model towards severe \TT{overfitting}, forcing it to memorize the local training set beyond what would be memorized in a honest execution. This amplifies the information leakage stemming from local generalization, {worsening the impact of the attacks presented in the previous sections.}\looseness=-1

To carry out an echo attack, at the broadcast step, instead of their own model, the adversary $\daa$ broadcasts echos of the victim's updates at the current or previous round. Such \textit{echo} updates artificially increase the relevance of the victim's local training set in the victim's local model. This diminishes the influence of other users in the system on the victim's model, magnifying the effect of local generalization.
More formally, during the echo attack, the adversary starts by collecting the neighbors' model updates and uses them to craft the echo update $\tilde{\Theta}$. In particular, the attacker uses the functionally marginalized version (Section~\ref{sec:fun_iso}) of the victim model update (\ie $\tilde{\Theta} \myeq \isom{\victm}{t}$) which approximately captures the isolated contribution of the victim. 
%As such, feeding this update to the victim should drive their local model towards the minimum defined by its local training set.} 
The adversary broadcasts $\tilde{\Theta}$ to all the attacker's neighbors including the victim, increasing overfitting at the next step of the victim's honest execution (line $8$ Algorithm~\ref{alg:dl}). The effect of echo attacks is magnified by the iterative interaction between the attacker and the victim and by the \TT{echo chamber effect} that results from the neighbors of the attacker also propagating the malicious echo update to the victim via second order interactions. We formalize echo attacks in Algorithm~\ref{algo:echo}. 
\par 

 Echo attacks are extremely efficient and easy to carry: the adversary does not require a local training set or any information on the learning task and they have very low computational cost, as it does not need to train a local model but only post-process the received updates.  Note that malicious FL users ($\fua$) cannot replicate echo attacks as they cannot isolate other users' individual contributions from the observable global state of the system and, thus, they cannot broadcast echo updates.
\par

While conceptually simple, echo attacks are extremely effective in practice. Figure~\ref{fig:gen_error_echo_iso} compares the generalization error of the victim's model against other non-target users in the system at different training iterations during the echo attack. While this gap is larger at the start of the training, on average, the generalization error of the target is about $10$ times more than the non-targets'. As seen in the previous section, the increase in the generalization error creates a massive privacy risk for the target node. 

\begin{figure*}[t!]
	\centering
	\includegraphics[trim=0mm 100mm 0mm 0mm, clip, width=1.5\columnwidth]{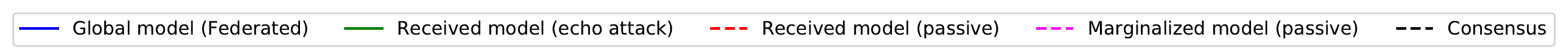}\\
	\begin{subfigure}{.5\columnwidth}
		\centering
		\includegraphics[trim = 0mm 0mm 0mm 0mm, clip, width=.95\columnwidth]{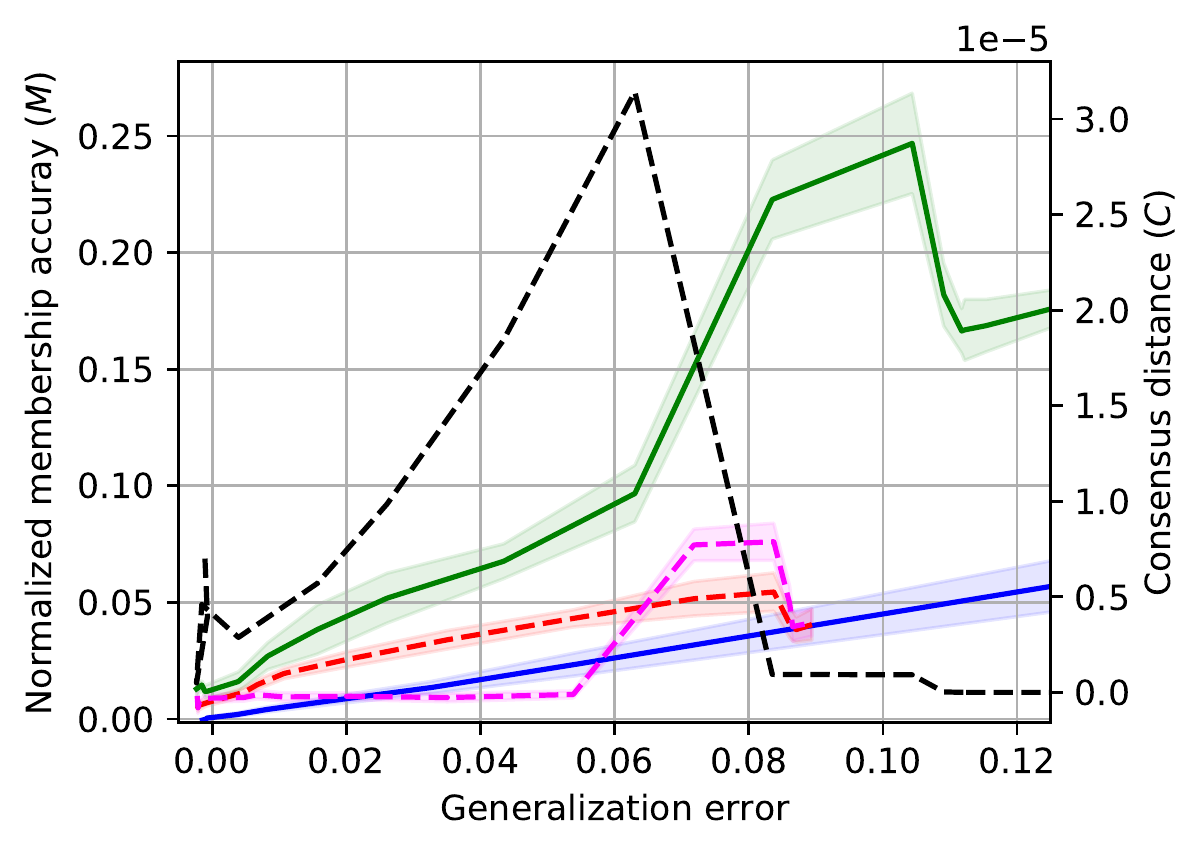}
		\caption{torus-36 \& CIFAR-10}
	\end{subfigure}\begin{subfigure}{.5\columnwidth}
		\centering
		\includegraphics[trim = 0mm 0mm 0mm 0mm, clip, width=.95\columnwidth]{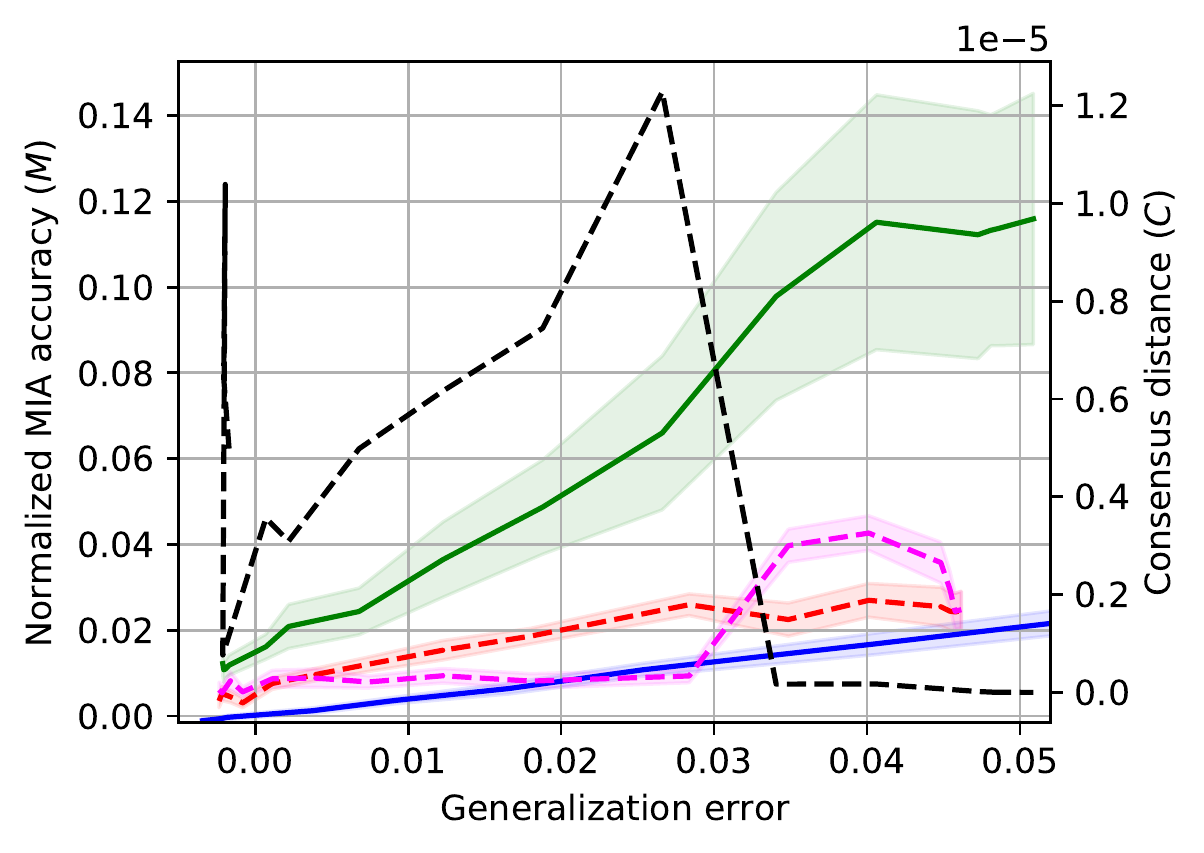}
		\caption{social-32 \& CIFAR-10}
	\end{subfigure}\begin{subfigure}{.5\columnwidth}
		\centering
		\includegraphics[trim = 0mm 0mm 0mm 0mm, clip, width=.95\columnwidth]{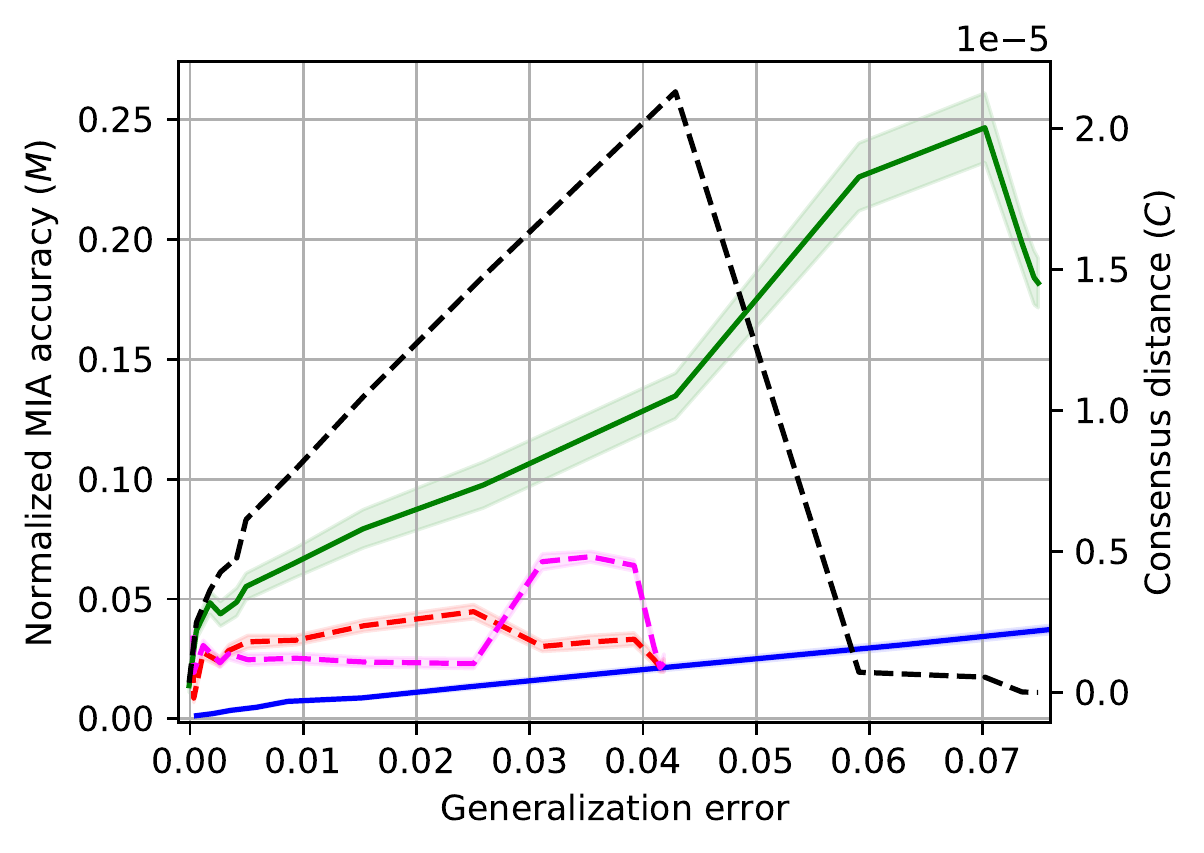}
		\caption{torus-36 \& CIFAR-100}
	\end{subfigure}\begin{subfigure}{.5\columnwidth}
		\centering
		\includegraphics[trim = 0mm 0mm 0mm 0mm, clip, width=.95\columnwidth]{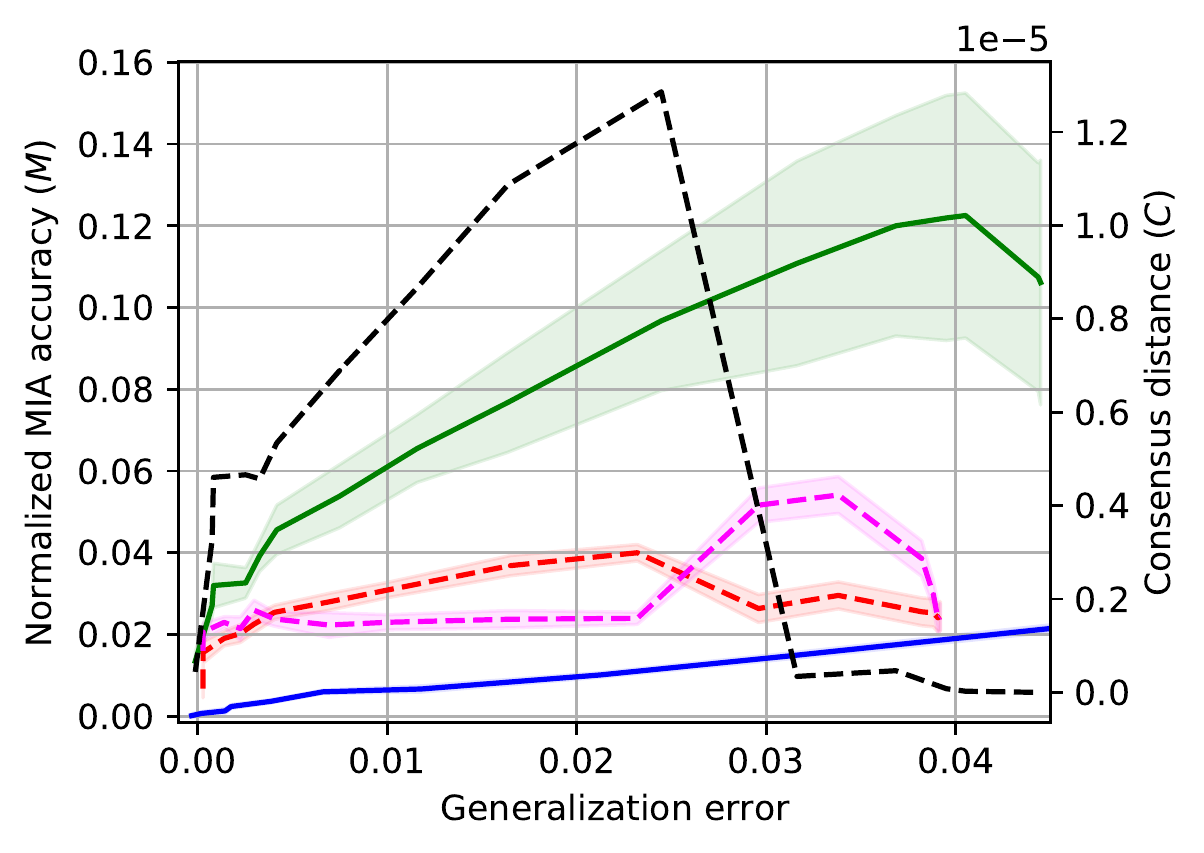}
		\caption{social-32 \& CIFAR-100}
	\end{subfigure}
	\caption{Average MIA vulnerability during an \textit{echo} attack, on four different combinations of communication topologies and training sets for DL and FL.% The halo surrounding the curves reports the standard deviation over the multiple runs.
	}
	\label{fig:mias_echo}
\end{figure*}

\begin{figure}[h]
	\centering
	\includegraphics[trim = 0mm 3mm 0mm 0mm, clip, width=.7\linewidth]{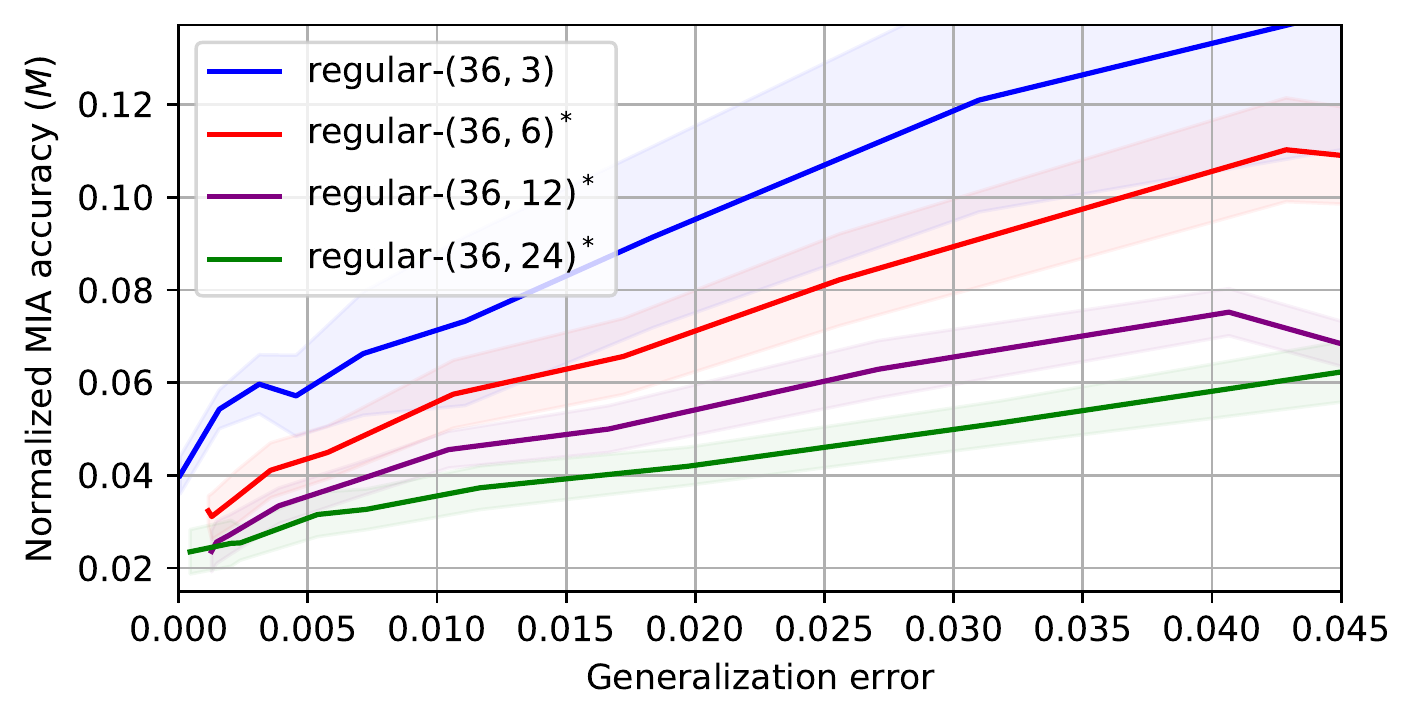}
	\caption{Effect of different numbers of neighbors for the target of the echo attack using CIFAR-100 as training set. }
	\label{fig:echo_diff_num_of_neig}
\end{figure}

We show the effectiveness of the echo attack in Figure~\ref{fig:mias_echo} on various configurations (green lines) and compare them to the results obtained with the passive inference attacks of Section~\ref{sec:passive_user_vs_user} (dashed lines). Even when the system finds consensus, the privacy risk for the target remains high. This is because the attacker's echo updates have actively influenced the global state of the system (not only the victims's one) by artificially increasing the relevance of the victims's contribution. 
For the \textit{social-32} topologies, we observe a large standard deviation. This is because the impact of the attack depends on the connectivity of the victim. Recall that the strength of an active attack is proportional to the influence factor of the attacker, which is inversely proportional to the number of neighbors of the victim (see Section~\ref{sec:influence_factor}). We illustrate this phenomenon in Figure~\ref{fig:echo_diff_num_of_neig}, where we evaluate the effect of the echo attack on targets with different number of neighbors on regular graphs with an increasing density. We keep the degree of the attacker fixed to $3$ in order to isolate the impact of the victim's connectivity on the privacy risk. We see that, as we predicted in Section~\ref{sec:influence_factor}, low degree boosts the impact of active attacks on users.

Also, attackers can improve their effectiveness by choosing their position in the communication topology to maximize their influence on the system. Like for gradient inversion, the best strategy is to maximize their number of neighbors. If this is not possible,  attackers can also aim to be in a position that maximizes the closeness centrality (or other centrality metrics) with the victim to strengthen the \TT{echo chamber effect}. However, adversaries can only use this strategy if they know the global topology. Finally, we note that if the attacker has the victim as sole neighbor or the marginalized model cannot be computed, the adversarial model update can be set to $\tilde{\Theta} \myeq \mup{\victm}{t+\oot}$ (\ie victim's model update), obtaining inferior but comparable performance; {we show this in Figure~\ref{fig:echo_rec_vs_mar} in Appendix~\ref{app:add_res}}.

\subsection{Echo attack on robust aggregation.}
\label{app:selfclip}
\begin{figure}[t]
	\centering
	\includegraphics[trim = 0mm 3mm 0mm 0mm, clip, width=.7\linewidth]{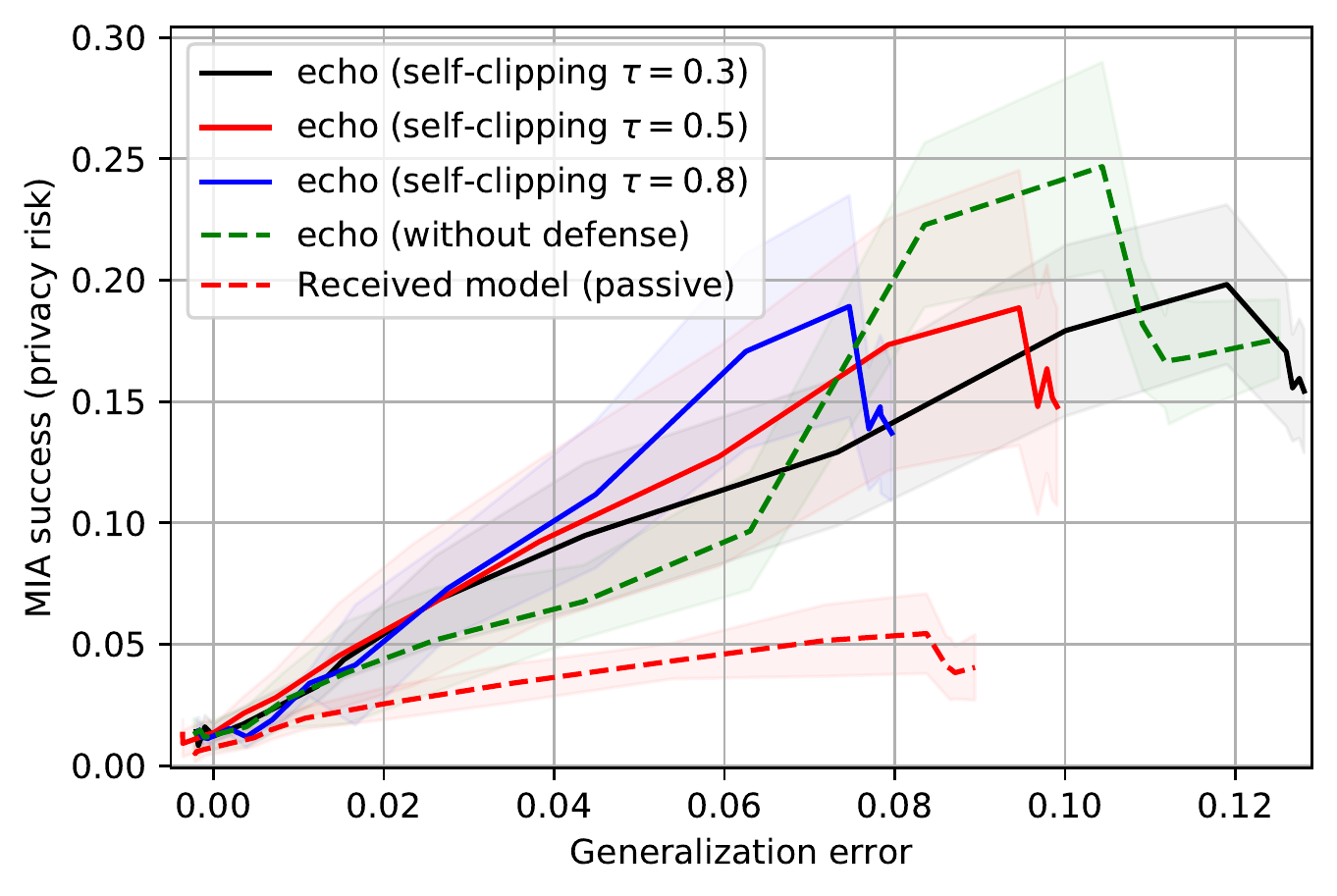}
	\caption{Effect of the self-centered clipping robust aggregation on the echo attack for \textit{torus-36} and \textit{CIFAR-10}. }
	\label{fig:self_clip}
\end{figure}
One common approach to reduce the adversarial influence of active attackers in both the federated and decentralized setting is to use robust aggregation methods~\cite{karimireddy2021learning}. An example for the decentralized setup is the work of He~\etal~\cite{he2022byzantine}. This work proposes to hamper the influence of byzantine nodes by using self-centered clipping. Nodes clip the received model updates in the $\tau$-sphere around their current local model before aggregating them:
\begin{equation}
	\mup{v}{t+1} \myeq \sum_{u \in \nn(v)} \biggl[w_{i,j} \cdot (\mup{v}{t+\oot} + \texttt{CLIP}(\mup{u}{t+\oot} - \mup{v}{t+\oot}, \tau)) \biggr],
\end{equation}
where $\texttt{CLIP}(x, \tau) \myeq \text{min}(1, \tau / ||x||)\cdot x$. 

This approach hides a trade-off between generalization and robustness. The clipping procedure simply degrades the information provided by the other users in the system in favor of the local one. 
{This successfully reduces the effectiveness of general active attacks.} However, it also reduces the generalization of the users' local models, magnifying the harmful effect of local generalization. Because there is less information from others, the local model updates retain more information about the local training set of the user. 

Eventually, self-centered clipping produces a very similar effect than an echo attack: the influence of local parameters is magnified. Therefore, this defense tends to amplify our attacks rather then defending from them. We show this effect in Figure~\ref{fig:self_clip}, where we compare the performance of echo attacks on systems  with and without self-centered-clipping~\cite{he2022byzantine}.
Of course, when $\tau$ gets closer to $0$, the system degenerates to non-collaborative learning (every node trains its model locally). In this case, active attacks such as echo does not offer much inference advantage to the adversary. %Then, active adversaries become as effective as passive ones.%and the echo attack would not offer any advantage. 

\begin{algorithm}
	%\tiny
	\KwIn{victim node: $\victm$}
	\For{$t \in [0, 1, \dots]$}{
		\tcc{Receive model updates from neighbors}
		\For{$u \in \nn(\attck) / \{\attck\}$}{
			\texttt{receive} $\mup{u}{t+\oot}$ \texttt{from} $u$\;
		}
		\tcc{Forge adversarial model update}
		$\tilde{\Theta} \myeq \isom{\victm}{t} \myeq  (| \nn(\attck)|-1) (\mup{\victm}{t+\oot} - \frac{\sum_{u\in \nn(\attck)/\{v, \attck\}} \mup{u}{t+\oot}}{| \nn(\attck)|-1})$\;
		
		\tcc{Broadcast the malicious model update}
		\For{$u \in \nn(\attck) / \{\attck\}$}{
			\texttt{send} $\tilde{\Theta}$ \texttt{to} $u$\;
		}
	}
	\caption{\textit{echo} attack for an active attacker node $\attck$.}
		\label{algo:echo}
\end{algorithm}
\begin{figure}
	\centering
	\includegraphics[trim = 0mm 3mm 0mm 0mm, clip, width=.8\linewidth]{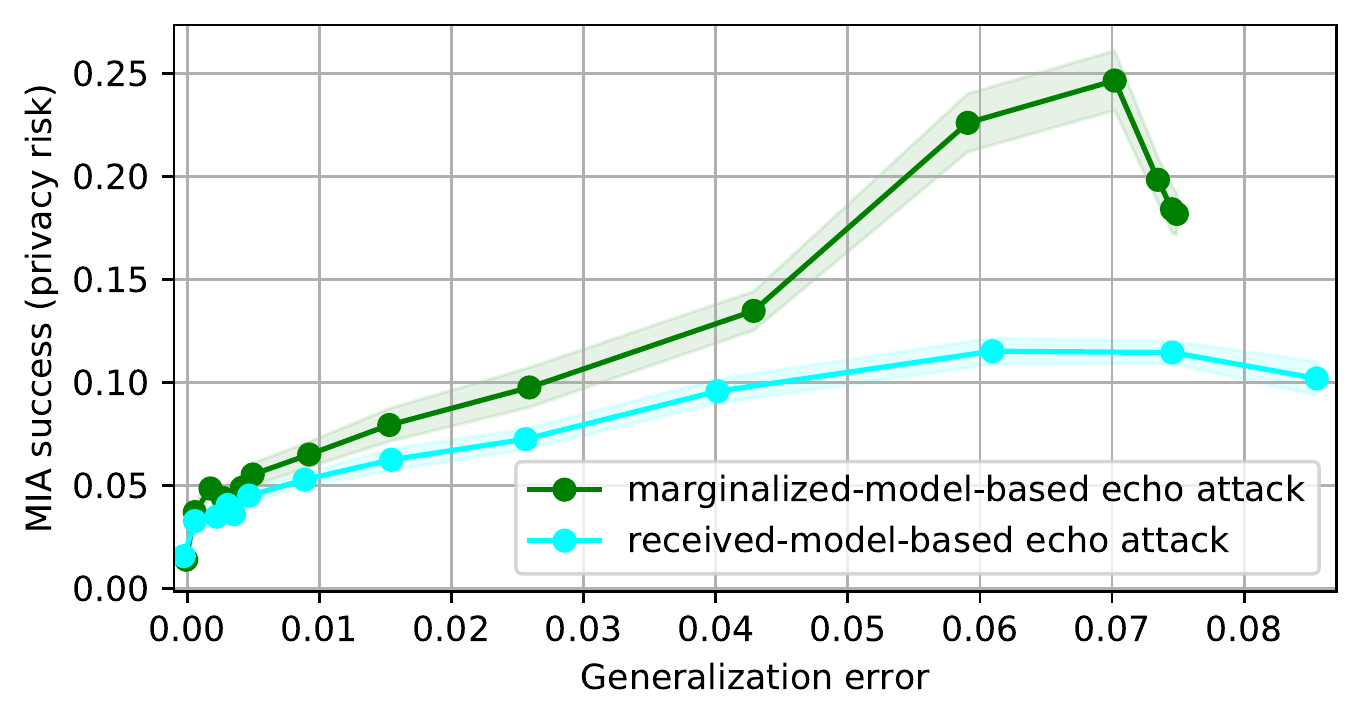}
	\caption{Comparison between echo attack based on marginalized model (green) and received model (cyan) for CIFAR-100, \textit{torus-36} and ResNet20.}
	\label{fig:echo_rec_vs_mar}
\end{figure}

\section{Details on model training}
\label{app:setup}

\begin{itemize}
	\item \textbf{Training set partition:} The training set is uniformly partitioned among users. Given a training set $X$: every user gets a disjointed sample from $X$ of size $\frac{|X|}{n}$, where $n$ is number of users in the system. No data augmentation is performed.
	\item \textbf{Optimizer:} We use SGD  with momentum ($\alpha \myeq 0.9)$.
	\item \textbf{Learning rate:} We anneal the learning rate during the training to speed up consensus for the decentralized systems. The initial learning rate is set to $0.1$, then we scale it by $0.1$ at iterations $200, 350$ and $450$ during the training. We do not schedule the learning for federated learning. 
	\item \textbf{Batch size:} $256$.
	\item \textbf{Stop condition:} We train the models with \textit{early-stopping}. We stop the training when the accuracy of the average of the local models on the validation set stops improving (with a patience of $3$).
\end{itemize}
Our code will be available upon publication.

\section{State-override attack with inexact information}
\label{app:override}
\begin{figure}
	\centering
	\includegraphics[trim = 0mm 3mm 0mm 0mm, clip, width=.6\linewidth]{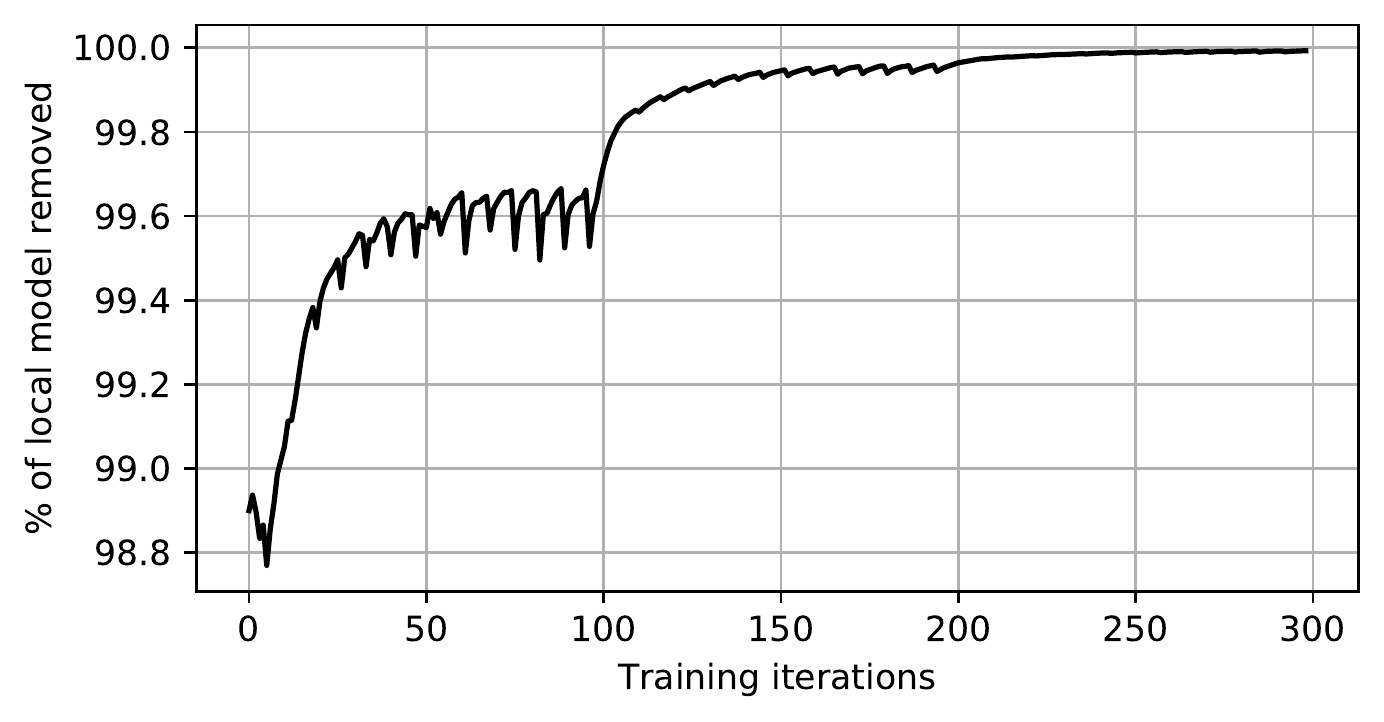}
	\caption{Performance of the state-override attack with inexact information for the topology in Figure~\ref{fig:gi_top}.}
	\label{fig:override}
\end{figure}
When the users are forced to send their updates synchronously in the decentralized protocol, an attacker will not receive the model updates of their neighbor before sending their own model update. To perform the state-override attack, the adversary then needs to rely on the model updates of the previous round. Of course, this results in an inexact suppression of the current state of the victim.
We show the result of the state-override attack using model updates received at the previous round in Figure~\ref{fig:override}. At worst, the attacker controls $98.7\%$ of the local state of the target. It approaches $100\%$ at the end of the training. 
We conclude that the state-override attack is effective even when adversaries cannot choose when to send their updates.

\section{Additional results}
\label{app:add_res}

\subsection{Shallower architectures}
\label{app:shallow}
\begin{figure}[t]
	\centering
	\includegraphics[trim=0mm 120mm 0mm 0mm, clip, width=.95\columnwidth]{./images/passive_legend}\\
	\begin{subfigure}{.4\columnwidth}
		\centering
		\includegraphics[trim = 0mm 0mm 0mm 0mm, clip, width=.95\columnwidth]{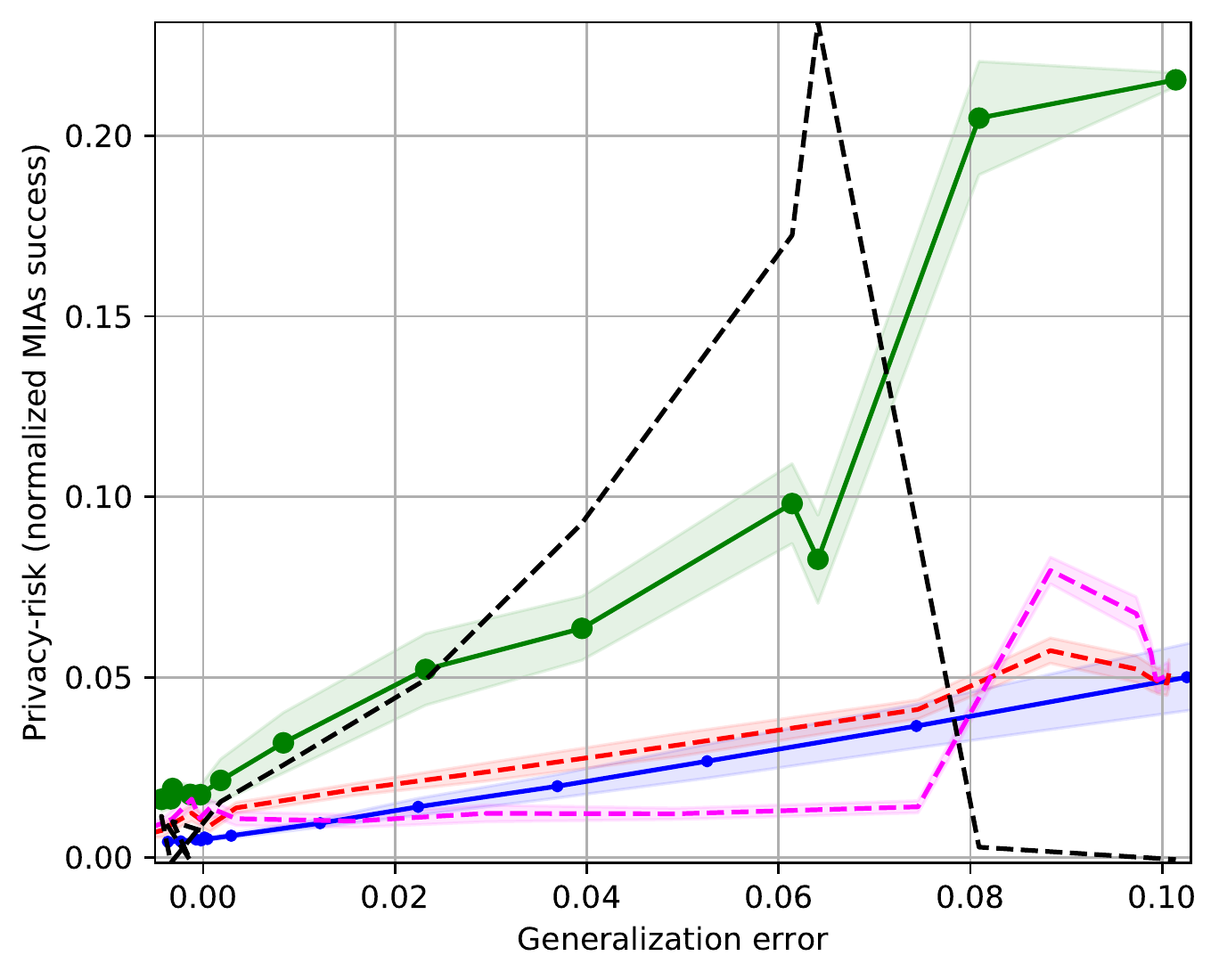}
		\caption{torus-36 \& CIFAR-10}
	\end{subfigure}\begin{subfigure}{.4\columnwidth}
		\centering
		\includegraphics[trim = 0mm 0mm 0mm 0mm, clip, width=.95\columnwidth]{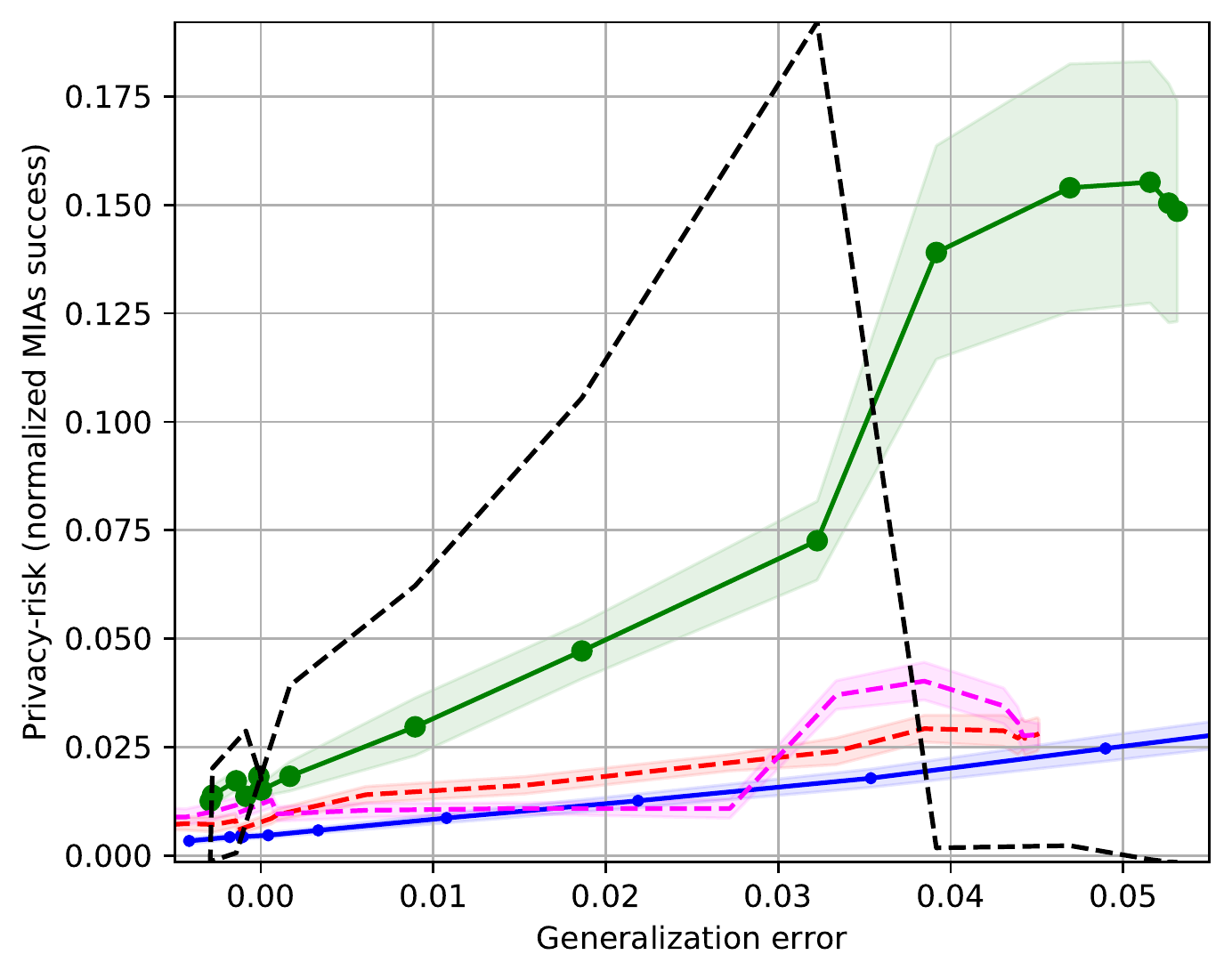}
		\caption{social-32 \& CIFAR-10}
	\end{subfigure}
	\caption{Average MIA vulnerability on two combinations of communication topologies and training sets for DL and FL when using a shallow CNN architecture.}
	\label{fig:mias_scnn_act}
\end{figure}

In Figure~\ref{fig:mias_scnn_act} we report the MIA accuracy for a shallow Convolution Neural Network (CNN) of $225,000$ parameters for both the passive and active attacks. 
While the privacy risk for the received model tends to be lower, the attacks match what was observed with the deeper ResNet20 model.

\subsection{Scaling-up number of users}
\label{app:torus64}
\begin{figure*}[t!]
	\centering
	\includegraphics[trim=0mm 100mm 0mm 0mm, clip, width=1.1\columnwidth]{./images/passive_legend}\\
	\begin{subfigure}{.5\columnwidth}
		\centering
		\includegraphics[trim = 0mm 0mm 0mm 0mm, clip, width=\columnwidth]{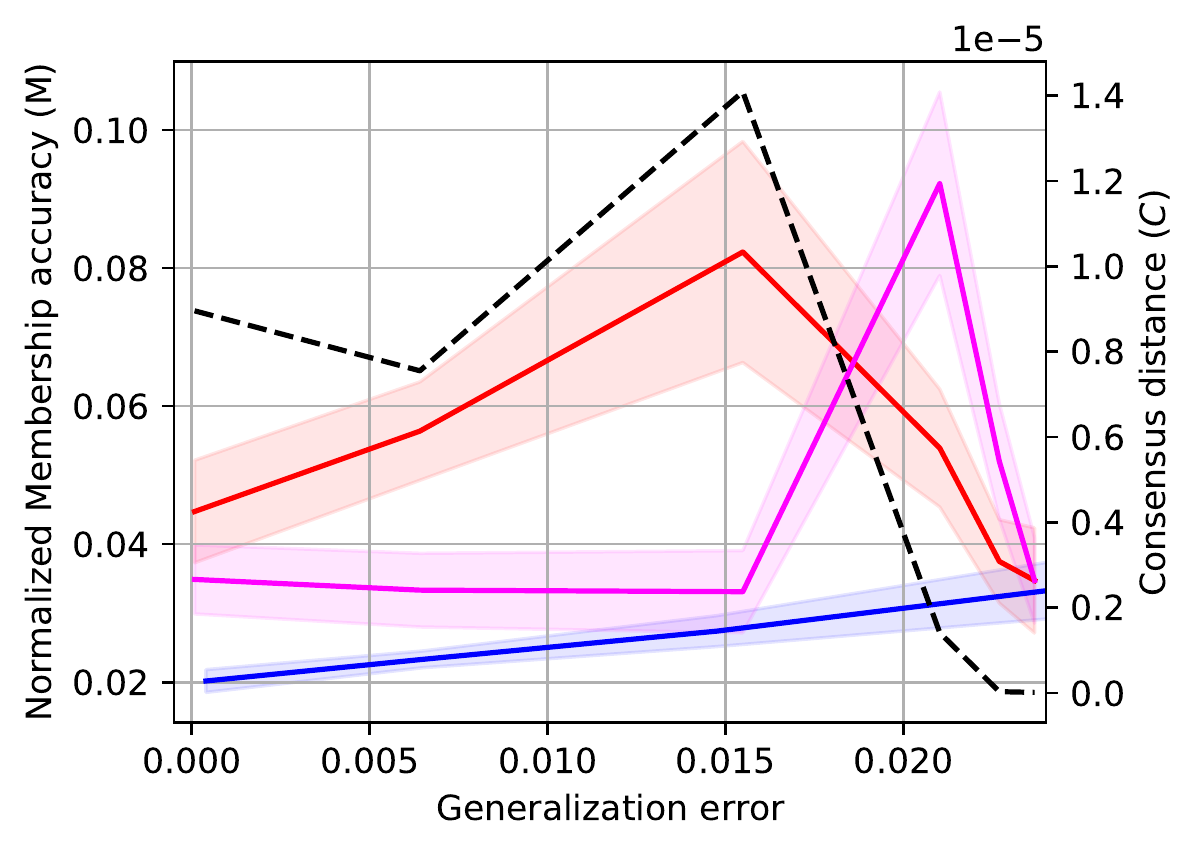}
		\caption{torus-128 \& CIFAR-10}
	\end{subfigure}\begin{subfigure}{.5\columnwidth}
		\centering
		\includegraphics[trim = 0mm 0mm 0mm 0mm, clip, width=\columnwidth]{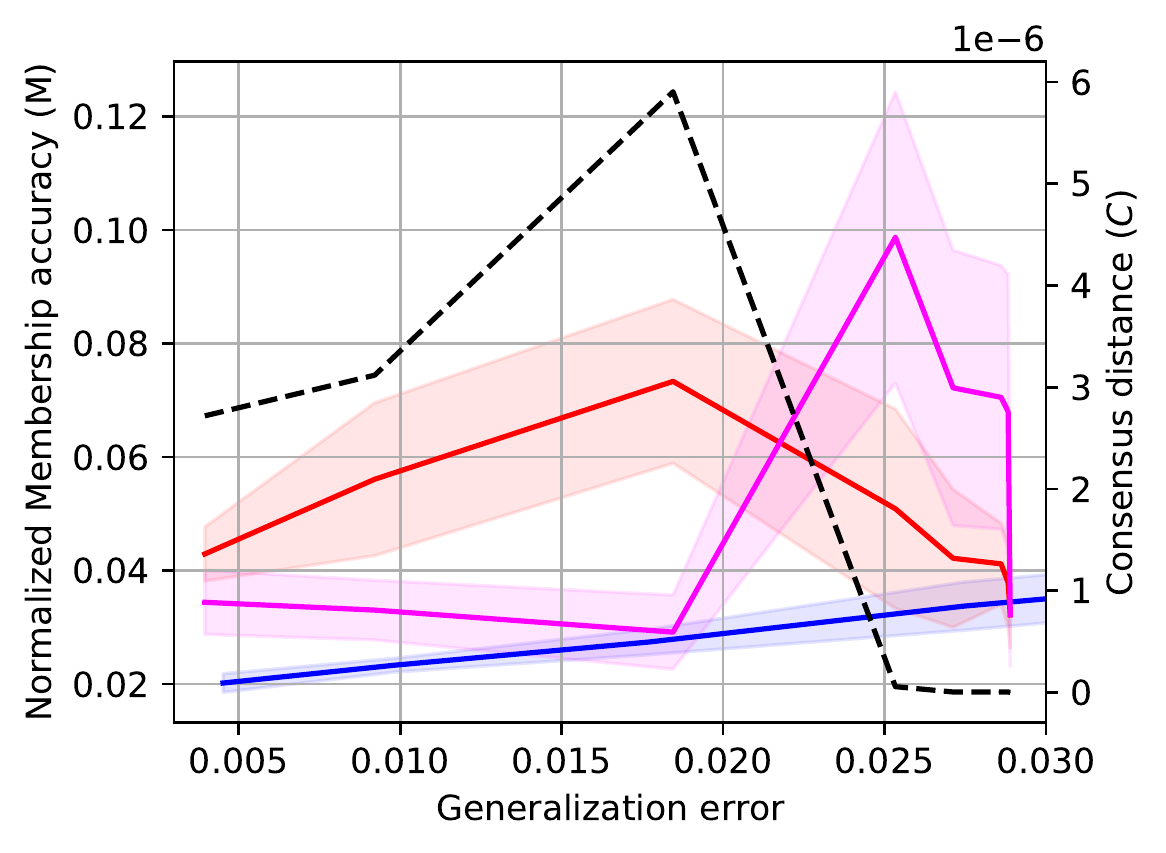}
		\caption{expander-128 \& CIFAR-10}
	\end{subfigure}\begin{subfigure}{.5\columnwidth}
		\centering
		\includegraphics[trim = 0mm 0mm 0mm 0mm, clip, width=\columnwidth]{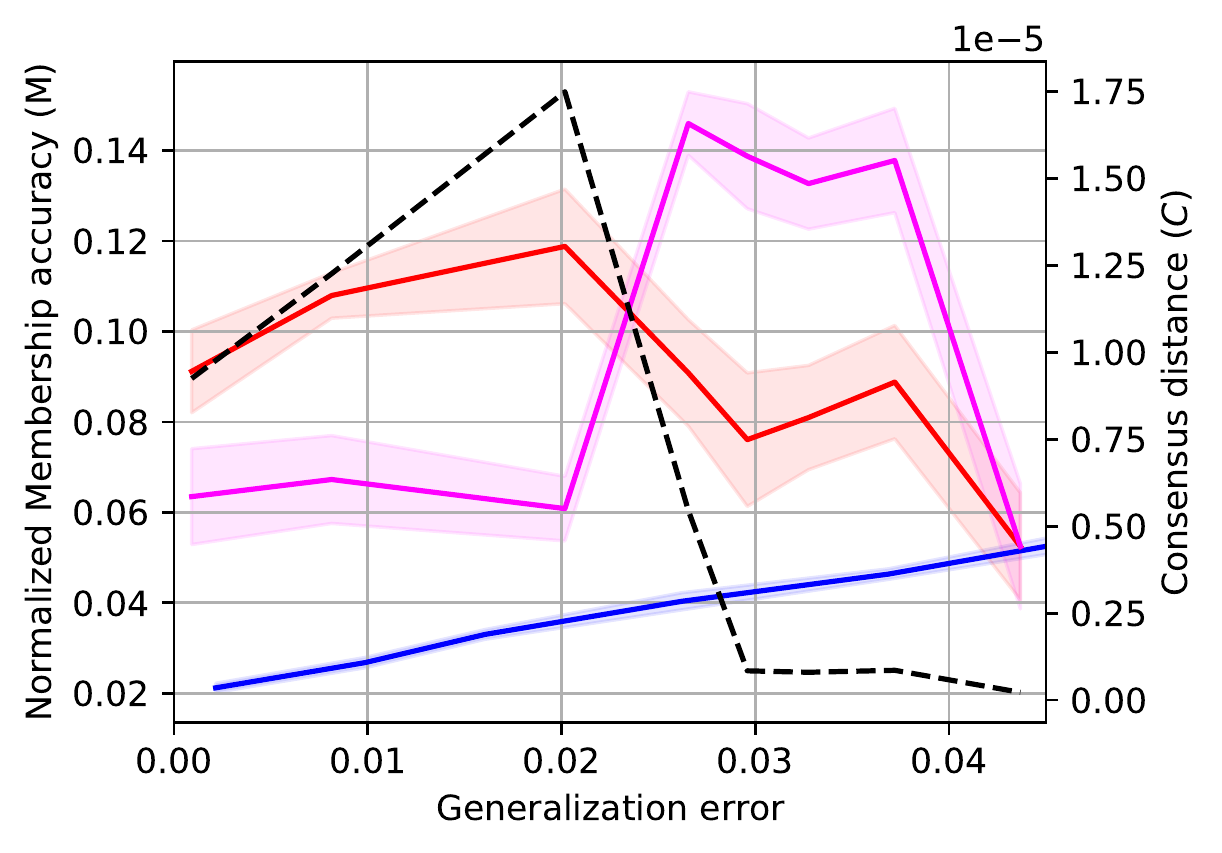}
		\caption{torus-128 \& CIFAR-100}
	\end{subfigure}\begin{subfigure}{.5\columnwidth}
		\centering
		\includegraphics[trim = 0mm 0mm 0mm 0mm, clip, width=\columnwidth]{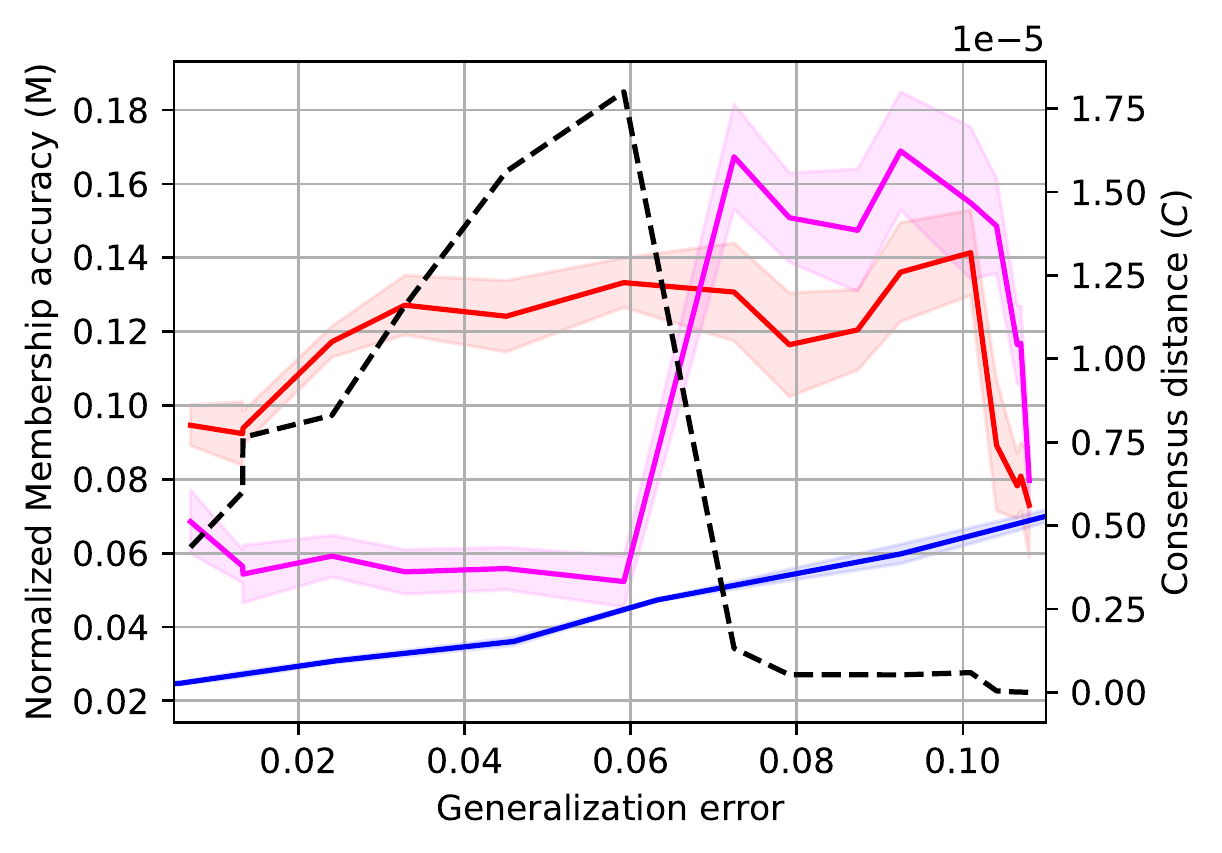}
		\caption{expander-128 \& CIFAR-100}
	\end{subfigure}
	\caption{\new{Average MIA vulnerability on four different combinations of communication topologies and datasets (DL in red and purple, and FL in blue). For each combination, we report the average results over $16$ runs.}}
	\label{fig:mias128}
\end{figure*}

\begin{figure*}[t!]
	\centering
	\includegraphics[trim=0mm 120mm 0mm 0mm, clip, width=.95\textwidth]{./images/passive_legend}\\
	\begin{subfigure}{.3\textwidth}
		\centering
		\includegraphics[trim = 0mm 0mm 0mm 0mm, clip, width=.95\textwidth]{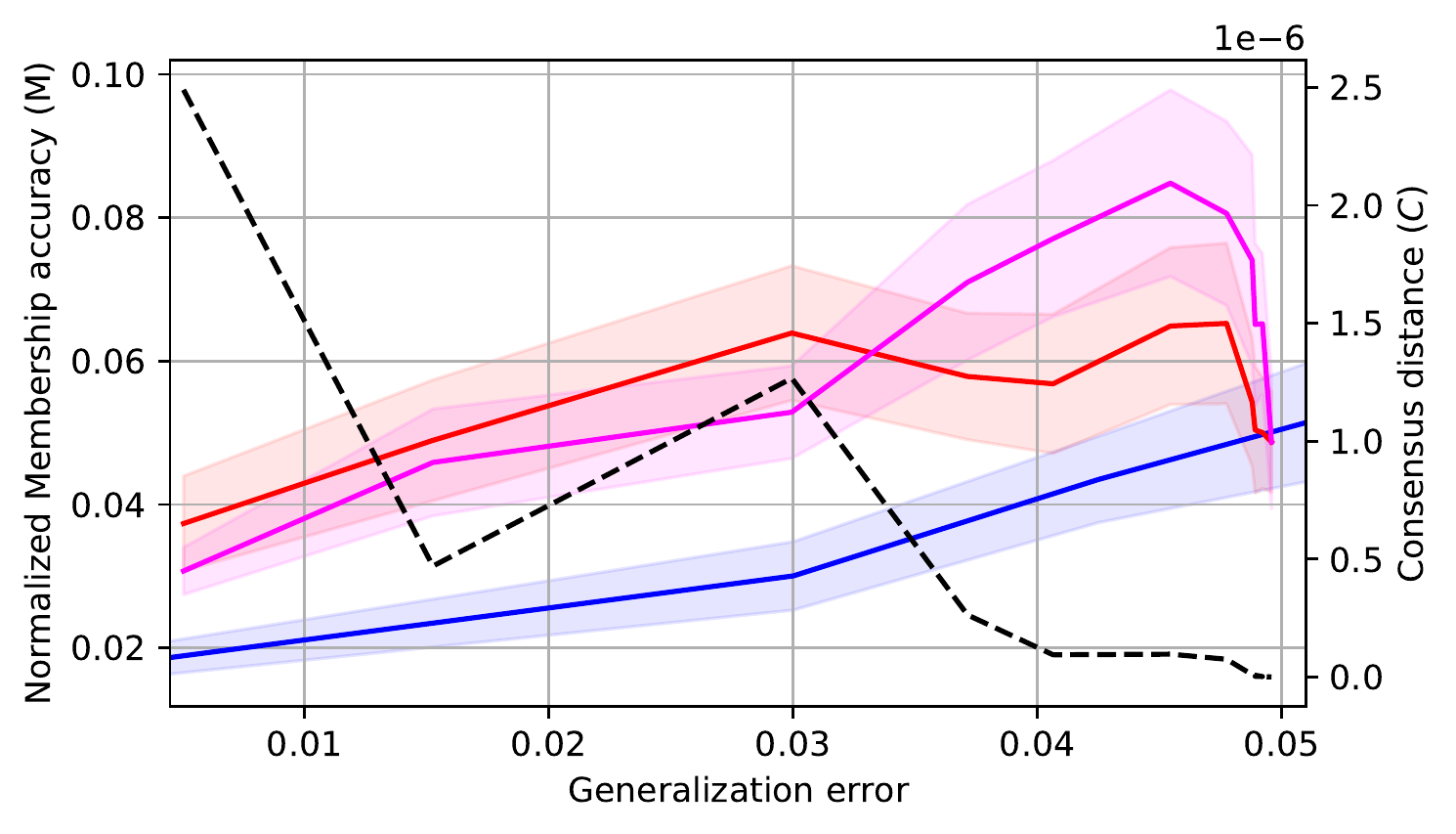}
		\caption{\footnotesize torus-36 \& imdb-reviews}
	\end{subfigure}\begin{subfigure}{.3\textwidth}
		\centering
		\includegraphics[trim = 0mm 0mm 0mm 0mm, clip, width=.95\columnwidth]{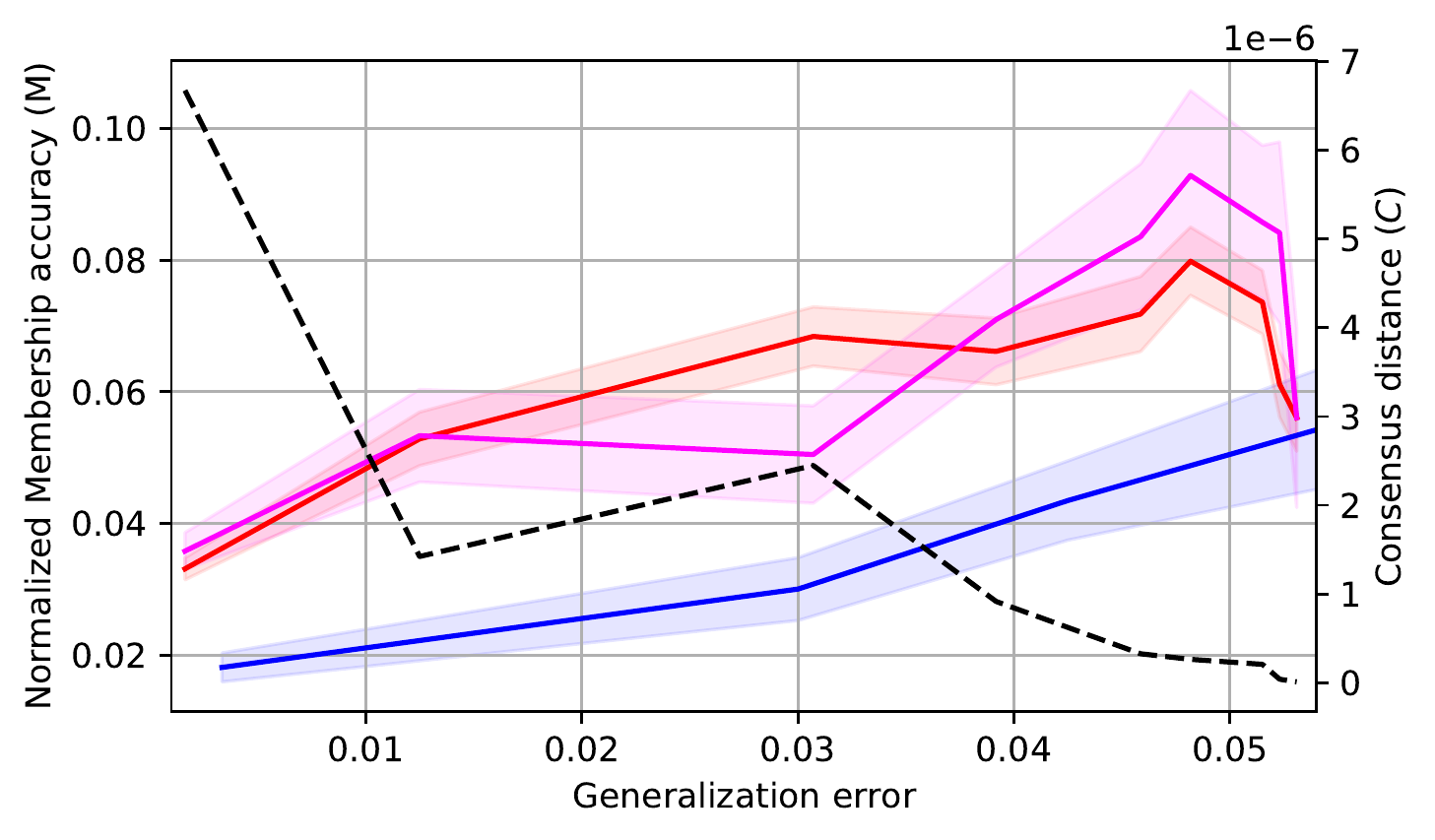}
		\caption{\footnotesize expander-36 \& imdb-reviews}
	\end{subfigure}\begin{subfigure}{.3\textwidth}
		\centering
		\includegraphics[trim = 0mm 0mm 0mm 0mm, clip, width=.95\textwidth]{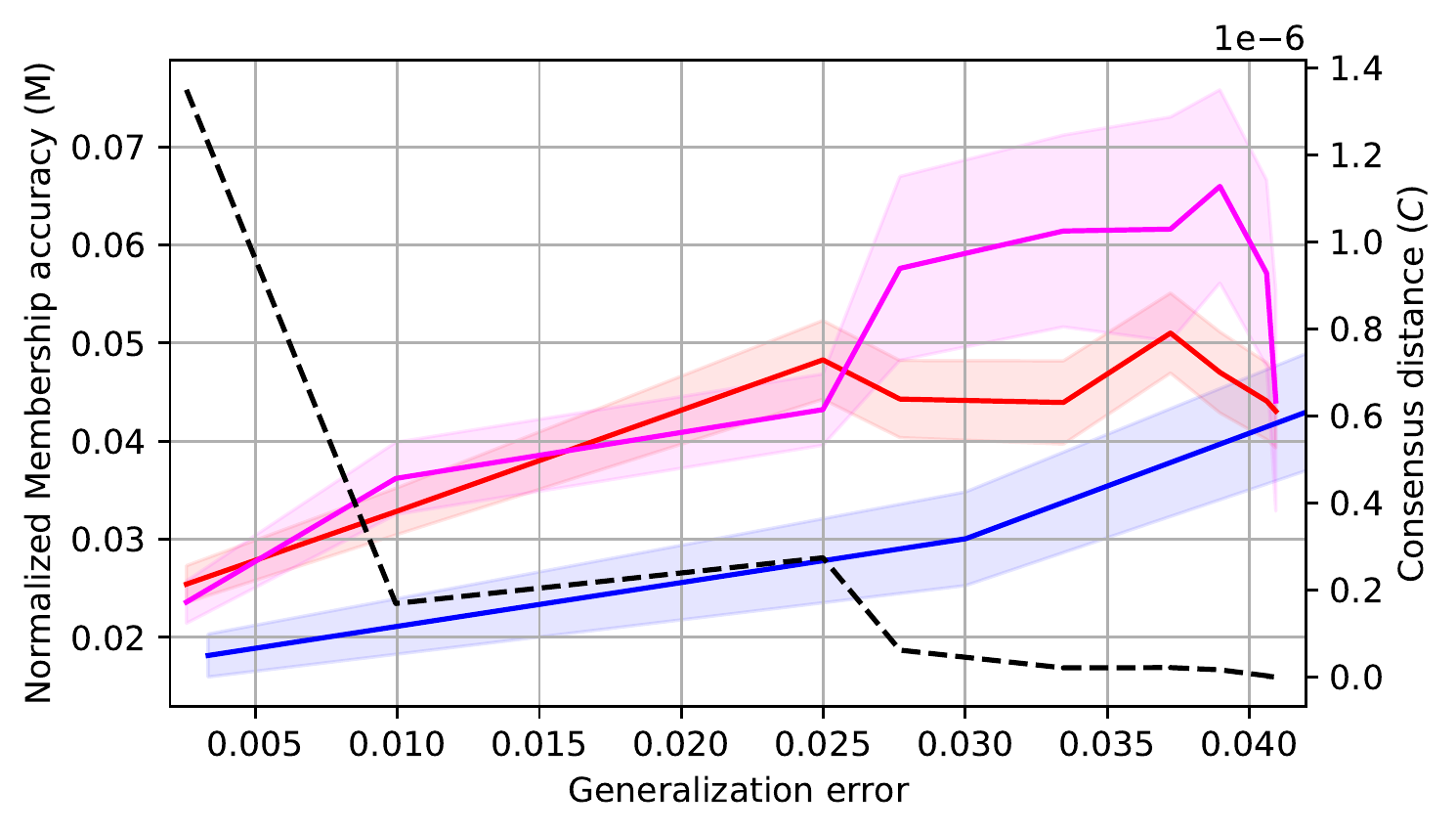}
		\caption{\footnotesize social-32 \& imdb-reviews}
	\end{subfigure}
	\caption{\new{Average MIA vulnerability on three DL setups on a text classification task.}}
	\label{fig:text}
\end{figure*}

\begin{figure}
	\centering
	\includegraphics[trim=0mm 120mm 0mm 0mm, clip, width=.95\columnwidth]{./images/passive_legend}\\
	\begin{subfigure}{.5\columnwidth}
		\centering
		\includegraphics[trim = 0mm 0mm 0mm 0mm, clip, width=.95\columnwidth]{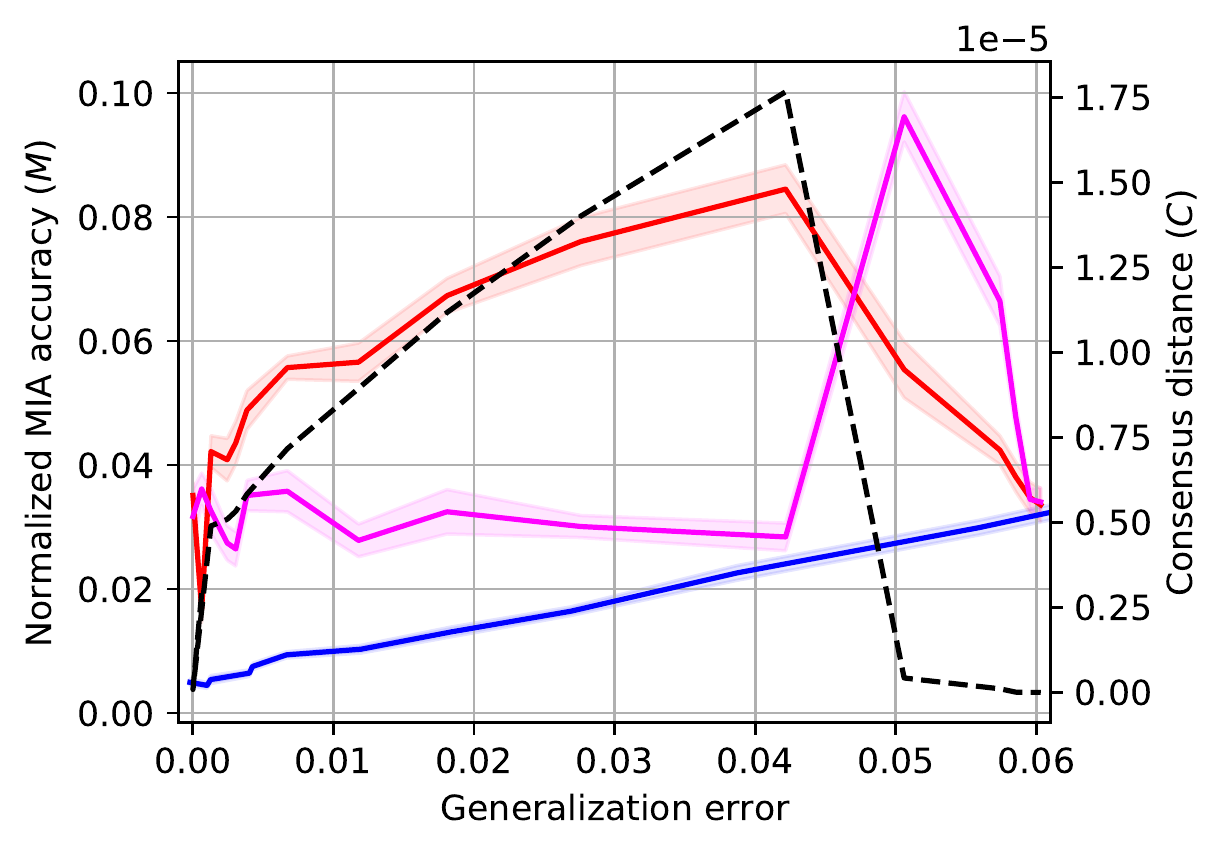}
		\caption{torus-64 \& CIFAR-10}
	\end{subfigure}\begin{subfigure}{.5\columnwidth}
		\centering
		\includegraphics[trim = 0mm 0mm 0mm 0mm, clip, width=.95\columnwidth]{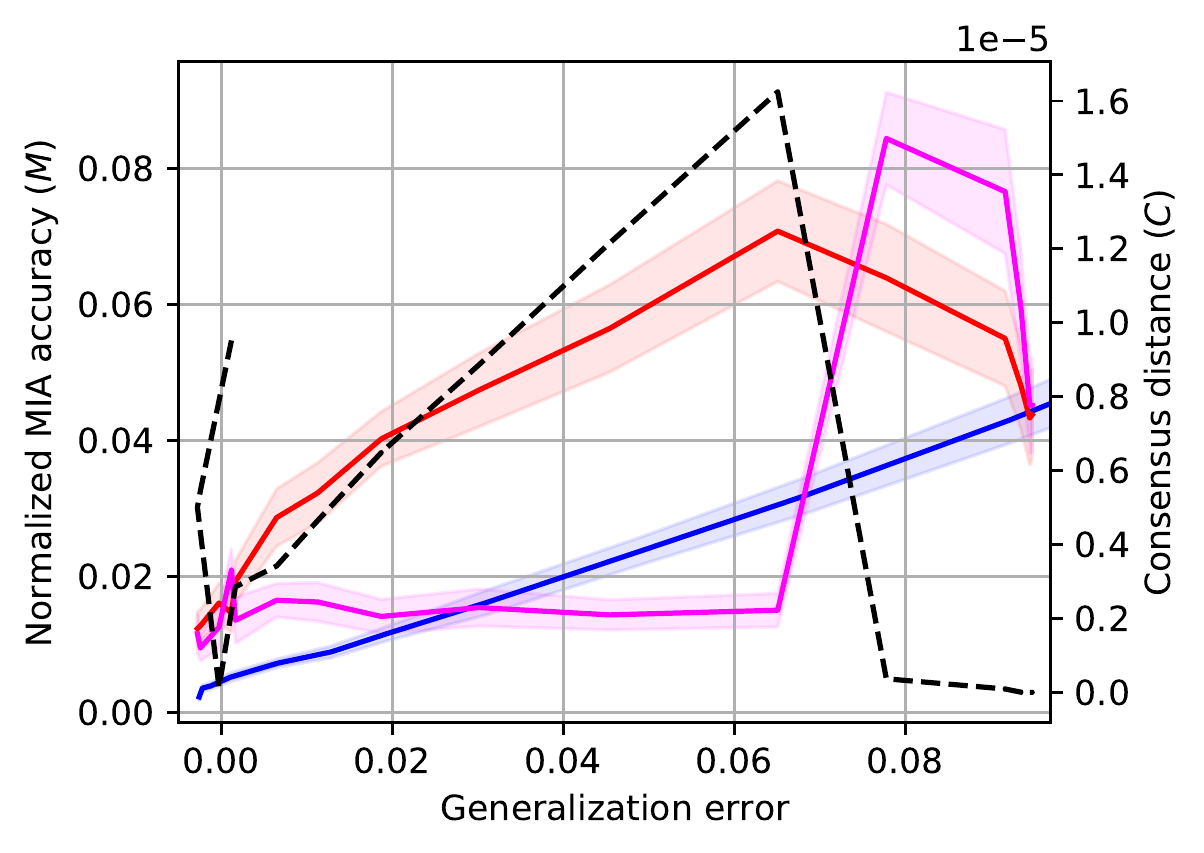}
		\caption{torus-64 \& CIFAR-100}
	\end{subfigure}
	\caption{Average MIA vulnerability on two different combinations of communication topologies and training sets for distributed and federated learning (ResNet20 architecture).}
	\label{fig:mias_torus64}
\end{figure}

In Figure~\ref{fig:mias_torus64} we report the MIA vulnerability of the model updates and their marginalized version in the \textit{torus-64} topology. Comparing these results to the \textit{torus-36} topology (Figure~\ref{fig:mias}), it is evident that the sparsity of the topology augments the vulnerability of the received model updates. 
As the number of users augments (but their number of neighbors stays 4), the average distance between users increases, which boosts the local generalization phenomenon and the inherent privacy risk associated (Section~\ref{sec:localgen_and_systemk}). This result reinforces our observation: sparse topologies reduce individual user’s privacy. To keep privacy risk constant, the topology density must adapt whenever new users join the protocol. \new{We observe similar results for \textit{torus-128} and \textit{expander-128} in Figure~\ref{fig:mias128}, where \textit{expander-128} is an \textit{Erdős–Rényi} random graph with edge probability~$\frac{log(n)}{n}$ as defined in~\cite{cyffers2022muffliato}}. 

%This fact challenges the claimed scalability property of the decentralized learning systems: growing comes at a cost.  

\subsection{Multiple local iterations}
\begin{figure}
	\centering
	\includegraphics[trim=0mm 120mm 0mm 0mm, clip, width=.95\columnwidth]{./images/passive_legend}\\
	\begin{subfigure}{.5\columnwidth}
		\centering
		\includegraphics[trim = 0mm 0mm 0mm 0mm, clip, width=.95\columnwidth]{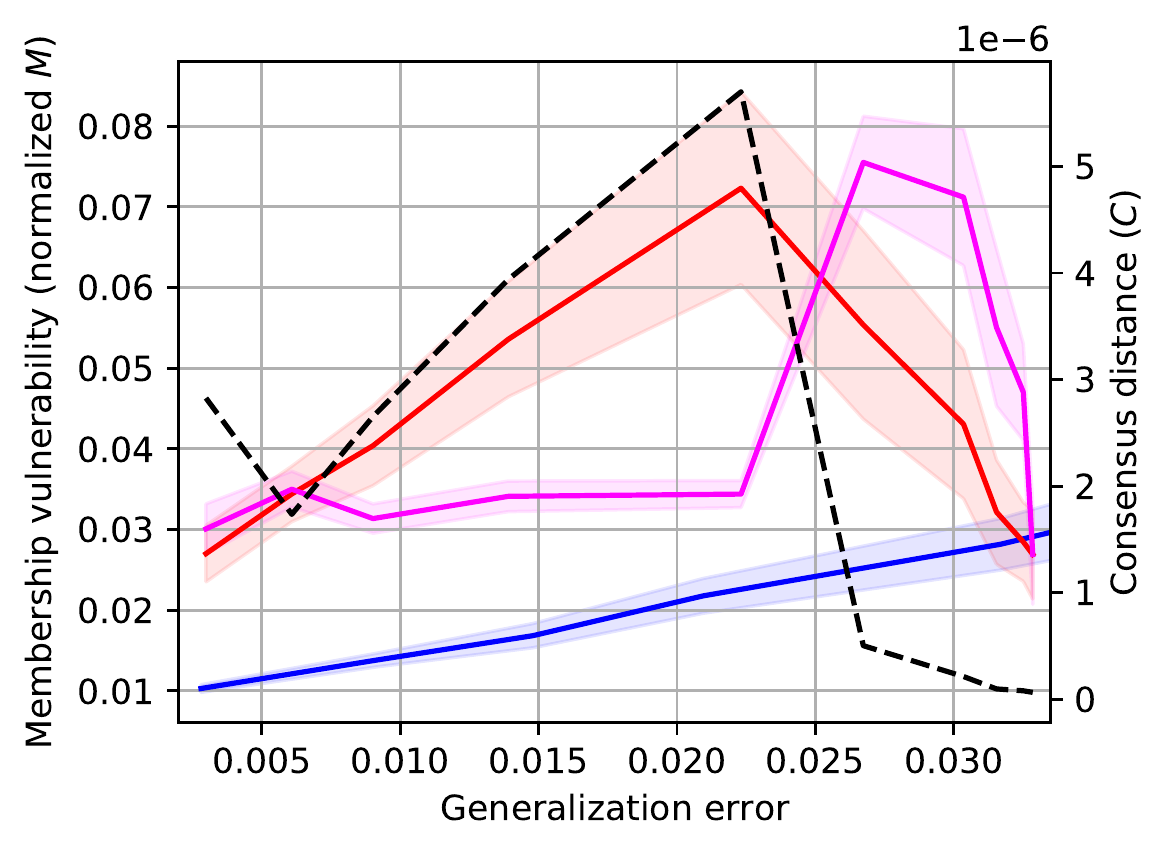}
		\caption{\footnotesize torus-36; CIFAR-10; $2$ steps}
	\end{subfigure}\begin{subfigure}{.5\columnwidth}
		\centering
		\includegraphics[trim = 0mm 0mm 0mm 0mm, clip, width=.95\columnwidth]{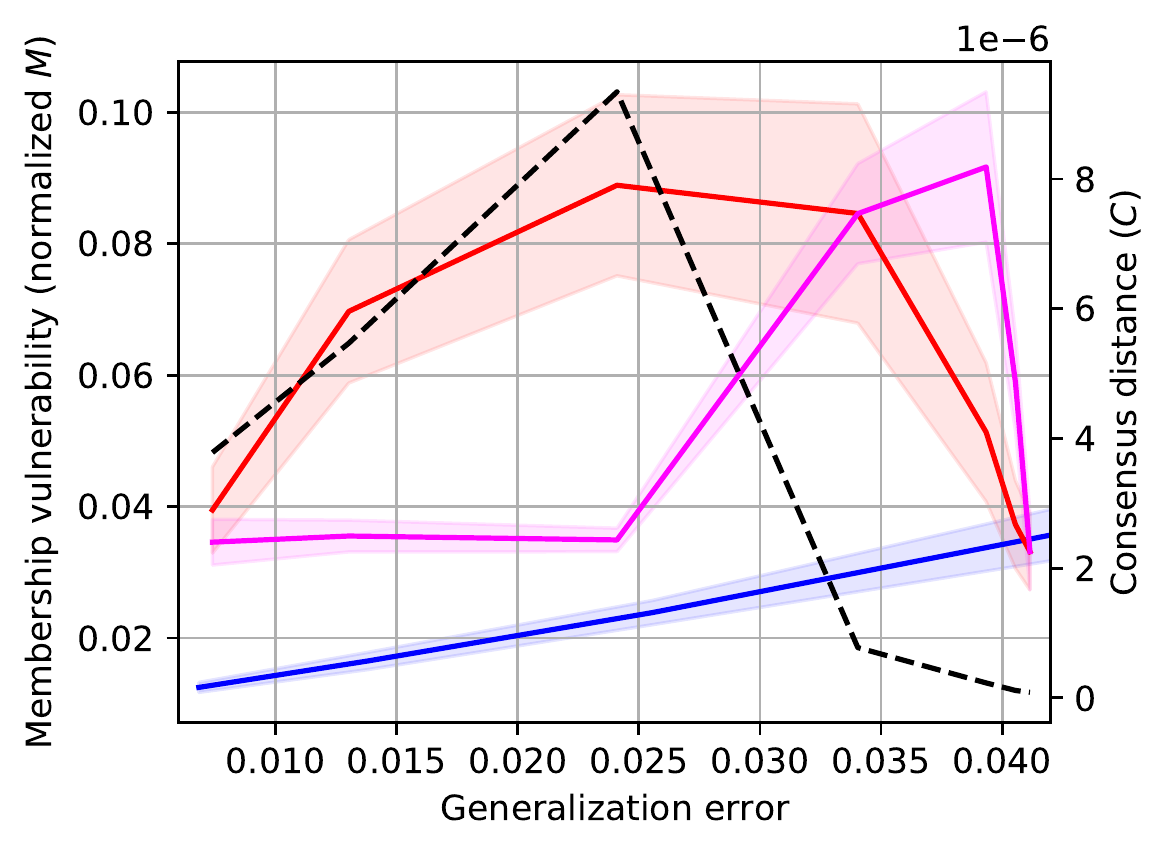}
		\caption{\footnotesize torus-36; CIFAR-10; $3$ steps}
	\end{subfigure}
	\caption{\new{Average MIA vulnerability on two DL setups with different number of local optimization steps.}}
	\label{fig:mulits}
\end{figure}

\new{We further investigate the effect of local generalization on different DL setups. In particular, we consider the impact of multiple optimization steps in the local optimization phase of clients (line~$3$ of Algorithm~\ref{alg:dl}). Results are reported in Figure~\ref{fig:mulits}, where $2$ and $3$ steps are considered for \textit{torus-36} and \textit{CIFAR-10}. Compared to the single step optimization setting (Figure~\ref{fig:mias}), running multiple optimization rounds boosts the local generalization phenomenon, magnifying the leakage inherent in the model updates. This effect is particularly marked for the MIAs based on the shared model updates (red lines).}

\subsection{Different input domains}
\label{app:nlp}
\new{In this section, we briefly explore data domains beyond the realm of vision, which is the primary focus of the decentralized learning literature. Specifically, we assess the effect of local generalization on a Natural Language Processing (NLP) task. To conduct our experiments, we utilize the widely-used \textit{imdb-reviews} dataset, which involves a movie review classification task. The experimental setup we employ is the one described in~\cite{tfimdb}. In particular, in this setting, our aim is to assess the ability of an attacker to infer the membership of an entire review (i.e., a sequence of words) in the local training set of other nodes that are part of the collaborative training process. In Figure~\ref{fig:text}, we report results for the topologies \textit{torus-36}, \textit{social-32}, and \textit{expander-36}. As evident from the results, although models exhibit varying behaviors due to the different learning task, the effectiveness of the inference attacks described in Section~\ref{sec:passive_user_vs_user} is consistent with the one observed in the vision domain.}

\end{document}